\documentclass[a4paper,11pt]{article}
\pdfoutput=1 % if your are submitting a pdflatex (i.e. if you have
\usepackage{jheppub} 
\usepackage{mathrsfs,graphicx,rotating,amsmath,amsfonts,mathtools,booktabs,amssymb,wasysym,caption}
\usepackage{hyperref}
\usepackage{slashed}
\usepackage{natbib}
\usepackage[table,xcdraw,dvipsnames]{xcolor}
\usepackage{graphicx}
\usepackage{bbold}
\usepackage[utf8x]{inputenc}
\usepackage[english]{babel}
\usepackage{multirow,multicol}
\usepackage{epstopdf}
\usepackage{bbm}
\usepackage{changepage}
\usepackage{appendix}
\usepackage{libertine}
\usepackage{braket}
\usepackage{wasysym}
\usepackage{empheq}
\usepackage{cancel}
\usepackage{enumitem}%
\usepackage{mathrsfs}
\usepackage{lmodern}
\usepackage{tabularx}
\usepackage{multicol}
\usepackage{color}
\usepackage{mathtools}
\usepackage{verbatim}
\usepackage{arydshln}
\usepackage{url}
\usepackage[normalem]{ulem}
\usepackage{youngtab}
\usepackage{float}

\newcommand{\SU}{\,{\rm SU}}

\newcommand{\U}{\,{\rm U}}
\newcommand{\GDC}{G_{DC}}
\newcommand{\GD}{G_D}
\newcommand{\GSM}{G_{SM}}

\newcommand{\SUNDC}{{\rm SU}(N_{DC})}
\newcommand{\SONDC}{{\rm SO}(N_{DC})}
\newcommand{\SUEW}{{\rm SU}(2)_{EW}}
\newcommand{\SUC}{{\rm SU}(3)_c}
\newcommand{\SUfiveGUT}{{\rm SU}(5)_{GUT}}
\newcommand{\UoneD}{{\rm U}(1)_D}
\newcommand{\UoneV}{{\rm U}(1)_{3V}}
\newcommand{\UoneB}{{\rm U}(1)_V}
\newcommand{\MPl}{M_{Pl}}
\newcommand{\LDC}{\Lambda_{DC}}
\newcommand{\GeV}{\,{\rm GeV}}
\newcommand{\MeV}{\,{\rm MeV}}

\newcommand{\NDC}{N_{DC}}
\newcommand{\CD}{\mathcal{C}_D}
\newcommand{\Gtwo}{\mathcal{G}_2}
\newcommand{\fD}{f}
\newcommand{\rhoD}{\rho}
\newcommand{\Tr}{\text{Tr}}
\newcommand{\Ocal}{{\mathcal O}}
\newcommand{\ms}{m_1}
\newcommand{\mt}{m_3}
\newcommand{\mgaD}{m_{\gamma_D}}

\title{\boldmath Composite Dark Matter from Strongly-Interacting Chiral Dynamics}

\author[a,b]{Roberto Contino,}
\author[a,b]{Alessandro Podo,}
\author[a,c]{Filippo Revello}

\affiliation[a]{Scuola Normale Superiore, Piazza dei Cavalieri 7, 56126, Pisa, Italy}
\affiliation[b]{Istituto Nazionale di Fisica Nucleare (INFN) - Sezione di Pisa \\ Polo Fibonacci Largo B. Pontecorvo, 3, I-56127 Pisa, Italy}
\affiliation[c]{Rudolf Peierls Centre for Theoretical Physics\\ Beecroft Building, Clarendon Laboratory, Parks Road, University of Oxford, OX1 3PU, UK}

\emailAdd{roberto.contino@sns.it}
\emailAdd{alessandro.podo@sns.it}
\emailAdd{filippo.revello@physics.ox.ac.uk}

\abstract{A class of chiral gauge theories is studied with accidentally-stable pseudo Nambu-Goldstone bosons playing the role of dark matter (DM). The gauge group contains a vector-like dark color factor that confines at energies larger than the electroweak scale, and a $\UoneD$ factor that remains weakly coupled and is spontaneously broken. All new scales are generated dynamically, including the DM mass, and the IR dynamics is fully calculable. We analyze minimal models of this kind with dark fermions transforming as non-trivial vector-like representations of the Standard Model (SM) gauge group. In realistic models, the DM candidate is a SM singlet and comes along with charged partners that can be discovered at high-energy colliders. The phenomenology of the lowest-lying new states is thus characterized by correlated predictions for astrophysical observations and laboratory experiments.}

\begin{document} 
\maketitle
\flushbottom

\section{Introduction}
\label{sec:intro}

In the Standard Model (SM) of particle physics the mass spectrum of the theory, with the exception of neutrinos, is explained in terms of just two fundamental scales, i.e. the scale of QCD confinement and the electroweak (EW) scale.  The QCD scale is generated through dimensional transmutation in terms of the QCD coupling and is thus dynamical. The EW scale, on the other hand, is set into the theory by hand through the Higgs mass term, and is moreover notoriously sensitive to UV corrections. This makes the SM description of the EW sector less predictive and compelling compared to the minimal and elegant characterization of the hadronic dynamics. Theories beyond the SM have been constructed -- such as Composite Higgs~\cite{Kaplan:1983fs,Banks:1984gj,Kaplan:1983sm,Georgi:1984ef,Georgi:1984af,Dugan:1984hq}, SUSY with radiative EW symmetry breaking~\cite{Ibanez:1982fr,Inoue:1982pi}, or Cosmological Relaxation~\cite{Graham:2015cka}-- where the electroweak scale is dynamically generated. The neutrino masses, which are generated by a dimension-5 operator in the SM effective theory, could be also set by a dynamical scale in a more complete theory, such as for example a Grand Unification Theory (GUT) or a theory with Dirac neutrinos.

If the Dark Matter (DM) in our Universe is made of one or more new particles, it is natural to ask whether its mass is generated dynamically
or if it entails yet another arbitrary fundamental scale. Theories where the DM mass is not dynamical (unless one considers a deeper layer of theoretical description), have been largely explored and include popular models like neutralino DM in Supersymmetry (see Ref.~\cite{Jungman:1995df} for a review) and Minimal DM models~\cite{Cirelli:2005uq}. Theories where the DM mass is dynamical, on the other hand, have been less thoroughly studied and classified.
While the possibility that the DM mass is set by the QCD or EW scales seems disfavored by data, it is still possible  --and in fact plausible-- that the GUT scale or the dynamical scale from a new strongly-coupled gauge theory may play this role. Strongly-coupled gauge theories can lead to a variety of DM candidates and have been intensively studied in the literature.
The new strongly-coupled dynamics may be at the origin of electroweak symmetry breaking, as in Technicolor and Composite Higgs theories, or be comprised in a `dark' or `mirror' sector. Recent results along these directions are reviewed in Refs.~\cite{Kribs:2016cew} and~\cite{Berezhiani:2003xm,Ciarcelluti:2010zz,Foot:2014mia}, see also Ref.~\cite{Chacko:2018vss} and references therein. 
Most of the work on dark sectors has focused on vectorlike confining theories~\cite{Kilic:2009mi}, where the dynamical scale is generated through dimensional transmutation. Their infrared (IR) behavior can be inferred from that of QCD, and their chiral symmetry breaking pattern preserves the SM group, in accord with the Vafa-Witten theorem~\cite{Vafa:1983tf} and vacuum alignment arguments~\cite{Peskin:1980gc,Preskill:1980mz}.  A classification of vectorlike dark sector models with fermions charged under the SM gauge group was presented in Ref.~\cite{Antipin:2015xia} (see also Refs.~\cite{Kilic:2009mi,Bai:2010qg,Buckley:2012ky,Appelquist:2015yfa} for previous studies of models in the same class),
where it was shown that many of them satisfy the current bounds and may lead to distinctive experimental signatures. 

In this paper we focus on confining, chiral dark gauge theories, in particular those with no fundamental dark scalar fields. 
Models of this kind with DM candidates appeared for example in Refs.~\cite{Harigaya:2016rwr,Co:2016akw,Hertzberg:2019bvt}.
In this class of theories the masses of all the dark states are generated dynamically, since no fermion mass term is allowed by gauge invariance. The same is not true in vectorlike theories, since in that case bare mass terms can be larger than the dynamical scale and determine the masses of some of the physical states, possibly including the DM candidate~\cite{Falkowski:2009yz,Mitridate:2017oky,Contino:2018crt}. Chiral theories are more constrained in general, hence less simple to construct, than vectorlike theories. The cancellation of gauge anomalies, for example, is a non-trivial requirement that significantly restricts the possible models. Interesting studies in this direction include techniques for finding anomaly-free sets of fermions in theories with a $U(1)$ chiral factor~\cite{Batra:2005rh,Costa:2019zzy,Allanach:2019gwp}, and a method to construct chiral theories starting from irreducible representations of a simple, anomaly-free gauge group~\cite{Berryman:2016rot}. 
Chiral theories are also notoriously difficult to simulate on the lattice~\cite{DeGrand:2019boy,Luscher:2000hn}, and their IR behavior is still not known for simple gauge groups. In that case bilinear fermion condensates cannot be singlets under both the gauge and the Lorentz group. It has been speculated that the theory may `tumble' into a Higgs phase with reduced gauge group~\cite{Raby:1979my} or even break spontaneously Lorentz invariance~\cite{Bais:1980wa}, although a reasonable possibility is that no bilinear fermion condensate forms at all at the non-perturbative level.

Lacking a clear theoretical understanding of chiral theories with simple groups, in this work we consider models where the dark gauge group is the product of a simple, vectorlike factor $\GDC$ (dark color), which gets strong and confines at a scale $\LDC$ larger than the EW scale, times a weak factor~$\GD$. In the context of these theories, we will specifically investigate those where the dark quarks are charged under the SM gauge group $\GSM = SU(3)_c \times SU(2)_{EW}\times U(1)_Y$, while the SM fermions are neutral under $\GDC\times\GD$. The full gauge group $\GDC\times\GD\times\GSM$ is assumed to be chiral. To avoid large corrections to the electroweak precision observables, we will require that the dark quarks transform as vectorlike representations of $\GDC\times\GSM$. This implies that, depending on whether $\GDC\times\GD$ is chiral and on the vacuum alignment, the dark condensate can break $\GD$ spontaneously and preserve~$\GSM$~\cite{Peskin:1980gc,Preskill:1980mz,toappear}. In absence of Yukawa couplings between the Higgs field and the dark quarks, this ensures negligibly small corrections to the electroweak precision observables, as in theories of vectorlike confinement~\cite{Kilic:2009mi}.
Having dark quarks charged under the SM may imply a further constraint if the theory is extended to include unification of the SM gauge forces into a simple group $G_{GUT}$. Assuming that the dark group $G_{DC} \times G_D$ commutes with $G_{GUT}$ and is not broken by the GUT dynamics, dark quarks will have to come in complete representations of  $G_{GUT}$, since no mass term is allowed to make some of them massive, differently from vectorlike theories.\footnote{A possible exception would be given by GUT multiplets where some of the SM components are chiral while others are vectorlike and can thus become massive below the GUT scale. In practice, we have found no realistic candidate of this kind in light of the constraints imposed by perturbativity and the requirement of DM neutrality discussed below.} Another difference between vectorlike and chiral gauge theories will be discussed at length in this work and concerns the accidental stability of the DM candidate: while in vectorlike theories dark baryons are generically more stable than dark pions~\cite{Antipin:2015xia}, the same is not true for chiral theories, where the operators responsible for the dark pion decay can easily have a very large dimension. 

Chiral theories with a product gauge group $\GDC \times \GD \times \GSM$ can be classified according to the number of irreducible representations of Weyl fermions. For a non-abelian subgroup~$\GD$, candidate theories can be constructed with just one or two representations. They will be studied in a forthcoming work. Here we focus on models with an abelian subgroup $\GD = U(1)_D$; in that case the minimum number of representations to have a chiral theory is 4. The three possible types of minimal theories are defined in Table~\ref{tab:minimal_theories}, where $R$ and $r$ are (generally complex) irreducible representations of respectively $\GDC$ and $\GSM$, and $a$ is a rational number.
%
%%%%%%%%%%
\begin{table}[t]
\centering
{\large  Type I}\\[0.2cm]
\begin{tabular}{lccc|cc}
       & $\GDC$ & $\UoneD$ & $\GSM$ & $\UoneV$ & $\UoneB$ \\
\cline{2-6}
$\psi_1$ & $R$    &  $+1$  & $r$ & $+1$  & $+1$ \\
$\psi_2$ & $R$    &  $-1$ & $r$ & $-1$ & $+1$ \\         
$\chi_1$ & $\bar R$ &  $-a$  & $\bar{r}$ & $-1$  & $-1$ \\        
$\chi_2$ & $\bar R$ &  $+a$ & $\bar{r}$ & $+1$  & $-1$ \\ 
\end{tabular}
\\[0.5cm]
{\large  Type II}\\[0.2cm]
\begin{tabular}{lccc|cc}
       & $\GDC$ & $\UoneD$ & $\GSM$ & $\UoneV$  & $\UoneB$ \\
\cline{2-6}
$\psi_1$ & $R$    &  $+1$  & $r$ & $+1$ & $+1$ \\
$\psi_2$ & $R$    &  $-1$ & $\bar r$ & $-1$ & $+1$ \\         
$\chi_1$ & $\bar{R}$ &  $-a$  & $\bar{r}$ & $-1$ & $-1$  \\        
$\chi_2$ & $\bar{R}$ &  $+a$ & $r$ & $+1$ & $-1$ \\ 
\end{tabular}
\\[0.5cm]
{\large  Type III}\\[0.2cm]
\begin{tabular}{lccc|cc}
       & $\GDC$ & $\UoneD$ & $\GSM$ & $\UoneV$ & $\UoneB$ \\
%\hline
\cline{2-6}
$\psi_1$ & $R$    &  $+1$  & $r$ & $+1$ & $+1$ \\
$\psi_2$ & $R$    &  $+a$ & $\bar r$ & $-1$ & $+1$  \\         
$\chi_1$ & $\bar{R}$ &  $-a$  & $\bar{r}$ & $-1$ & $-1$ \\        
$\chi_2$ & $\bar{R}$ &  $-1$ & $r$ & $+1$ & $-1$ \\ 
\end{tabular}
\caption{The three minimal classes of chiral theories with 4 irreducible representations of the gauge group $\GDC\times\UoneD\times\GSM$.
All fields are left-handed Weyl fermions. Charges under the accidental global $\UoneV\times\UoneB$ are indicated in the last column, separated by a vertical line. The parameter $a$ is an arbitrary rational number in the interval $[-1,1)$ for Type II and III theories, and in the interval $[0,1)$ for Type I theories.}
\label{tab:minimal_theories}
\end{table}
%%%%%%%%%%
The case with fermions transforming as singlets under $\GSM$ ($r = 1$) and as fundamental representations of $\GDC = \SUNDC$ ($R=F$) was studied in Refs.~\cite{Harigaya:2016rwr,Co:2016akw}. There the dark color gauge group confines in the infrared forming three dark pions, according to the pattern of global symmetry breaking $SU(2)\times SU(2) \to SU(2)_V$. The $U(1)_D$ is spontaneously broken and one of the dark pions is eaten to form a massive dark photon. The residual global symmetry is $\UoneV\times\UoneB$, where $\UoneB$ is the dark baryon number and $\UoneV \subset SU(2)_V$ is an additional vectorial factor. The two uneaten dark pions have $\UoneV$ charge $\pm 2$ and are thus accidentally stable, providing a viable DM candidate. Interactions with the Standard Model sector are achieved through a kinetic mixing between the $\UoneD$ gauge field and hypercharge~\cite{Holdom:1985ag}.

The purpose of this work is to study the theories of Table~\ref{tab:minimal_theories} with non-trivial SM representations $r$. Compared to the analysis of  Refs.~\cite{Harigaya:2016rwr,Co:2016akw}, these models are characterized by less suppressed interactions between the dark and SM sectors, and lead to experimental signatures that are more easily testable. In particular, even though the DM candidate is a singlet under the SM, it has charged partners that can be discovered at colliders. 

The paper is organized as follows. Section~\ref{sec:minimalmodels} makes an analysis of the minimal models of Tab.~\ref{tab:minimal_theories} and identifies those that are fully realistic. In these theories $r$ is a fundamental of $\SUEW$ or a fundamental of $\SUC$. The rest of the paper focuses on the model with $\SUEW$ doublets. Its Lagrangian and symmetries are discussed in Section~\ref{sec:themodel}, the spectrum and the lifetimes of the lowest-lying states are analyzed in Section~\ref{sec:phenomenology}, while its cosmological history is studied in Section~\ref{sec:cosmology}. Sections~\ref{sec:DMsearches} and~\ref{sec:coll} discuss the constraints set by direct and indirect DM searches, and by collider data respectively. We present our summary and outlook in Section~\ref{sec:summary}. 
Finally, the content of the Appendices is as follows: \ref{app:darkpions}~reports useful formulas on dark pions used in the text;~\ref{app:modelwithtriplets} gives a short description of the model with $\SUC$ triplets;~\ref{app:boltzmann} includes a detailed analysis of the Boltzmann equations relevant for the evolution of the dark sector during freeze out;~\ref{app:gammad} discusses some aspects of the phenomenology of dark photons.

\section{Analysis of minimal models}
\label{sec:minimalmodels}

The minimal models of Table~\ref{tab:minimal_theories} are all free of gauge anomalies except when $r$ is an irreducible (non-trivial) representation of hypercharge, in which case Type I and III have $U(1)_Y [U(1)_D]^2$ anomalies, and the only anomaly-free choice is Type~II.~\footnote{For $a=-1$ Type III theories with hypercharge are also anomaly free, since in this limit they are equivalent to Type II.}  They have fermion representations that are chiral under $\GDC\times\UoneD\times\GSM$ and vectorlike under $\GDC\times\GSM$. For $R$ complex, representations under $\GDC\times\UoneD$ are chiral in Type~I and II theories, and vectorlike in Type III. The physical domain of the parameter $a$ can be restricted to the interval $[0,1)$ for Type I and to the interval $[-1,1)$ for Type II and III.~\footnote{ One can always redefine $a\to 1/a$ by rescaling the $\UoneD$ charge and relabeling $\psi_i \leftrightarrow \chi_i$. In Type I theories it is possible to further restrict $a$ to $[0,1)$ by relabeling $\chi_1 \leftrightarrow \chi_2$.}  If $R$ is real or pseudoreal, then Type II theories are overall vectorlike and therefore not interesting for our purposes; Type I and III instead become physically equivalent and remain chiral as long as $r$ is complex. Similarly, if $r$ is real or pseudoreal then Type III theories are overall vectorlike and not interesting; Type I and II instead become physically equivalent and remain chiral if $R$ is complex.

The choices for the representations $R$ and $r$ can be restricted by requiring that the dark color group $\GDC$ is asymptotically free (as needed to have confinement), and that the SM gauge couplings remain perturbative (with no Landau poles) until the Planck scale. The corresponding conditions are:
\begin{align}
\label{eq:AF}
& T(R) \dim(r) < \frac{11}{8} C_2(Adj) \\
\label{eq:LandauPoles}
& T(r) \dim(R) \leq \frac{3}{8} \left(\log\frac{\MPl}{\LDC}\right)^{-1} \left[ \frac{2\pi}{\alpha_i(m_Z)} + b^{SM}_i \log\frac{\MPl}{m_Z} \right] , 
\qquad i = 1,2,3\, , 
\end{align}
where $\LDC$ is the dark confinement scale, while $T({\mathcal R})$, $\dim({\mathcal R})$ and $C_2({\mathcal R})$ are respectively the Dynkin index, the dimension and the quadratic Casimir of a representation $\mathcal R$. For the dark color gauge group to confine, one needs also to require that the respective number of Weyl flavors ($n_f$) is below the lower end of the conformal window ($n_f^c$); this adds the following condition
\begin{equation}
\label{eq:confinement}
n_f \equiv 4 \dim(r) < n_f^c\, .
\end{equation}
Focusing on the case of irreducible representations $r$, Eqs.~(\ref{eq:AF}) and (\ref{eq:LandauPoles}) can be satisfied only if $r$ is a fundamental of one of the SM simple group factors. Furthermore, the only possible dark color representations turn out to be:
\begin{itemize}
\item $R =$ fundamental of $\SUNDC$ for $2 \leq N_{DC} \leq 6$ 
\item $R =$ fundamental of $\SONDC$ for $4 \leq N_{DC} \leq 6$ ($N_{DC} = 4$ only if $\dim(r) \leq 2$)
\item $R =$ fundamental of ${\rm USp}(4)$ or ${\rm USp}(6)$.
\end{itemize}
In those cases where an estimate of $n_f^c$ is available from lattice simulations, Eq.~(\ref{eq:confinement}) is satisfied and does not impose further restrictions.

In absence of the weak $\UoneD\times\GSM$ gauging, the pattern of global symmetries of the theories listed above is constrained by the Vafa-Witten theorem, which ensures that the vectorial subgroup is linearly realized~\cite{Vafa:1983tf,Kosower:1984aw}. Assuming maximal symmetry breaking~\cite{Peskin:1980gc,Preskill:1980mz}, one has the following three patterns and corresponding numbers of Nambu-Goldstone bosons (NGBs):
\begin{itemize}
\item ${\rm SU}(2\dim(r))_L \times {\rm SU}(2\dim(r))_R \times \UoneB \to {\rm SU}(2\dim(r))_V \times \UoneB$ \\[0.1cm]
in $\text{SU}(N_{DC}>2)$ dark color theories with vectorlike representations \\[0.1cm]
{\small \#} NGBs = $4\dim^2(r) -1$
\item ${\rm SU}(4\dim(r))_R \to {\rm SO}(4\dim(r))$ \\[0.1cm]
in $\SONDC$ dark color theories with real representations  \\[0.1cm]
{\small \#} NGBs = $(4\dim(r) -1)(2\dim(r)+1)$
\item ${\rm SU}(4\dim(r))_R \to {\rm USp}(4\dim(r))$ \\[0.1cm]
in ${\rm USp}(4)$, ${\rm USp}(6)$ and $\text{SU}(2)$ dark color theories with pseudoreal representations \\[0.1cm]
{\small \#} NGBs = $(4\dim(r) +1)(2\dim(r)-1)$.
\end{itemize}
The weak gauging reduces this global invariance to a subgroup and determines the vacuum alignment.~\footnote{A thorough analysis of the vacuum alignment in theories with a weak chiral gauging will be reported in a forthcoming work~\cite{toappear}.}
 In Type I theories with $R$ complex there exists no bilinear dark condensate that leaves $\UoneD$ unbroken, and (up to a field redefinition) the vacuum aligns in the SM-preserving direction with $\langle \psi_1 \chi_1\rangle = \langle \psi_2 \chi_2\rangle \not = 0$. In the other theories the vacuum also preserves $\GSM$ and has the same orientation, at least for values of the dark coupling $e_D$ smaller than a certain critical value $e_D^c$. Assuming $e_D < e_D^c$ and focusing on irreducible representations $r$, the residual global symmetry is 
\begin{equation}
\label{eq:globalwithweakgauging}
{\rm U}(1)_{3L}\times {\rm U}(1)_{3R} \times \UoneB = \UoneD \times \UoneV \times \UoneB\,  \to \UoneV \times \UoneB\, ,
\end{equation}
where ${\rm U}(1)_{3L}$ (${\rm U}(1)_{3R}$) acts on the relative phase between $\psi_1$ and $\psi_2$ ($\chi_1$ and $\chi_2$), and the charges under the accidental vectorlike subgroup $\UoneV \times \UoneB$ are shown in Tab.~\ref{tab:minimal_theories}. Notice that while $\UoneV$ is exact both at the classical and quantum level, the dark baryon number $\UoneB$ is anomalous with respect to $\UoneD$. The NGB implied by the symmetry breaking in Eq.~(\ref{eq:globalwithweakgauging}) is eaten to make the dark photon massive, and there remain no massless NGBs.

The quantum numbers of all the NGBs under $\GSM \times \UoneV \times \UoneB$ are summarized in Table~\ref{tab:NGBs}. 
%
%%%%%%%%%%%%%%%%%%%%%%%%%
\begin{table}[t]
\centering
\begin{tabular}{lccccc}
& \multicolumn{3}{c}{$R$ complex} & & \\
& Type I & Type II & Type III & $\UoneV$ & $\UoneB$ \\
\hline
$\psi_1\chi_1 + \psi_2\chi_2$ & Adj & Adj & Adj & 0 & 0 \\
$\psi_1\chi_1 - \psi_2\chi_2$ & 1+Adj & 1+Adj & 1+Adj & 0 & 0 \\
$\psi_1 \chi_2$ & 1+Adj & S+A & S+A & $+2$ & 0 \\
$\psi_2 \chi_1$ & 1+Adj & $\bar{\rm S}$+$\bar{\rm A}$ & $\bar{\rm S}$+$\bar{\rm A}$ & $-2$ & 0
\end{tabular}
\\[0.8cm]
\begin{tabular}{lcccc}
& $R$ real & $R$ pseudoreal  & $\UoneV$ & $\UoneB$ \\
\hline
$\psi_1\psi_1$ & S & A & $+2$ & $+2$ \\
$\psi_1\psi_2$ & S & A & $0$ & $+2$ \\
$\psi_2\psi_2$ & S & A & $-2$ & $+2$ \\
$\chi_1\chi_1$ & $\bar{\rm S}$ & $\bar{\rm A}$ & $+2$ & $-2$ \\
$\chi_1\chi_2$ & $\bar{\rm S}$ & $\bar{\rm A}$ & $0$ & $-2$ \\
$\chi_2\chi_2$ & $\bar{\rm S}$ & $\bar{\rm A}$ & $-2$ & $-2$ 
\end{tabular}
\caption{Standard Model quantum numbers of the NGBs in the minimal theories of Table~\ref{tab:minimal_theories}. We assume that $r$ is a fundamental representation of one of the simple SM factors, and denote the corresponding singlet, adjoint, symmetric and antisymmetric representations respectively with 1, Adj, S and A. If the SM simple factor is $U(1)_Y$, then those NGBs transforming as the adjoint must be removed. Each NGB corresponds to a dark color-singlet fermion bilinear. Theories with $R$ complex have their NGBs listed in the upper panel. The NGBs of theories with $R$ real or pseudoreal are those listed in the Type I column of the upper panel plus those in the lower panel. The last two columns report the quantum numbers under the global $\UoneV\times\UoneB$.}
\label{tab:NGBs}
\end{table}
%%%%%%%%%%%%%%%%%%%%%%%%%
%
The lightest dark particles charged under $\UoneV \times \UoneB$ are accidentally stable. In Type I theories with complex $R$, these are two dark pions (with charge $\pm 2$ under $\UoneV$) and the lightest dark baryon (with charge $N_{DC}$ under $\UoneB$). The stable dark pions are neutral under the SM and thus potentially good DM candidates. If $r$ is a doublet of $\SUEW$, then the lightest baryon can be SM singlet only for an even number of dark colors. Theories with $\NDC =3,5$ are thus excluded by the severe constraints that exist on the fraction of DM component with non-vanishing electromagnetic charge (see Sec.~\ref{sec:chargedrelics} and references therein). If $r$ is a triplet of $\SUC$, the lightest dark baryon is a SM singlet for $\NDC =3,6$. Theories with $\NDC =4,5$ are problematic if not excluded by the current bounds on exotic matter (see Ref.~\cite{Perl:2009zz}). 
Type II and III theories, as well as theories with real or pseudoreal dark color representations~$R$, do not seem to lead to any additional realistic model. Indeed, if $r$ is a doublet of $\SUEW$ then: Type II are physically equivalent to Type I; Type I with $R$ (pseudo) real and Type III are vectorlike. If $r$ is a triplet of $\SUC$, instead, Type I theories with $R$ (pseudo) real, as well as Type II and III ones contain stable colored states that form exotic bound states at QCD confinement; they are therefore very constrained by data and most likely excluded. Finally, a particular example of Type~II theories are those where $r$ is an irreducible representation of hypercharge only.
In such theory the accidentally stable pions have both $\UoneV$ charge and hypercharge, and as such are not an acceptable DM candidate. The impossibility of gauging hypercharge in theories with irreducible $r$ directly follows from the fact that the only anomaly-free, unbroken subgroup of the dark global invariance is $\UoneV$ itself. This means that the accidentally stable NGBs have necessarily non-zero hypercharge.

In light of the above difficulty with hypercharge and in order to build models compatible with Grand Unification of the SM gauge couplings, it is interesting to analyze the case where the representation $r$ is reducible. This possibility is severely constrained by the condition of Eq.~(\ref{eq:LandauPoles}) implied by the request of perturbativity of the SM gauge couplings. In practice, apart from adding SM singlets or doubling the matter content of minimal models, the only possibility is to have $r$ equal to the direct sum of a doublet of $\SUEW$ plus a triplet of $\SUC$. This choice implies two additional anomaly-free vectorial $U(1)$ subgroups, which are unbroken by the dark condensate and can be chosen to gauge hypercharge. One such choice corresponds to a standard assignment of hypercharges for the weak doublet and color triplet fermions contained in a fundamental of $\SUfiveGUT$. One can thus consider a GUT chiral model where quantum numbers are assigned as in Type I of Tab.~\ref{tab:minimal_theories} with $r =5$ of $\SUfiveGUT$. Its main difficulty is given by the presence of two very light pseudo NGBs, that are subject to strong phenomenological and cosmological constraints. 
The existence of potentially massless NGBs is in fact a general issue in chiral gauge theories, where unwanted accidental symmetries cannot be lifted by mass terms or Yukawa couplings. In the theories of Tab.~\ref{tab:minimal_theories}, if $r$ is made of $\kappa$ irreducible components, then one has $(2\kappa -2)$ NGBs that are complete gauge singlets (i.e. neutral under both $\GSM$ and $\UoneD$). Indeed, there are $(2\kappa-1)$ axial $U(1)$’s changing the phases of the dark fermions that are free from dark color anomalies, commute with the weak gauging, and are spontaneously broken by the dark condensate; one (linear combination) of them is $U(1)_D$. For $\kappa =2$, as in the case of the GUT model under discussion, one predicts two such NGBs. One of them has $\SUC$ anomalous interactions and receives a mass from the QCD dynamics, $\delta m_\phi^2 = m_\pi^2 f_\pi^2/f_\phi^2$. A further contribution to the mass of both NGBs comes from GUT gauge interactions, which explicitly break the two axial $U(1)$ factors. This effect scales naively as $\delta m_\phi^2 \sim (\alpha_{GUT}/4\pi) \LDC^4/M_{GUT}^2$. For $\LDC \lesssim 10^5\,$GeV, the QCD contribution dominates, one NGB behaves as the QCD axion and the model is excluded by current constraints. Larger values of $\LDC$ give the chance of evading the bounds on axions, but  are challenging for cosmology, since the abundance of the stable (massive) NGBs is naively too large in presence of a standard cosmological history. While this model is potentially very interesting, assessing its relevance requires a dedicated analysis that we defer to a future work. Finally, we notice that a similar, though different, chiral GUT model was considered in Ref.~\cite{Bai:2016vca}. There, the quantum number assignments are of Type II (Type III was also mentioned as a possibility), and a dark scalar field is added whose Yukawa couplings give an additional contribution to the NGBs masses. In order to let the dangerous dark pions decay, the authors use higher-dimensional operators assuming a low cutoff scale. As a consequence, the model has no DM candidate.

\section{The model with $\SUEW$ doublets}
\label{sec:themodel}

The analysis of the previous section suggests that, if one restricts to irreducible representations $r$, the only chiral theories of Tab.~\ref{tab:minimal_theories} with realistic DM candidates are $\SU(\NDC>2)$ Type I models with $\SUEW$ doublets or $\SUC$ triplets. In the following we will focus on the model with EW doublets and analyze in detail its phenomenology and cosmological history.

At the renormalizable level, the Lagrangian of the model can be written as
\begin{equation}\label{lafond}
\mathcal{L} = \mathcal{L}_{SM}+\mathcal{L}_{DS}+\mathcal{L}_{mix}\, ,
\end{equation}
where $\mathcal{L}_{DS}$ describes the dark fermions and their minimal couplings, while
\begin{equation}\label{eq:mix}
\mathcal{L}_{mix} = \frac{\varepsilon }{2} B^{\mu \nu}F_{\mu \nu}^D
\end{equation}
is a mixing term between hypercharge and the dark photon. We assume that $\SUNDC$ confines at a scale $\LDC$ higher than the electroweak scale, at which all the other interactions are weak. We anticipate that, in order to obtain a viable DM candidate, $\LDC$ will be of order $1\!-\!50\,$TeV.

The low-energy dynamics of the theory can be characterized in terms of its continuous and discrete global symmetries. Let us first consider the case $0<a<1$. In absence of the weak gauging, there is a global $ \SU(4)_L \times \SU(4)_R \times \UoneB$ symmetry (with $4=2\times dim(r)$, where $r$ are $SU(2)_{EW}$ doublets), spontaneously broken to $\SU(4)_V \times \UoneB$. After turning on the weak gauging, the dark photon acquires a mass and the residual global symmetry is $\UoneV \times \UoneB$. The $\UoneB$ dark baryon number is actually anomalous with respect to $\U(1)_D$, whereas $\UoneV$ is a genuine accidental symmetry. For $a\neq 1$, the model also possesses two approximate discrete symmetries with interesting phenomenological implications:

\paragraph{Dark Charge Conjugation}
The transformation
\begin{equation}
\label{eq:c_dark}
\CD: \,\,\,
\begin{cases}
A_{\mu}^D \longrightarrow -A_{\mu}^D \\
\psi_{1} \longleftrightarrow \psi_{2} \\
\chi_{1} \longleftrightarrow \chi_{2}
\end{cases} ,
\end{equation}
dubbed dark charge conjugation in the following, leaves $\mathcal{L}_{DS}$ invariant and is not broken by the dark condensate. It is explicitly violated by the mixing term $\mathcal{L}_{mix}$, as it cannot be extended to the full SM sector. In analogy with QED, one can state a generalized version of Furry's theorem: any Green function of $\CD$-invariant operators with an odd number of dark photon fields vanishes identically. Two important consequences are:
\begin{itemize}
 \item The decay of the dark photon to a $\CD$-even state, such as any combination of SM particles, is forbidden for $\varepsilon = 0$. This implies that the dark photon is stable if its decays to dark sector particles are kinematically forbidden.
 \item For $\varepsilon = 0$, the mixing term between hypercharge and the dark photon is not radiatively generated. Hence, $\varepsilon$ receives quantum corrections proportional to itself. A small or vanishing $\varepsilon$, depending on the UV dynamics, is therefore technically natural.
\end{itemize}
\paragraph{$G$-parity.} 
In analogy with QCD, a generalized $G$-parity transformation acting only on dark sector fields can be defined as~\cite{Bai_2010}:
\begin{equation}
\Gtwo: \,\,\,\,
\begin{cases}
\psi \longrightarrow e^{i\pi T_2} \mathcal{C} \psi \\
G_{\mu}^a \lambda_a \longrightarrow -G_{\mu}^a \lambda_a^*  \\
\end{cases} , 
\end{equation}
where $T_2$ and $\lambda^a$ are respectively $\SUEW$ and $\SUNDC$ generators, $G_{\mu}^b$ is the dark gluon field, and $\psi_i \leftrightarrow \chi_i$ under the charge conjugation $\mathcal{C}$. For $a\neq 1$, $\Gtwo$ is an exact symmetry in the absence of $\UoneD$ gauge interactions. Under the combined action of $\CD$ and $\Gtwo$, all the NGBs have a definite parity, and this will be useful to analyze their properties.

\subsubsection*{Accidental stability and higher-dimensional operators}

The accidental $\UoneV$ symmetry is an exact invariance at the renormalizable level. In order to estimate the lifetime of the lightest dark states with non-vanishing $\UoneV$ charge, it is important to identify the lowest-dimensional operators that violate this symmetry. A simple analysis reveals that, for any given dimension, $\UoneV$-violating operators can be built only for a discrete set of rational values of the parameter $a$. In particular, we find that:
\begin{itemize}
 \item For $D=5$, no $\UoneV$-violating operator exists, for any value of $a$.
 \item For $D=6$, the only possibility is $a=0$. The operators are of the form:
\begin{equation}
\psi_1\psi_2 \chi_i \chi_i \quad \text{or} \ \ \chi_i\chi_i\chi_i\chi_j \ (\text{for } \NDC=4)\qquad \forall \, i,j = 1,2\,.
 \end{equation}
\item For $D=7$, it is possible to build $\UoneV$-violating operators only for $a = \pm 3$ and $a = \pm 1/3$. For example one has ($\forall\, j=1,2$):
 \begin{equation}
\begin{alignedat}{3}
\psi_1^{\dag}i\slashed{D} \psi_2 \psi_2 \chi_j \quad  &\text{and} \quad \psi_2^{\dag}i\slashed{D} \psi_1 \psi_1 \chi_j
\qquad & \text{for } a&=\pm 3 \, ,\\[0.1cm]
\chi_1^{\dag}i\slashed{D} \chi_2 \chi_2 \psi_j \quad  &\text{and} \quad \chi_2^{\dag}i\slashed{D} \chi_1 \chi_1 \psi_j
\qquad & \text{for } a&=\pm \frac{1}{3}\, .
\end{alignedat}
 \end{equation}
\end{itemize}
This shows that the accidental stability of the lightest NGBs charged under $\UoneV$ is a robust prediction of our chiral theory, as $\UoneV$-violating operators can have naturally very high dimension. This has to be compared with vectorlike theories, where accidental symmetries acting on NGBs are typically violated at the $D=5$ level~\cite{Antipin:2015xia}.

A similar analysis shows that in our theory $\UoneB$-violating operators first appear at the $D=6$ level for $N_{DC}=4$; in this case they have the form $\psi_i \psi_j \psi_k \psi_l$ or $\chi_i \chi_j \chi_k \chi_l$ ($\forall\, i,j,k,l = 1,2$).

\subsubsection*{The case a = 0}
In the limit $a=0$ the theory possesses an enhanced global symmetry at the renormalizable level that is left unbroken by the weak gauging:
\begin{equation}
{\rm U}(1)_{3L}\times {\rm SU}(2)_{R} \times \UoneB \to \UoneV \times \UoneB\, ,
\end{equation}
where $\SU(2)_{R}$ acts on the $\chi$ fields. This pattern of symmetry breaking gives three exact NGBs: one of them is eaten by the dark photon, the other two are the SM-singlet NGBs, which become massless for $a\to 0$. They can acquire a mass only through $\SU(2)_{R}$-breaking operators. The first such operators appear at the $D=6$ level, for example of the form $\psi_1 \psi_2 \chi_i \chi_j$. The corresponding NGB mass squared is of order
\begin{equation}
 m^2 \sim \frac{\bar g^2}{16 \pi^2}\frac{\Lambda_{DC}^4}{\Lambda^2_{UV}}\, ,
\end{equation}
where $\bar g^2/\Lambda^2_{UV}$ is the coefficient of the $D=6$ operator. For $\Lambda_{UV}/{\bar g} \gtrsim 10^{16}\,$GeV and $\LDC \sim 1\!-\! 50\,$TeV, this implies a very light and long-lived pair of NGBs. Such light degrees of freedom are relativistic at the epoch of neutrino decoupling, and can give a sizable contribution to the number of additional relativistic species $\Delta N_{eff}$. The relevance of this scenario will be discussed in Sec.~\ref{sec:alternatives}.

\section{Phenomenological profile}
\label{sec:phenomenology}

In this section we sketch the phenomenological profile of the model with $\SUEW$ doublets, discussing its spectrum and the dynamics of its NGBs.

\subsection{Dark Baryons}

The spectrum of dark hadrons contains baryonic states, made of the antisymmetric product of $\NDC$ dark quarks, with mass of order $\LDC$. They have $\UoneB$ baryon number $\NDC$ and the lightest among them are accidentally stable. They are organized in multiplets of the $\SU(4)$ flavour group. Due to Fermi statistics, their wave function is completely symmetric under the combined action of flavour and spin symmetries.

For $\NDC=3$ the lightest baryons have spin $1/2$ and flavour structure $\tiny \yng(2,1)$, corresponding to a $20$ of the global $\SU(4)$, which decomposes into $2_{\pm 3}\oplus 2\times 2_{\pm 1} \oplus 4_{\pm 1}$ of $\SUEW\times\UoneV$.
Since this contains no SM singlet, the lightest, accidentally stable baryons will have non-vanishing electromagnetic charge. As discussed in Sec.~\ref{sec:cosmology-baryons}, their relic density is never small enough to satisfy the stringent constraints on the charged fraction of DM, and the model is thus excluded. Similar conclusions hold for any odd number of dark colors, in particular $\NDC=5$.

For $\NDC=4$ the lightest baryons have spin $0$ and flavour structure $\tiny \yng(2,2)$, corresponding to a $20'$ of the global $\SU(4)$, which decomposes into $1_{\pm 4} \oplus 1_{\pm 2} \oplus 2\times 1_{0}\oplus 3_{\pm 2}\oplus 3_{0} \oplus 5_{0}$  of $\SUEW\times\UoneV$. The SM singlet components neutral under $\UoneV$ are expected to be the lightest, accidentally stable baryons. They contribute a small fraction of the DM abundance and are electromagnetically neutral. Similar conclusions hold for any even number of dark colors. Taking into account the constraints from Landau poles, we conclude that models with $\NDC=4, 6$ are viable and we will focus on them in the following.

\subsection{Dark Pions and low-energy effective theory}
\label{ssc:dkpions}

At energies much lower than $\LDC$, the dynamics of the lightest states in the spectrum can be characterized by making use of Chiral Perturbation Theory (ChPT). The pattern of spontaneous symmetry breaking implies the existence of 15 pseudo NGBs in the adjoint of $\SU(4) $, one of which eaten by the dark photon. Their quantum numbers can be derived from the transformation properties of the associated conserved currents; there are:
\begin{itemize}
\item Two $\SUEW$ triplets charged under $\U(1)_D$, the $\mathbf{3_{\pm}}$.
\item Two $\SUEW$ triplets neutral under $\U(1)_D$, the $\mathbf{3_0}$ and the $\mathbf{3_0'}$, with dark conjugation charge equal to $+1$ and $-1$, respectively.
\item Two SM singlets, the $\mathbf{1_{\pm}}$. These are the lightest particles charged under $\U(1)_V$, and will constitute our primary DM candidate.
\item A global singlet, the $\mathbf{1_0}$, that is eaten to form the longitudinal polarization of the dark photon. With an appropriate choice of the gauge fixing, this can be removed from the spectrum.
\end{itemize}
A summary of the transformation properties of these particles under the relevant global and gauge symmetries is provided in Tab.~\ref{tab:pions}.
%%%%%%%%%%%%%%%%%%%%%%%%%%%%%%%%%%%%%%%%%
\begin{table}[t]
\begin{center}
\begin{tabular}{l|cccc}
   & $\SU(2)_{EW}$  &  $\U(1)_V$ & $\mathcal{C}_D$ & $\mathcal{C}_D\!\cdot\mathcal{G}_2$\\[0.1cm]
\hline\\[-0.4cm]
$3_{\pm}$  &  $3$ & $\pm{2}$ & $3_{\mp}$ & $-$\\[0.05cm]
$3'_0$      & $3$ & $0$      & $-$ & $+$\\[0.05cm]    
$3_0$      & $3$ & $0$      & $+$  & $-$\\[0.05cm] 
$1_{\pm}$ & $1$ & $\pm{2}$ & $1_{\mp}$  & $+$\\[0.05cm] 
$\gamma_D$ & $1$ & $0$    & $-$ & $\times$ 
\end{tabular}
\caption{Transformation properties of the NGBs and of the dark photon under the symmetries of the model.}
\label{tab:pions}
\end{center}
\end{table}
%%%%%%%%%%%%%%%%%%%%%%%%%%%%%%%%%%%%%%%%

We construct the effective chiral Lagrangian by adopting a standard non-linear representation for the NGB fields
\begin{equation}
\Sigma(x) = \exp\left(\frac{2 i \pi^a(x) T^a}{\fD}\right)\, ,
\end{equation}
and write the covariant derivative as
\begin{equation}
 D_{\mu} \Sigma  = \partial_{\mu} \Sigma -\frac{ig W^{a}_{\mu}}{2} \left( T_{EW}^a\Sigma - \Sigma T_{EW}^a\right)
 -ie_D A^D_{\mu} \left( T_D\Sigma - a \Sigma T_D \right)\, ,
\end{equation}
where $\SU(4)$ generators are normalized to $\text{Tr}(T^a T^b) = \delta^{ab}/2$ and
\begin{equation}
T_{EW}^a = \begin{bmatrix}
    \sigma_a & 0 \\
    0 & \sigma_a 
\end{bmatrix} , \qquad
T_D = \begin{bmatrix}
    \mathbbm{1} & 0 \\
    0 & -\mathbbm{1}
\end{bmatrix} .
\end{equation}
At lowest order in the derivative expansion there is only the NGB kinetic term
\begin{equation}\label{eq:kin}
\mathcal{L}_0 = \frac{\fD^2}{4} \Tr[D_{\mu} \Sigma D^{\mu} \Sigma^{\dagger}]\, ,
\end{equation}
together with the NGB potential generated by 1-loop radiative effects (see Sec.~\ref{ssec:spectrum}). The kinetic term, in particular, contains the lowest-order interactions of $3_{\pm}$ and $1_{\pm}$ with the dark photon (the other NGBs do not interact at this order with $\gamma_D$):
\begin{equation}\label{eq:kinetic}
\mathcal{L}_0 \supset -ie_D(1+a) A^D_{\mu}(\pi_- \partial_{\mu} \pi_+ -\pi_+ \partial_{\mu} \pi_- )
+4 a e_D^2 A^D_{\mu} A^D_{\mu} \pi_+ \pi_-.
\end{equation}
While the cubic term is that of scalar QED with charge $(1+a)$, the quartic term has a modified (unless $a=1$) coefficient. This comes as a consequence of the spontaneous breaking of $\UoneD$, and leads to a non-trivial dependence on $a$ of the cross section for the process $1_+ 1_- \rightarrow \gamma_D \gamma_D$, which sets the DM abundance in our model. Another feature of~${\cal L}_0$ due to the chiral gauging is the appearance of vertices with three NGBs and one dark photon. They are proportional to $(1-a)$ and thus vanish in the vectorlike limit $a=1$. Their analog with the photon in the QCD chiral Lagrangian is forbidden by parity. Their expressions are reported in Appendix~\ref{app:darkpions}.

At $\Ocal(p^4)$, the chiral Lagrangian reads
\begin{equation}
\label{eq:Lp4}
\begin{split}
\mathcal{L}_1 = & \, C_1 \,\Tr[D_{\mu} \Sigma D^{\mu} \Sigma^{\dag}] \Tr[D_{\nu} \Sigma D^{\nu} \Sigma^{\dag}]
 + C_2 \,\Tr[D_{\mu} \Sigma D_{\nu} \Sigma^{\dag}] \Tr[D^{\mu} \Sigma D^{\nu} \Sigma^{\dag}] \\[0.1cm]
 &+C_3 \, \Tr[D_{\mu} \Sigma D_{\mu} \Sigma^{\dag} D^{\nu} \Sigma D^{\nu} \Sigma^{\dag}]
 + C_4 \,\Tr[D_{\mu} \Sigma D_{\nu} \Sigma^{\dag} D^{\mu} \Sigma D^{\nu} \Sigma^{\dag}]\\[0.1cm]
 &+C_5 \, \Tr[W^L_{\mu \nu}D^{\mu} \Sigma D^{\nu} \Sigma^{\dag}+ W^R_{\mu \nu}D^{\mu} \Sigma^{\dag} D^{\nu} \Sigma]
 +C_6 \, \Tr[W^L_{\mu \nu} \Sigma W_R^{\mu \nu} \Sigma^{\dag}]\, ,
\end{split}
\end{equation}
where
\begin{equation}
\begin{split}
W_{\mu\nu}^L  & = T^a_{EW} W_{\mu\nu}^a + T_D F^D_{\mu\nu} \\[0.1cm]
W_{\mu\nu}^R  & = T^a_{EW} W_{\mu\nu}^a + a\, T_D F_{\mu\nu}^D \, .
\end{split}
\end{equation}
The size of the chiral coefficients is estimated to be $C_i \sim 1/(16 \pi^2)$ from Naive Dimensional Analysis (NDA).~\footnote{The form of the last two terms in Eq.~(\ref{eq:Lp4}) is schematic, since terms with different physical field strengths will have different coefficients. Also, the appropriate powers of the couplings have been omitted in the NDA estimate of the chiral coefficients for simplicity.} Working at $\Ocal(p^4)$ one also has to include the Wess-Zumino-Witten term, which encodes the effects of anomalies. By explicit calculation, all squared anomalies (both $\U(1)_D^2$ and $\SU(2)_{EW}^2$ ones) turn out to vanish, while mixed ones under $\SUEW\times\UoneD$ do not. In particular, the axial current that interpolates the $3_0'$ has an anomaly
\begin{equation}
\label{mixanoma}
 \bra{0} \partial^{\mu} J_{\mu}^{(3_0') i}\ket{0} = -\frac{ g e_D(1+a)}{16 \sqrt{2} \pi^2} \varepsilon^{\mu \nu \rho \sigma} W^i_{\mu \nu} F^D_{\rho \sigma}\, ,
\end{equation}
which can mediate the decay $3_0' \rightarrow W \gamma_D$ as discussed below.

\subsection{Decay channels}
\label{sec:decaychannels}

As discussed in the previous section, the $1_\pm$ are accidentally stable at the renormalizable level and their decay is induced only by higher-dimensional operators. The other NGBs and the dark photon instead can decay through the following channels: 

\begin{itemize}
\item \textbf{Dark Photon: $\gamma_D$} 

The dark photon lifetime is suppressed by a factor $\varepsilon^2$, as $\CD$ makes $\gamma_D$ stable in absence of the mixing term (\ref{eq:mix}). As shown in the next section, this can have striking implications on the cosmological history of the model, leading to dark matter dilution if $\varepsilon$ is small enough. The dominant decay channel is into SM fermions, $f \bar{f}$, but $Z h$ and $W^+ W^-$ are also relevant if kinematically allowed. 
The decay rate has the form
\begin{equation}
\Gamma = \alpha_{em} \varepsilon^2 m_{\gamma_D} C(m_i,g_i)\, ,
\end{equation}
where $C(m_i,g_i)$ is a dimensionless function of the masses and the couplings, see Appendix D for details.

\item \textbf{Charged Triplets: $3_\pm$} 

Due to $\UoneV$ invariance, the $3_\pm$ decays to final states that contain at least one $1_\pm$. The main channels are~\footnote{The decay $3_\pm \to 1_{\pm} \gamma_D V$ is also allowed but its width is parametrically smaller than that of the other channels.}
\begin{equation}
3_{\pm} \longrightarrow  1_{\pm} V  \quad \quad \text{and} \quad \quad 3_{\pm} \longrightarrow  1_{\pm}  \gamma_D \gamma_D V
\qquad (V=W,Z,\gamma)\, .
\end{equation}
The first one is forbidden if $\Gtwo \!\cdot \CD$ is unbroken (see Tab.~\ref{tab:pions}) and thus occurs only through a loop of dark photons with the anomalous $3_0' W\gamma_D$ vertex, see Fig.~\ref{fig:decays}a. 
%
%%%%%%%%%%%%%%%%%%%
\begin{figure}
\centering
\begin{minipage}[b]{0.37\textwidth}
\centering
\includegraphics[width=\textwidth]{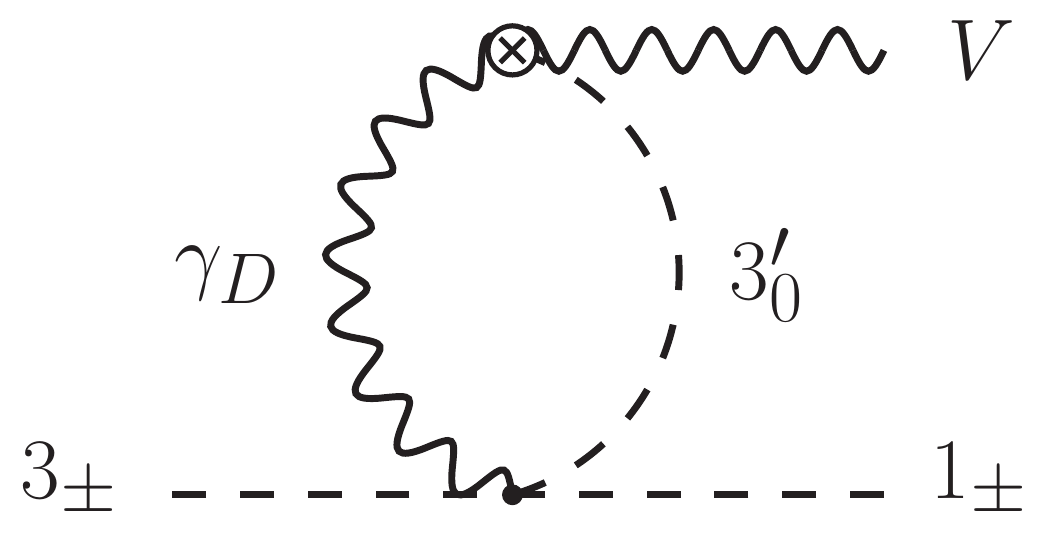}\\[0.1cm] (1a)
\end{minipage}
\hspace{1.5cm}
\begin{minipage}[b]{0.37\textwidth}
\centering
\includegraphics[width=\textwidth]{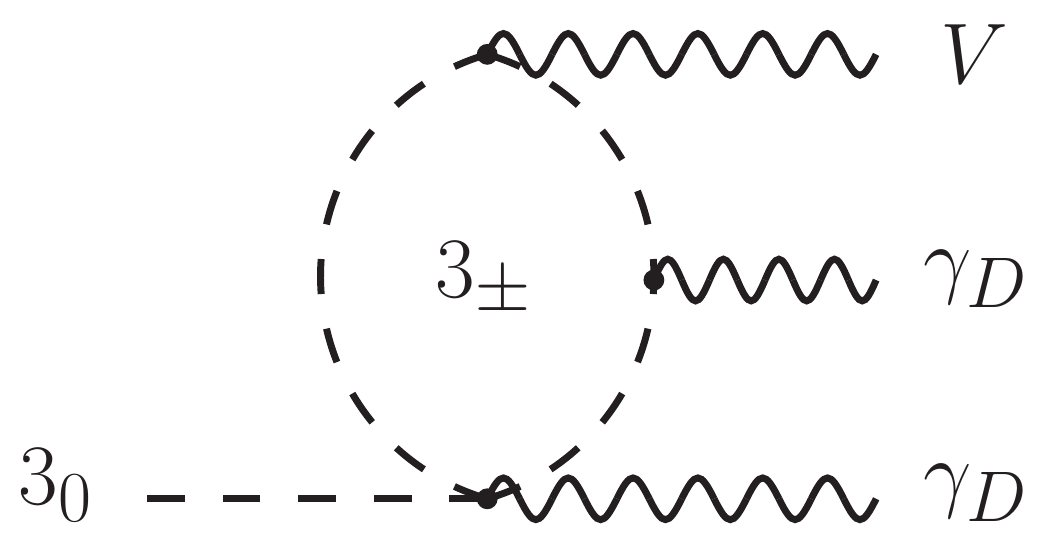}\\[0.1cm] (1b)
\end{minipage}
\caption{On the left (1a): Feynman diagram for the decay $3_\pm \to 1_\pm V$. The crossed vertex denotes the anomalous interaction. On the right (1b): One of the Feynman diagrams mediating the decay $3_0\to \gamma_D\gamma_D V$.}
\label{fig:decays}
\end{figure}
%%%%%%%%%%%%%%%%%%%
%
The second occurs through a tree-level diagram also featuring one anomalous vertex. The estimated rate is the same for both channels:
\begin{equation}
\Gamma \sim \frac{\alpha_W \alpha_D^2}{(16 \pi^2)^3}  (1-a^2)^2  \frac{m^5_{3_\pm}}{\fD^4}\, .
\end{equation}
Numerically, this implies 
\begin{equation}
c \tau \sim 6.6  \,\Bigg( \frac{500 \GeV}{m_{3_\pm}} \Bigg) \Bigg( \frac{2\cdot 10^{-3}}{\alpha_D} \Bigg)^2 \times 10^{-3} \,\text{cm}\, ,
\end{equation}
which does not give rise to displaced vertices at high-energy colliders for any value of $e_D$ allowed by cosmology.

\item \textbf{Neutral Triplet: $3'_0$}

The decay modes of $3_0'$ are restricted by $\SUEW$ and $\CD$ invariance. If kinematically allowed, the main channel is
\begin{equation}
3'_0 \longrightarrow  V \,\, \gamma_D .
\end{equation}
This process proceeds through the anomaly and its rate is~\footnote{Here and in the following, $\mt$ denotes the mass of $3_0$, which is equal to the mass of $3_0'$ at the 1-loop level.}
\begin{equation}
\label{eq:anomrate}
\Gamma = \frac{\alpha_D \alpha_W N_c^2(1+a)^2 }{8 \pi^3} 
\frac{\mt^3}{\fD^2} \, g \!\left( \frac{m_W^2}{\mt^2} ,\frac{m_{\gamma_D}^2}{\mt^2} \right)\, ,
\end{equation}
where
\begin{equation}
g(x,y) = \big[ (1-x-y)^2-xy \big] \sqrt{1-2(x+y)+(x-y)^2} \, .
\end{equation}
For small $\mt$ this channel is kinematically forbidden and $3_0'$ decays to $V \, \bar{f} f$ through the $\CD$-violating mixing term~(\ref{eq:mix}). The decay rate receives a suppression of order $\varepsilon^2 (\alpha_D/4 \pi)$ with respect to Eq.~(\ref{eq:anomrate}), and the $3_0'$ can become long-lived on collider scales. In the limit of $\varepsilon$ small, decays between different electroweak components also become important, though suppressed by phase space. For $\varepsilon \lesssim 3 \times 10^{-6} \, (2 \mt/\ms)^{3/2}\, (1\,\text{TeV}/ \ms)^{1/2}$, the electromagnetically-charged components $3_0^{\prime\,\pm}$ mostly decay into the neutral component $3_0^{\prime\, 0}$ by emitting a soft pion: $3_0^{\prime\,\pm}\rightarrow 3_0^{\prime\,0} \pi^{\pm}$. The decay width for this de-excitation process is~\cite{Cirelli:2005uq}:
\begin{equation}
\label{eq:deexcitation}
\Gamma_{\rm EW} \simeq \dfrac{2}{\pi} \,G_{\rm F}^{2}\, V_{ud}^{2}\, \Delta m^{3}\, \fD^{2} \,\sqrt{1-\dfrac{m_{\pi}^{2}}{\Delta m^{2}}},
\end{equation}
where $\Delta m \simeq 166\,$MeV is the mass splitting between the charged and neutral components (see Eq.~(\ref{eq:splitting})).

\item \textbf{Neutral Triplet: $3_0$} 

Invariance under $\SUEW$ and $\CD$ also restricts the decays of $3_0$. The main channel is
\begin{equation}
3_0 \longrightarrow  \gamma_D \gamma_D V\, ,
\end{equation}
which proceeds through 1-loop diagrams with $\Ocal(p^2)$ interactions, see Fig.~\ref{fig:decays}b. Its amplitude is finite, consistently with the absence of a counterterm in the $\mathcal{O}(p^4)$ Lagrangian~(\ref{eq:Lp4}). The NDA estimate of the decay rate is therefore
\begin{equation}
\Gamma \sim \frac{\alpha_D^2 \alpha_W (1-a)^2}{(16 \pi^2)^2} \frac{\mt^3}{\fD^2} \, .
\end{equation}
When this channel is forbidden by kinematics, the two dark photons can be closed to form a loop and the $V$ becomes virtual, so that the decay is to a SM pair ($f\bar f$, $VV$ or $Vh$). The estimated width in this case is
\begin{equation}
\Gamma \sim \frac{\alpha_D^2\alpha_W^2 }{(4\pi)^5} \frac{\mt^3}{\fD^2}.
\end{equation}
For the benchmark point $m_{1} = 2  m_{\gamma_D}$, this translates numerically into 
\begin{equation}
c \tau \sim 6.5 \,   \Bigg( \frac{500 \GeV}{\mt} \Bigg) \Bigg( \frac{2\cdot 10^{-3}}{\alpha_D} \Bigg)^2 \times 10^{-5} \,\text{cm}\, ,
\end{equation}
implying that the decay can be still considered as ``prompt" on collider scales.

\end{itemize}

\subsection{Spectrum} 
\label{ssec:spectrum}

When the weak $\SUEW\times\UoneD$ gauging is switched on, 1-loop radiative effects generate a potential, hence a mass term, for the NGBs.  It is a peculiar feature of our model that all the states receive mass from the weak gauging (unless $a= 0$). Using the Weinberg sum rules and saturating them with the lowest-lying spin-1 resonances, the effective potential can be computed at leading order in $1/\NDC$ using a standard approach, see for example~\cite{Contino:2010rs}. We find the mass terms
\begin{align}
\label{eq:ms}
\ms^2 & = 24 \ln(2) a \dfrac{e_D^2}{16\pi^2} m_\rhoD^{2} \\ 
\mt^2 &= 12 \ln(2) \dfrac{g^2}{16\pi^2}  m_\rhoD^{2}\\
m^2_{3_{\pm}} &= \left( 24 \ln(2) \dfrac{e_D^2}{16\pi^2} + 12 \ln(2) \dfrac{g^2}{16\pi^2} \right) m_\rhoD^{2}\, ,
\end{align}
where $m_{\rhoD} \simeq 4 \pi \fD/\sqrt{\NDC}$ is the mass of the first vector resonance. For $a=1$ Eq.~(\ref{eq:ms}) agrees with the charged-neutral pion mass difference in QCD, taking into account that the dark pions have $\UoneD$ charge $2 e_D$ in this limit.~\footnote{Notice however that there is a factor of 4 discrepancy with the result of Refs.~\cite{Harigaya:2016rwr,Co:2016akw}.} As regards the dark photon, inserting $\Sigma = \mathbb{1}$ in (\ref{eq:kin}) yields
\begin{equation}
m^2_{\gamma_{D}} =2 e_{D}^{2}(1-a)^{2} \fD^2 \sim \dfrac{\NDC}{8\pi^{2}}e_{D}^{2}(1-a)^{2} m^{2}_\rhoD \, .
\end{equation}

At sub-leading order, the degeneracy between the electromagnetically charged and neutral components of the triplets is lifted by electroweak and custodial symmetry breaking. While the mass squared of the triplets are of $\Ocal(g^2)$, the mass splitting arises at $\Ocal(g^4)$. One should therefore take into account the NGB wavefunction renormalization, which also contributes at $\Ocal(g^4)$ to the physical mass. Computing the NGB effective potential is thus not enough to fully capture the mass splitting. On the other hand, the loop integral relevant for the mass splitting is finite and converges at around the electroweak scale, i.e. much earlier than the onset of the dark strong dynamics. The same calculation valid for elementary scalars is therefore valid also in our case and gives~\cite{Cirelli:2005uq}:
\begin{equation}
\label{eq:splitting}
\Delta m|_{EW} = \frac{g}{4 \pi} m_W \sin^2\frac{\theta_W}{2} = 166 \pm 1 \MeV\, . 
\end{equation}

\section{Cosmology}
\label{sec:cosmology}

Depending on its spectrum, the model can undergo different thermal histories. In this section we will describe the cosmological evolution and compute the DM abundance predicted in the bulk of the theoretical parameter space, and then comment how the physics is modified in special limits of the parameters.

In this spirit we will first discuss the case where:
\begin{itemize}
\item The dark photon is lighter than the SM-singlet dark pions ($m_{\gamma_D} < \ms$) but has a mass larger than their freeze-out temperature ($m_{\gamma_D} > T_\text{f.o} \sim \ms/20$). This can be expressed as a condition on the parameter~$a$, implying $0.26 \lesssim a \lesssim 0.93$ for $\NDC=4$;
\item Both $3_\pm$ and $3_0, 3_0'$ are heavier than $1_\pm$, and the mass splitting $\Delta m = m_3-m_1$ is larger than the temperature relevant for the freeze-out of the singlets, \emph{i.e.} $\Delta m \gtrsim m_{1}$. This condition corresponds to a dark coupling of order $\alpha_D \lesssim \alpha_2/4a$;
\item The dark coupling $\alpha_D$ is sufficiently large that the singlets annihilate efficiently in dark photons at temperatures of order $T \sim m_{1}$, and triplets decay promptly to lighter states. This condition requires a dark coupling
\begin{equation}
 \alpha_D \gtrsim 10^{-9}\dfrac{1}{(1+a)^2} \left( \dfrac{m_{1}}{\rm TeV} \right)^{\frac{1}{2}}.
\end{equation}
\end{itemize}
The decay width of the dark photon is controlled by an additional independent parameter: the~$\varepsilon$ coefficient. We shall first discuss the case in which the dark photon decays promptly at temperatures relevant for the computation of the dark matter relic density, $T \sim m_{1}$. This corresponds to requiring $\Gamma _{\gamma_{D}} > H(T = m_{1})$, which can be expressed as a condition on $\varepsilon$:
\begin{equation}
\label{eq:metastability}
\varepsilon \gtrsim 10^{-7} \left( \dfrac{a}{{\NDC} (1-a)^{2}} \right)^{\frac{1}{4}} \left( \dfrac{m_{1}}{\rm TeV} \right)^{\frac{1}{2}}.
\end{equation}
We will then consider the case where the dark photon decays after the freeze-out of dark matter but before Big Bang Nucleosynthesis (BBN), and study the possible entropy injection that it gives into the SM bath with the subsequent dilution of the DM relic abundance. Finally, we will comment on the case of cosmologically stable dark photons as a possible dark matter candidate. In the latter scenario, requiring the dark photon lifetime to be larger than $10^{25}\,$s implies
\begin{equation}
\label{eq:gamma_stability}
\varepsilon \lesssim 10^{-24} \left( \dfrac{\rm GeV}{m_{\gamma_{D}}} \right)^{\frac{1}{2}}.
\end{equation}

\subsection{Dark baryons}
\label{sec:cosmology-baryons}

Dark baryons are produced at the confinement temperature and are accidentally stable. They are thus expected to be a dark matter component. Their relic abundance is set by the freeze out of the process of annihilation into dark pions and other mesons. Since their mass is parametrically larger than the mass of the NGBs, baryons freeze out at temperatures where dark pions are still in chemical equilibrium with the SM thermal bath; therefore, assuming thermalization of the decay products, the computation of the dark pion relic abundance is unaffected by the baryons.

The baryon-antibaryon annihilation cross section is difficult to compute from first principles, but can be estimated as 
\begin{equation}
\label{eq:BBxsec}
\sigma_{B\bar{B}\rightarrow \rm mesons} v = c\, \dfrac{4\pi}{m_{\rhoD}^2}\, ,
\end{equation}
where $c$ is an $\Ocal(1)$ proportionality factor. This estimate can be checked in the case of QCD by making use of nuclear physics data~\cite{Bertin:1997gn}. For anti-neutrons annihilating on a proton target, the data display the expected $1/v$ dependence and are well reproduced for $c \simeq 3$.
Using Eq.~(\ref{eq:BBxsec}) with $c=3$ and comparing with the pion annihilation cross section~\eqref{eq:pion_ann}, one finds that the energy density of dark baryons relative to dark pions is suppressed by a factor
\begin{equation}
\label{eq:ratioOmegas}
\dfrac{\Omega_{DB}}{\Omega_{D\pi}} \simeq 
\frac{\langle \sigma_{\pi\pi \rightarrow \gamma_{D} \gamma_{D}} v \rangle}{\langle \sigma_{B\bar{B}\rightarrow \rm mesons} v\rangle} 
\sim \dfrac{1}{3} \dfrac{e_D^2 (1+a)^4}{24 \ln(2) a}\, .
\end{equation}
Therefore, the dark matter is expected to have two components, with dark pions giving the dominant contribution in the majority of the parameter space. 
For example, for $a=1/2$ and $\alpha_D \leq 0.04$, the DM fraction made of dark baryons is of order $10\%$ or smaller. Finally, requiring that dark baryons do not overclose the Universe sets an upper bound on the confinement scale $\Lambda_{DC} \lesssim 100 \,$TeV (as pointed out in Ref.~\cite{Griest:1989wd}), and correspondingly on the mass of the scalar triplets $\mt\lesssim 10 \, \rm TeV$.

\subsection{Short-lived dark photons}

Let us first focus on the case in which dark photon decays are efficient at temperatures relevant for the computation of the relic abundance, i.e. when $\varepsilon \gtrsim 10^{-7}$.

For these values of the mixing parameter, the dark and visible sectors are in thermal and chemical equilibrium at temperatures larger than the pion mass scale, thanks to processes mediated by dark photons. As the Universe cools down, pions become non relativistic. If the splitting between EW triplet and singlet pions is larger than the temperatures relevant for the freeze out (as in the parameter space region under consideration), then the abundance of triplets is suppressed by an exponential Boltzmann factor and the dynamics can be described by considering singlets and dark photons only.~\footnote{If this is not the case, i.e. for $\alpha_D \gtrsim \alpha_2 / 4a$, then processes such as $1_+ 1_- \rightarrow 3_+ 3_-$ can give a non zero population of triplets. A complete analysis, in such a scenario, would require a careful study of the system of coupled Boltzmann equations for singlets and triplets.}

The relevant process for the calculation of the DM relic abundance is the annihilation of dark pions, dominated by the dark photon channel $1_+ 1_-\rightarrow \gamma_D \gamma_D$. For dark photon prompt decays, the process is described by a standard Boltzmann equation for a single DM species ($Y_\pi = (n_{\pi_+} + n_{\pi_-})/s$, $x=\ms/T$):
\begin{equation}
\label{eq:boltzmann}
\dfrac{\mathrm{d}Y_{\pi}}{\mathrm{d}x}=
-\dfrac{1}{2x^{2}} \dfrac{s(\ms)}{H(\ms)} \langle \sigma_{\pi\pi \rightarrow \gamma_{D} \gamma_{D}} v \rangle \left(Y_{\pi}^{2}-Y^{2}_{\pi,eq} \right)\, .
\end{equation}
In the case of light dark photons, the thermally-averaged annihilation cross section is~\footnote{This formula differs from the corresponding result in Eq.~(8) of Ref.~\cite{Co:2016akw}.}
\begin{equation}
\label{eq:pion_ann}
\langle \sigma_{\pi\pi \rightarrow \gamma_{D} \gamma_{D}} v \rangle= \dfrac{1}{8\pi} \dfrac{e^4_D (1+a)^4}{\ms^2} f(y) + \mathcal{O}(v^2), \quad \quad y = \frac{\mgaD^2}{\ms^2},
\end{equation}
with
\begin{equation}\label{eq:pion_f}
f(y) = \sqrt{1-y} \left[\frac{y^2}{(y-2)^2}+\frac{(y-2)^2}{y^2}(K-1)^2+2K^2+2K-2 \right],  \quad K= \frac{4a}{(1+a)^2}.
\end{equation}
The limit of scalar QED is recovered by first setting $K=1$ in Eq.~(\ref{eq:pion_f}), and then letting \mbox{$y\to 0$} to obtain a massless dark photon.~\footnote{If one takes the limit $a\to 1$ by varying $K$ and $y$ together through their functional dependence on $a$, one finds a cross section different from that of scalar QED. This is because in this limit the contribution from the dark photon longitudinal polarization does not decouple. By virtue of the Goldstone Boson Equivalence Theorem (since the mass of the dark photon goes to zero for $a\to 1$), the latter can be computed in terms of the annihilation into $1_0$'s, which are part of the physical spectrum for $a=1$. Once added to the contribution from the transverse dark photon polarizations, this result correctly reproduces the $a=1$ limit of Eq.~(\ref{eq:pion_ann}). Notice that computing correctly the annihilation cross section into $1_0$'s requires to include the contributions from both the four-pion derivative interactions from the Chiral Lagrangian and the radiatively-generated quartic potential.}
In the regime of light dark photons, the Sommerfeld enhancement factor can be important, especially at the low values of the velocity relevant for indirect detection and CMB constraints. However, for $0.26\lesssim a\lesssim 0.93$ the dark photon and the dark matter have comparable masses and the Sommerfeld enhancement is negligible.

\subsection{Long-lived dark photons}
\label{sec:longlived}

The cosmological history can evolve differently if the dark photon is metastable, i.e. has a lifetime longer than the inverse Hubble rate at $T\sim m_{1}$ and decays before the present era. According to Eq.~\eqref{eq:metastability}, this happens for $10^{-24} \lesssim \varepsilon\lesssim 10^{-8}$.

In this case, the following effects can take place and change the DM abundance:
\begin{itemize}
\item The SM and dark sectors could become thermally decoupled at temperatures larger than the freeze-out temperature of the pions. 
The computation of the relic abundance of DM can be modified, especially if number-changing interactions in the dark sector are efficient.
\item If dark photon annihilations into SM particles are out of equilibrium, then the evolution of the dark pion density with temperature is modified with respect to the standard freeze-out scenario, as previously noted in Ref.~\cite{Berlin:2016vnh}. The abundance of dark photons can also be affected in this case.
\item If sufficiently long-lived, dark photons can give rise to an early phase of matter domination. Their subsequent decay and entropy injection into the SM thermal bath suppress the abundance of DM relics.
\end{itemize}
We discuss each of these effects in the following.

\subsubsection{Kinetic equilibrium}
\label{sec:kineq}

The SM and dark sectors are kept in thermal equilibrium by interactions involving the dark photon-hypercharge mixing or mediated by loops of NGB triplets.

After diagonalising the kinetic mixing of Eq.~(\ref{eq:mix}), one finds that electromagnetically-charged particles also have a coupling, of order $\varepsilon e$, to the dark photon. The leading process controlling the kinetic equilibrium between dark and SM sectors is thus the elastic scattering, mediated by the dark photon, of a charged pion on SM particles. Its cross section is of order
\begin{equation}
\sigma \sim 2\pi \varepsilon^{2} \dfrac{\alpha_{\rm em} \alpha_{D}}{T^{2}}\, .
\end{equation}
Comparing the rate of this process with the Hubble rate, we see that for $\varepsilon \lesssim 10^{-7}$ it becomes inefficient at a temperature of order $m_{1}$, in the region relevant for the relic abundance computation.

When dark photon interactions become inefficient, those mediated by a loop of pion triplets between pion singlets and EW gauge bosons can still be effective in maintaining thermal equilibrium. The low-energy theory obtained by integrating out the triplets contains two operators that can give the leading contribution, depending on the value of the coupling~$e_{D}$. The first has dimension 8 and involves two derivatives of the pions, the second has dimension 6 and breaks the NGB shift symmetry.  Their coefficients are estimated to be:
\begin{equation}
\label{eq:effective operators}
\begin{split}
&\mathcal{O}_{8} \sim \dfrac{1}{\fD^{2}} \dfrac{1}{\mt^{2}} \dfrac{g_{2}^{2}}{16\pi^{2}} \,\left( \partial \pi \right)^{2} \left( W_{\mu\nu}^{i} \right)^{2}  \\[0.1cm]
&\mathcal{O}_{6} \sim  \dfrac{m_{\rhoD}^{2}}{16\pi^2 \fD^{2}} \dfrac{1}{\mt^{2}}  \dfrac{e_{D}^{2}g_{2}^{2}}{16\pi^{2}} \, \left(\pi \right)^{2} \left( W_{\mu\nu}^{i} \right)^{2} \, ,
\end{split}
\end{equation}
where $m^2_\rhoD/(16\pi^2 \fD^2) \sim 1/\NDC$ in the large $\NDC$ limit.

We find that these interactions are efficient in maintaining kinetic equilibrium during the pion freeze out in the majority of the relevant parameter space. There is only a small corner, corresponding to triplets with masses in the range $(500 \div 1500) \,\rm GeV$, where the kinetic decoupling occurs at a temperature of order $T_{kd} \sim (1 \div 0.1) m_{1}$. In the following we will compute the DM abundance assuming that the temperature in the dark sector scales as the temperature in the visible sector. We thus neglect possible additional effects related to the kinetic decoupling in the region with $m_{3} \lesssim 1500 \, \rm GeV$. While these effects can play a role during the freeze-out epoch, they are not expected to give qualitative changes in the dilution factor computed in Sec.~\ref{sec:dilution}.

\subsubsection{Chemical equilibrium}
\label{sec:chemeq}

At temperatures of order $m_{1}$ or smaller, the evolution of the dark sector is described by a system of two coupled Boltzmann equations for the two species $\pi_{\pm}$ and $\gamma_{D}$. As explained in more details in Appendix~\ref{app:boltzmann}, for metastable dark photons and at temperatures higher than $T_{\rm decay}$ (where the latter is defined by the condition $\Gamma_{\gamma_{D}}\sim H(T_{\rm decay})$), these equations can be approximately written as
\begin{equation}
\label{eq:boltzmann_simplified}
\begin{split}
\dfrac{\mathrm{d}Y_{\pi}}{\mathrm{d}x} & =-\dfrac{1}{2x^{2}} \dfrac{s(\ms)}{H(\ms)}  \langle \sigma_{\pi\pi \rightarrow \gamma_{D} \gamma_{D}} v \rangle \left(Y_{\pi}^{2}-\frac{Y_{\pi , eq}^2}{Y_{\gamma_{D} , eq}^2}\, Y_{\gamma_{D}}^2\right)\,, \\[0.5cm]
\dfrac{\mathrm{d}Y_{\gamma_{D}}}{\mathrm{d}x} & = \dfrac{1}{x^{2}} \dfrac{s(\ms)}{H(\ms)} \Bigg[ 
\dfrac{1}{2} \langle \sigma_{\pi\pi \rightarrow \gamma_{D} \gamma_{D}} v \rangle \left(Y_{\pi}^{2}-\frac{Y_{\pi , eq}^2}{Y_{\gamma_{D} , eq}^2}\, Y_{\gamma_{D}}^2\right) \\
& \hspace{2.4cm} -2 \langle \sigma_{\gamma_{D}\gamma_{D}\rightarrow SM} v \rangle \left( Y_{\gamma_{D}}^{2}- Y_{\gamma_{D} , eq}^{2}\right)
\Bigg] 
\,.
\end{split}
\end{equation}
If dark photon annihilations into SM particles are efficient to keep dark photons in chemical equilibrium until the freeze-out temperature, the abundance of dark pions follows the usual evolution. On the other hand, if $\sigma_{\gamma_{D}\gamma_{D}\rightarrow SM} \ll \sigma_{\pi\pi \rightarrow \gamma_{D} \gamma_{D}}$, then dark photons go out of chemical equilibrium before dark pions and have a much larger abundance. In this case the total density in the dark sector can be approximated as $Y_{\rm D, tot}\simeq Y_{\gamma_{D}}$ and taking the sum of the two equations in (\ref{eq:boltzmann_simplified}) we obtain, after an initial transient,
\begin{equation}
\label{eq:boltzmann_simplified2}
\begin{split}
\dfrac{\mathrm{d}Y_{\pi}}{\mathrm{d}x} & =-\dfrac{1}{x^{2}} \dfrac{s(\ms)}{H(\ms)} \dfrac{1}{2} \langle \sigma_{\pi\pi \rightarrow \gamma_{D} \gamma_{D}} v \rangle \left(Y_{\pi}^{2}-\frac{Y_{\pi , eq}^2}{Y_{\gamma_{D} , eq}^2}\, Y_{\gamma_{D}}^2\right) \\[0.3cm]
\dfrac{\mathrm{d}Y_{\gamma_{D}}}{\mathrm{d}x} & = -
2 \dfrac{1}{x^{2}} \dfrac{s(\ms)}{H(\ms)}  \langle \sigma_{\gamma_{D}\gamma_{D}\rightarrow SM} v \rangle \left( Y_{\gamma_{D}}^{2}- Y_{\gamma_{D} , eq}^{2}\right) \,.
\end{split}
\end{equation}
From the first equation we see that the abundance of pions traces $(Y_{\pi , eq}/Y_{\gamma_{D} , eq})\, Y_{\gamma_{D}}$ once the dark photons are out of equilibrium and this modifies the freeze-out temperature for dark pions.

By solving the system (\ref{eq:boltzmann_simplified}) numerically we have verified that the approximation given by (\ref{eq:boltzmann_simplified2}) is accurate. We find that the evolution of the dark pion and dark photon energy densities is modified with respect to the standard freeze out when $\sigma_{\gamma_{D}\gamma_{D}\rightarrow SM} \ll \sigma_{\pi\pi \rightarrow \gamma_{D} \gamma_{D}}$. However, in all the relevant parameter space, the asymptotic value for the DM relic abundance is within 50\% of the naive estimate obtained from the simplified description of Eq.~(\ref{eq:boltzmann}). A similar conclusion was reached in Ref.~\cite{Berlin:2016vnh}. We refer the reader to Appendix~\ref{app:boltzmann} for more details.

\subsubsection{Dilution by entropy injection}
\label{sec:dilution}

If dark photons dominate the energy density of the Universe in some early phase of its evolution, they can modify the relic abundance of any pre-existing relic species by decaying and injecting entropy in the SM bath. More in detail, the different evolution of the temperature during such early epoch of matter domination (EMD) translates into an effective dilution factor~\cite{Co:2015pka} (see also~\cite{Contino:2018crt}),
\begin{equation}
\mathcal{F} \simeq \dfrac{4.8}{(g_*^{\rm SM}(T_{\rm EMD}))^{1/4}} \dfrac{\sqrt{M_{\rm Pl}\Gamma_{\gamma_{D}}} }{T_{\rm EMD}}\, ,
\end{equation}
that characterizes the prediction of the DM abundance today compared to the naive calculation. Here $T_{\rm EMD}$ has been defined as the temperature at the onset of the early matter domination era, while $g_*^{\rm SM}(T)$ is the number of SM relativistic degrees of freedom at the temperature $T$.

\begin{figure}[t]
\centering
\includegraphics[width=0.6\textwidth]{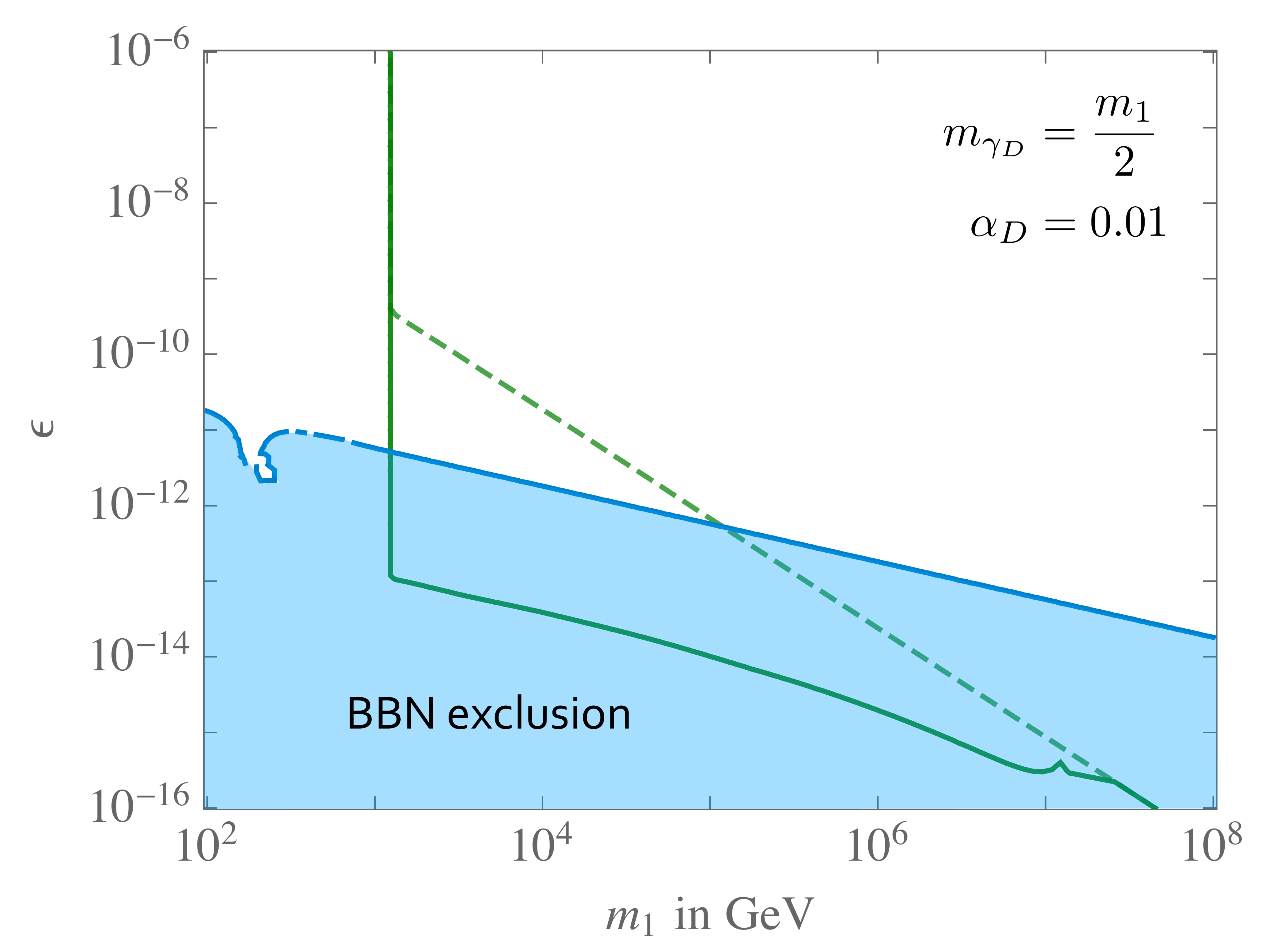}
\caption{Isocurve of observed DM abundance (solid green line) and BBN bounds on the lifetime of the dark photon (shaded blue area) in the plane $(\ms,\varepsilon)$. The dashed green curve is obtained by neglecting annihilations mediated by NGB triplets and using Eq.~(\ref{eq:TEDMnaive}). The non-trivial dependence on $\varepsilon$ is due to the dilution of the relic density caused by the early phase of matter domination. The plot assumes $\alpha_D=0.01$ and $m_{\gamma_D} = \ms/2$.}
\label{fig:dilution_BBN}
\end{figure}

We computed $T_{\rm EMD}$ as the temperature at which the energy density of dark photons before their decay, $\rho_{\gamma_{D}}$, and the energy density of the SM radiation, $\rho_{SM}$, are equal. Since $\rho_{\gamma_{D}}(T)=Y_{\gamma_{D}}(T) s(T) m_{\gamma_{D}}$ and $\rho_{SM}(T)=(3/4) s(T) T$, one has
\begin{equation}
\label{eq:TEMD}
T_{\rm EMD}=\dfrac{4}{3} Y_{\gamma_{D}}(T_{\rm EMD})\, m_{\gamma_{D}}\, .
\end{equation}
The dilution from dark photon decays was computed in previous studies in the context of models where $\gamma_D$ annihilates into SM particles only through its kinetic mixing with hypercharge, see Refs.~\cite{Berlin:2016vnh,Berlin:2016gtr,Cirelli:2018iax}. In that case, for $\varepsilon \lesssim 10^{-8}$ the dark photons have a chemical decoupling while they are still relativistic, and their abundance $Y_{\gamma_{D}}(T_{\rm EMD})$ is very large. In our model, even for very small $\varepsilon$, dark photons can annihilate into pairs of SM vector bosons through loops of NGB triplets, and undergo a standard freeze out at $T=T_{f.o.,\gamma_{D}}$. This gives a smaller abundance $Y_{\gamma_{D}}(T_{\rm EMD})$ hence a lower temperature $T_{\rm EMD}$. Using Eq.~(\ref{eq:TEMD}) and extracting $Y_{\gamma_{D}}(T_{\rm EMD})$ from an approximate analytic solution of the second equation in~(\ref{eq:boltzmann_simplified2}) gives
\begin{equation}
\label{eq:TEDM}
T_{\rm EMD}\simeq \dfrac{30}{\pi^{2}} \sqrt{\dfrac{4\pi^{3}g_*^{\rm SM}(T_{f.o.,\gamma_{D}})}{45}} \dfrac{1}{g_*^{\rm SM}(T_{f.o.,\gamma_{D}})} \dfrac{1}{M_{\rm Pl}} \dfrac{x_{f.o.,\gamma_{D}}}{2 \langle \sigma_{\gamma_{D}\gamma_{D}\rightarrow SM} v \rangle}\, ,
\end{equation}
which agrees well with the result obtained from a numerical integration of the system~(\ref{eq:boltzmann_simplified}), and should be compared with the estimate in absence of dark photon annihilations:
\begin{equation}
\label{eq:TEDMnaive}
\left(T_{\rm EMD}\right)_{\rm naive}\simeq \zeta(3)\dfrac{30}{\pi^{4}}\dfrac{3}{g_*^{\rm SM}(T_{\rm EMD})}\; m_{\gamma_{D}}\, .
\end{equation}
Figure~\ref{fig:dilution_BBN} shows, for benchmark values of $\alpha_D$ and $a$, the isocurve in the $(\ms,\varepsilon)$ plane that reproduces the observed DM abundance in our model (solid line) and in models where Eq.~(\ref{eq:TEDMnaive}) applies (dashed line). 
%%%%%%%%%%%%%%%%%%%%%%%%%%
%%%%%%%%%%%%%%%%%%%%%%%%%%
While dilution is an important effect in the latter scenario, in our model it occurs for values of $\varepsilon$ excluded by the bounds on the dark photon lifetime discussed below. The same conclusion holds in the majority of the interesting parameter space.

\subsubsection*{Bounds on the dark photon lifetime}
\enlargethispage{\baselineskip}
The evolution of the Universe at temperatures below $1 \,\rm MeV$ is accurately described by the standard cosmological model starting with BBN until the current epoch. Decays of long-lived particles during or after BBN can alter this picture and are generally excluded. Strong and robust constraints come in particular from observations of light elements abundances produced during the BBN (see~\cite{Kawasaki:2017bqm}) and from the CMB spectrum (see~\cite{Poulin:2016anj,Slatyer:2016qyl}). To comply with these bounds, we require that metastable dark photons decay before BBN, i.e. we impose $T_{\rm decay} \gtrsim 10\,\rm MeV$.

In presence of an era of early matter domination, the decay temperature $T_{\rm decay}$ can be expressed in terms of the fundamental parameters by taking into account that the relation between time and temperature is modified with respect to the standard radiation-dominated case.
The equations governing the system are the Friedmann equations plus the equation for the evolution of the radiation and matter densities (see for instance~\cite{Contino:2018crt}). An approximate analytic solution can be obtained in the early matter dominated era; in particular, the time interval comprised between the freeze-out and a moment in time well into the early matter domination era can be expressed as:
\begin{equation}
\label{eq:timedecay}
\begin{split}
t-t_{\rm f.o.,\gamma_{D}} \simeq \,
& \sqrt{\dfrac{90}{32\pi^{3}g_{\rm f.o.}}} M_{\rm Pl}  \left( \dfrac{r^{2/3}}{T_{\rm EMD}^{2}} - \dfrac{1}{T_{\rm f.o.,\gamma_{D}}^{2}}\right) 
 + \sqrt{\dfrac{5}{\pi^{3}g_{\rm EMD}}} \dfrac{M_{\rm Pl}}{T_{\rm EMD}^{2}} \times \\[0.2cm]
& \times \left[ \dfrac{T_{\rm EMD}^{4}}{T^{4}} \left( \dfrac{\Gamma_{\gamma_{D}} M_{\rm Pl}}{{\displaystyle\frac{5}{2}\left( \frac{4\pi^{3}}{45}g_{\rm EMD}\right)^{1/2}}T_{\rm EMD}^{2}+\Gamma_{\gamma_{D}} M_{\rm Pl}} \right)-1 \right] ,
\end{split}
\end{equation}
where $g_{\rm f.o.} \equiv g_*^{\rm SM}(T_{\rm f.o.,\gamma_{D}})$, $g_{\rm EMD} \equiv g_*^{\rm SM}(T_{\rm EMD})$, $r\equiv g_{*S}^{\rm SM}(T_{\rm f.o.,\gamma_{D}})/g_{*S}^{\rm SM}(T_{\rm EMD})$, and $g_{*S}^{\rm SM} (T)$ is the effective number of `entropic' relativistic degrees of freedom in the SM at the temperature $T$. 
The decay temperature $T_{\rm decay}$ is obtained by equating this time interval to the lifetime of the dark photon. In the limit $T_{\rm f.o.}, T_{\rm EMD} \gg T_{\rm decay}$, this reduces to the relation $ T_{\rm decay} = \sqrt{ \Gamma_{\gamma_{D}} M_{Pl}} \sqrt{3/(\pi^3 g_{\rm EMD})}$.
The constraint $T_{\rm decay} \gtrsim 10\,\rm MeV$ derived in this way excludes the blue region in the plot of Fig.~\ref{fig:dilution_BBN}. In practice, we find that  all the relevant parameter space is excluded for $10^{-24}\lesssim\varepsilon\lesssim 10^{-12}$.

\subsection{Cosmologically stable dark photons}
\label{sec:stablegaD}

In the limit of very small $\varepsilon$, i.e. for $\varepsilon \lesssim 10^{-24}$, the dark photons are cosmologically stable. Such small values of the mixing parameter are technically natural even though the dark sector comprises particles with SM charges, thanks to the dark charge conjugation symmetry~\eqref{eq:c_dark}. On the other hand, operators with two dark photons are not forbidden by $\CD$ and will be generated at loop-level, inducing dark photon annihilations into SM particles. For instance, loops of NGB triplets mediate annihilations into pairs of $W$ bosons with a cross section of order
\begin{equation}
\label{eq:gaDgaDtoWW}
\langle \sigma_{\gamma_{D}\gamma_{D}\rightarrow WW} v  \rangle \sim \dfrac{1}{8\pi} \alpha_{D}^{2} \alpha_{2}^{2} \frac{m_{\gamma_{D}}^{2}}{m_{3}^{4}}\, .
\end{equation}
These processes could set the abundance of dark photons to reproduce the observed DM one. The lower bound on triplets from collider searches derived in Sec.~\ref{sec:coll}, $m_{3} \gtrsim 300\, \rm GeV$, implies however that the cross section of Eq.~(\ref{eq:gaDgaDtoWW}) is always too small if $m_{\gamma_{D}}<m_3$, and the energy density of dark photons would overclose the Universe. Conversely, for $m_{\gamma_{D}}>m_3$ the dark photon is no longer stable, as it can decay to triplets and $W$ bosons.

\subsection{Alternative choices of parameters and cosmological scenarios}
\label{sec:alternatives}

So far we have described the cosmological evolution predicted in the bulk of the parameter space of our model. We now analyze some interesting limits where the thermal history is significantly different. These are obtained by varying the parameters $\alpha_D$ and $a$.

\subsubsection*{Varying $\alpha_{D}$}

As the strength of the dark coupling $\alpha_D$ increases, the abundances of dark pions and dark baryons become comparable, see Eq.~(\ref{eq:ratioOmegas}). For $\alpha_{D}\gtrsim 0.4 $ dark baryons become the dominant component of dark matter, with a dynamical scale of order $\Lambda_{DC} \sim 50\div 100\,$TeV.

Moreover, for couplings $\alpha_D \gtrsim \alpha_2$ the neutral triplets $3_{0},3'_{0}$ become lighter than the singlets and the dark photons. In this scenario the kinematics is reversed with respect to section~\ref{sec:decaychannels}. The dark photon decays to one $3'_{0}$ plus one electroweak boson $V$ through the anomaly, while the triplet $3_{0}$ can decay to SM particles at the two-loop order with a rate independent of $\varepsilon$. On the other hand, the $3'_{0}$ is metastable thanks to the $\mathcal{C}_{D}$ symmetry and can decay only through $\varepsilon$ suppressed interactions. For very small or zero kinetic mixing, $\varepsilon \lesssim 10^{-24}$, the $3'_{0}$ is an interesting example of scalar triplet candidate of Minimal Dark Matter, with improved accidental stability and a dynamical mass. The dark matter would be multicomponent, with triplets $3'_{0}$, singlets $1_{\pm}$ and dark baryons, each protected by a symmetry. For larger kinetic mixings, $\varepsilon > 10^{-24}$ , the triplets are too short lived to be a DM candidate and the scenario is similar to the one described previously with now the $3'_0$ playing the role of metastable species.

In the case of small mass splitting $(\mt-\ms)/ \ms \lesssim 0.1$, corresponding to dark couplings of order $\alpha_D \sim \alpha_2/2a$, co-annihilations could also play a role in determining the relic abundance. We do not attempt an analysis of these effects, leaving it to a future work.

\subsubsection*{Varying $a$}

The other parameter that can be varied and has a large impact on the dynamics of the model is the chiral charge $a$.  For $a>0.9$ the dark photon becomes much lighter than the singlet, approaching the massless limit as $a \rightarrow 1$. For light dark photons, the dark sector interaction can become long-ranged and effects such as Sommerfeld enhancement and bound state formation in annihilation processes should be properly taken into account to have reliable predictions.

For sufficiently small $a$ (e.g. $a\lesssim 0.26$ for $N_{DC}=4$), the dark photon becomes heavier than the NGB singlet. The latter will thus annihilate into SM particles either through virtual dark photon exchange, with an $\varepsilon$-suppressed cross section, or though loops of triplets. In the model with SM-neutral dark fermions of Ref.~\cite{Harigaya:2016rwr}, only the first process is possible and reproducing the correct DM abundance requires low values of the dynamical scale $\LDC$ for small $\varepsilon$. In our model, on the other hand, annihilations mediated by dark photons turn out to be inefficient for $\varepsilon \lesssim 10^{-2}$, and in this limit the DM annihilates into $W$ bosons through the operators of Eq.~\eqref{eq:effective operators} induced by loops of triplets. At the same time, the dynamical scale cannot be too small in light of the collider bounds on NGB triplets discussed in Sec.~\ref{sec:coll}. Enforcing these bounds and requiring the correct DM abundance implies a lower bound on the mass of the singlets $m_{1} \gtrsim 130 \, \rm GeV$.
For $\varepsilon \lesssim 10^{-24}$ and dark photon masses in the range $\ms < m_{\gamma_D} < 2\ms$ (corresponding to $0.10 \lesssim a \lesssim 0.26$ for $\NDC=4$), the dark photon is cosmologically stable and can annihilate efficiently into NGB singlets. Its energy density is thus small enough not to overclose the Universe (compare with Sec.~\ref{sec:stablegaD}), and gives an additional subdominant contribution to the DM abundance.

Finally, in the limit $a\rightarrow 0$, the NGB singlets become very light and are a component of dark radiation, rather than dark matter.

\subsubsection*{The case $a=0$}

As explained in section~\ref{sec:themodel}, for $a=0$ the model has an enhanced global symmetry. The~$1_{\pm}$'s become exact Nambu-Goldstone bosons (up to the effect of higher-dimensional operators), with only shift-symmetric interactions, and are massless. The analysis is qualitatively similar to the one performed in Ref.~\cite{Co:2016akw} for the model with SM-neutral fermions, with some qualifications.

The thermal abundance of dark baryons must account for all of the dark matter, and this fixes the dynamical scale to $\LDC \sim 50\div 100 \, \,$TeV. The massless singlets $1_{\pm}$, on the other hand, behave as a component of dark radiation. Their abundance can be expressed in terms of an effective number of neutrinos:
\begin{equation}
\Delta N_{eff} = \dfrac{8}{7} \left( \dfrac{g_*^{SM}(1 \, \rm MeV)}{g_*^{SM}(T_{dec})}\right)^{\frac{4}{3}},
\end{equation}
where $T_{dec}$ is the temperature at which the SM bath and dark radiation thermally decouple from each other. The number of SM degrees of freedom saturates at $(g_*^{SM})_{\rm max}= 106.75$ for large enough decoupling temperatures, implying $\Delta N_{eff} \gtrsim 0.05$.

There are two categories of processes that can keep the $1_\pm$ in thermal equilibrium: elastic scatterings mediated by dark photons -- with a cross section suppressed by $\varepsilon$ -- and processes mediated by loops of triplets through the effective operators~\eqref{eq:effective operators}. The second class of interactions breaks the correlation between the cross section for DM direct detection and the number of relativistic degrees of freedom described in Ref.~\cite{Co:2016akw}. Moreover, it implies an upper bound on the decoupling temperature, valid also in the limit of $\varepsilon$ small:
\begin{equation}
\label{eq:tdecbound}
T_{dec} < 2.5 \left( \dfrac{\Lambda_{DC}}{\rm TeV} \right)^{\frac{8}{7}} \, \rm GeV.
\end{equation}
This differs from model of Ref.~\cite{Co:2016akw}, where the decoupling temperature can be arbitrarily high depending on $\varepsilon$. 

For dynamical scales $\Lambda_{DC} \sim 50 \div 100 \,$TeV,  the upper bound (\ref{eq:tdecbound}) is not strong enough to further constrain $\Delta N_{eff}$. Furthermore, the observational bound $\Delta N_{eff} <0.3$~\cite{Aghanim:2018eyx} leaves unconstrained a wide region of the parameter space, corresponding to $\alpha_{D}\geq 10^{-8}$, and $\varepsilon \leq 10^{-3}$. In any point of such region, $\Delta N_{eff}$ falls in the interesting range that will be probed by future CMB experiments, see for example~\cite{Abazajian:2016yjj}.
Complementary observables can be dark photon searches and direct detection signals from scattering of dark baryons on nucleons, depending on $\alpha_{D}$ and $\varepsilon$~\cite{Co:2016akw}. 

Finally, we notice that in this model the dark radiation and the dark baryon component of dark matter are tightly coupled due to dark meson-dark baryon interactions. This effect can have important implications on structure formation, suppressing structures on small scales and potentially alleviating the $\sigma_{8}$ and $H_{0}$ problem~\cite{Lesgourgues:2015wza,Buen-Abad:2017gxg}, see also~\cite{Chacko:2016kgg,Ko:2017uyb}. We leave a detailed analysis of this scenario to the future.

\section{Constraints from direct and indirect DM searches}
\label{sec:DMsearches}

This section discusses the constraints set by direct and indirect DM searches on the model with $\SUEW$ doublets. Bounds from experiments at high-energy colliders are analysed in Sec.~\ref{sec:coll}.

\subsection{Direct detection}

The elastic scattering of DM particles on nuclei gives rise to recoil signals that are being looked for in dedicated high-precision experiments. For values of $\varepsilon$ large enough, the main contribution to the elastic cross section comes from the tree-level exchange of the $Z$ and the dark photon. This is spin independent in the non-relativistic limit and strongly constrained by conventional direct-detection experiments. We performed a calculation valid for arbitrary values of $\varepsilon$ and of the vector boson masses.~\footnote{Our result is valid also in the case of mass resonance $m_{Z}=m_{\gamma_{D}}$. We assume that the on-shell dark photon mass is larger than the typical momentum exchanged.} We find that, once the non-relativistic limit is taken, the spin-independent cross section per nucleon has a very simple form: 
\begin{equation}
\label{eq:SIxsec}
\sigma_{\pi N}^{\rm S.I.}= \varepsilon^{2}  \dfrac{\mu_{\pi N}^{2}}{\pi} \dfrac{ Q_D^{2} e^{2} \cos^{2}\theta_{W}}{m_{\gamma_{D}}^{4}} \left( \dfrac{Z}{A} \right)^{2} + \mathcal{O}\!\left(\dfrac{|\vec p|^{2}}{m_{Z,\gamma_{D}}^{2}}\right),
\end{equation}
where $Q_{D}=(1+a)e_{D}$ for dark pions, $\theta_{W}$ is the weak mixing angle, $Z,A$ are respectively the atomic and mass number of the target nuclei, $e$ is the electromagnetic coupling, and $\mu_{\pi N}$ is the reduced mass of the dark matter-nucleon system.
Up to higher-order corrections in the momentum expansion, DM particles interact only with protons (i.e. the contribution from scattering off neutrons vanishes), and the result scales as $\varepsilon^{2}$. Our formula is valid for generic models where the dark matter is a Dirac fermion or a complex scalar that couples with charge $Q_D$ to kinetically-mixed dark photons or $Z'$ bosons.~\footnote{Notice that the spontaneous breaking of $\UoneD$, which is a direct consequence of its being chiral, leads to  modified quartic interactions which are however not relevant for the tree-level cross section.}
It agrees with the result of Ref.~\cite{Co:2016akw} and with previous literature on dark photons, see for example Refs.~\cite{Cline:2014dwa,Escudero:2017yia,Evans:2017kti}. We checked the correctness of Eq.~(\ref{eq:SIxsec}) by performing the calculation in two different ways: first, by diagonalising the kinetic and mass terms and deriving the modified couplings reported in Appendix~\ref{app:gammad}; second, by working in the non-diagonal basis and computing the propagators including gauge-boson mixing. Both methods agree and give the simple result of Eq.~(\ref{eq:SIxsec}).

An additional contribution to the elastic scattering of dark pions off nuclei comes from effective operators generated by loops of NGB triplets. Those of Eq.~(\ref{eq:effective operators}), for example, arise at the 1-loop level and mediate the DM scattering through 1-loop diagrams. Other operators like $\pi^\dagger i\!\overleftrightarrow{\partial}_{\!\!\mu} \pi J_{SM}^\mu$, where $J_{SM}^\mu$ is a SM quark or Higgs current, are generated at two loops and mediate the DM scattering at tree level. Using the analysis of Ref.~\cite{Kavanagh:2018xeh}, we estimate that the effect of any of these operators in our model is too small to be detected and does not lead to any bound.

\begin{figure}[t]
\centering
\includegraphics[width=0.7\textwidth]{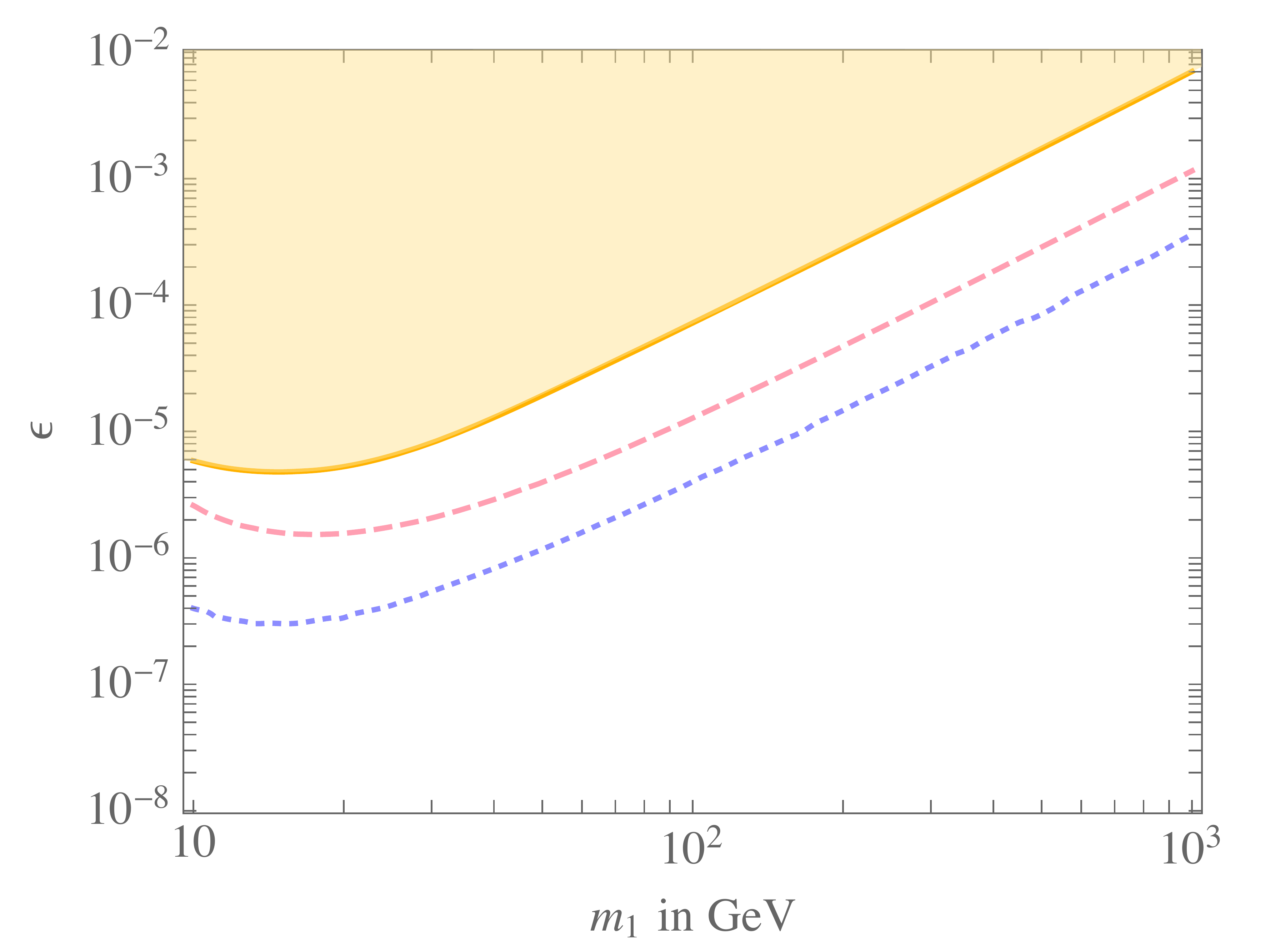}
\caption{Direct detection bounds on the kinetic mixing parameter $\varepsilon$ as a function of the dark matter mass $m_{1}$. We set $a=1/2$ and fixed $\alpha_{D}$ such that the thermal relic abundance of dark pions reproduces the correct DM density. The solid yellow line corresponds to the spin-independent bound set by Xenon1T, while the pink dashed and purple dotted lines are respectively the projected sensitivity of the LZ experiment and the neutrino floor.}
\label{fig:direct detection}
\end{figure}

We thus focus on the scattering of dark pions mediated by the $Z$ and the dark photon, and use Eq.~(\ref{eq:SIxsec}) to derive the constraints from direct-detection experiments in the $(\varepsilon,\ms)$ plane. In the mass range of interest, the strongest bounds currently come from Xenon1T~\cite{Aprile:2018dbl}, while PandaX-II~\cite{Cui:2017nnn} and LUX~\cite{Akerib:2016vxi} give comparable though weaker limits. We show the corresponding exclusion curve in Fig.~\ref{fig:direct detection}, together with the projected sensitivity of the future LZ experiment~\cite{Akerib:2015cja} and the neutrino floor curve. 
%%%%%%%%%%%%%%%%%%%%%%%%
%%%%%%%%%%%%%%%%%%%%%%%%
We focus on dark pion masses up to $1\,$TeV, as already for $\ms \sim 10\,$TeV dark baryons contribute significantly to the relic density and deriving an accurate bound would require knowing precisely their fraction. Notice also that the data of Ref.~\cite{Aprile:2018dbl} extend up to $1\,$TeV, though a linear extrapolation to higher masses seems reasonable and could be done.

\subsection{Limits on charged relics}
\label{sec:chargedrelics}

Theories with odd $\NDC$ contain dark baryons with weak isospin equal to $1/2$ and $3/2$ and with zero hypercharge, thus carrying half-integer (hence non-vanishing) electromagnetic charge. Due to their accidental stability, the dark baryons form an electrically-charged subdominant component of the thermal relic abundance. We denote this charged dark matter component as cDM in the rest of this section. 

Even though the cDM fraction can be parametrically suppressed, this scenario is subject to very stringent constraints. 
The analysis of Refs.~\cite{Dolgov:2013una,dePutter:2018xte} makes use of CMB data to derive limits on the fraction of charged relics, but assumes that the Compton scattering of these particles is negligible. While this is a good approximation in the case of milli-charged dark matter, it is not so in our theories, where the cDM-baryon scattering is efficient. Therefore, the bounds of Refs.~\cite{Dolgov:2013una,dePutter:2018xte} cannot be applied in our scenario.
On the other hand, cosmic rays with order-one charge can produce ionization signals that can be detected by experiments looking for ionizing particles. Bounds on charged stable particles with $q\sim |e|$ are subject to uncertainties in the mass window $(10^{5} \div 10^{11})\,\rm GeV$, due to the large impact that supernovae shock waves can have on their galactic and momentum distribution (see for instance Refs.~\cite{Chuzhoy:2008zy} and~\cite{Dunsky:2018mqs}). In our theories, however, dark baryons are thermal relics and there is an upper bound on their mass of order $10^{5}\, \rm GeV$. For these values of masses, uncertainties should be under control.
The analysis of Ref.~\cite{Dunsky:2018mqs} constrains the mass fraction of charged relics to be extremely small: $\Omega_{\rm cDM} \lesssim (10^{-10} \div 10^{-14}) \,\Omega_{\rm DM}$, depending on the experiment and on the mass. Combining this result with the collider bounds of Sec.~\ref{sec:coll} excludes the whole parameter space of interest for odd $\NDC$.

\subsection{Dark matter annihilation signals}

Residual annihilations of relic particles can produce cosmic rays and energetic photon signals, tested by indirect detection experiments.
 
Observations of the positron and antiproton flux rates by AMS-02 can provide strong limits on DM annihilations in the parameter space of our model. However, these constraints are subject to large uncertainties associated with cosmic ray propagation and the estimate of astrophysical backgrounds, and for this reason we will not use them in the following.

The observation of gamma ray signals from clean DM-dominated environments, such as dwarf spheroidal galaxies, can also set stringent bounds provided that DM annihilations produce a large flux of photons. In our case, the dark matter mostly consists of  SM-singlet dark pions that annihilate into dark photons. The dark photons decay in flight to SM final states. We do not attempt here a detailed analysis and a computation of the produced photon spectrum. 
We use the bounds of Ref.~\cite{Profumo:2017obk} on dark matter annihilations to short-lived mediators with mass $m_{\rm mediator}\sim m_{\rm DM}$, obtained from the recast of the FERMI-LAT results of Ref.~\cite{Ackermann:2015zua}. Since dark photons decay to hadronic final states with a large branching ratio~\cite{Cirelli:2016rnw}, we make use of the bounds that assume fully hadronic decays of the mediators.
In most of the parameter space of our model, the dark pion and the dark photon have comparable mass and effects from Sommerfeld enhancement and bound state formation are negligible (see however the discussion in Sec.~\ref{sec:alternatives}). With this assumption, we find that the measured flux of gamma rays from dwarf spheroidal galaxies of Ref.~\cite{Ackermann:2015zua} excludes masses $m_{1}\lesssim 100 \,\rm GeV$.~\footnote{This result is valid both if dark pions have a thermal abundance (hence they do not reproduce the observed DM density in generic points of the parameter space), and if they have the observed DM abundance (hence in general they are not thermal relics). In the latter case, an additional portion of the parameter space can be excluded.} This limit is expected to improve in the future thanks to the discoveries of new dSphs galaxies from LSST combined with continued Fermi-LAT observations~\cite{Drlica-Wagner:2019xan}, reaching masses $\ms \lesssim 400 \,\rm GeV$.

For completeness we have also analyzed the CMB limits on dark matter annihilation from the 2018 Planck release~\cite{Aghanim:2018eyx}, that are less stringent but have independent uncertainties. We find that these data exclude masses $\ms\lesssim 10 \,\rm GeV$.

In the future, the Cherenkov Telescope Array (CTA) observatory will start probing thermal dark matter in the $\rm TeV$ mass range through observations of the Galactic Center. Assuming an Einasto dark matter profile, the expected sensitivity~\cite{Acharyya:2020sbj} for thermal dark matter candidates is in the range $400 \,{\rm GeV} \lesssim \ms \lesssim 10 \,{\rm TeV}$. CTA will thus be able to probe a large part of the parameter space relevant for the dark matter models presented in this work.

\section{Constraints from collider searches}
\label{sec:coll}

The dark sector can be probed at high-energy colliders through SM gauge interactions  in a way that is complementary to direct and astrophysical searches. Unlike kinetic mixing, whose effects strongly depend on the value of~$\varepsilon$, a gauge portal to the SM does not introduce any unknown couplings and provides sharper predictions. For values of the dark confinement scale under consideration, the production of the lightest SM-charged particles in the model is within the reach of the LHC or of one of its future extensions. In the rest of this section we will be concerned with the analysis of the constraints arising from existing collider data, and we will discuss the expected reach of a Future Circular Collider (FCC) in Sec.~\ref{sec:summary}. Given the rich phenomenology predicted by our model at colliders, our study should be considered as an exploratory one, to be completed in a future work.
See Refs.~\cite{Barducci:2018yer,Kribs:2018ilo} for related studies.

\subsection{Production and decays of NGB triplets}\label{ssc:triprod}

\begin{figure}[t]
\centering
\includegraphics[width=0.7\textwidth]{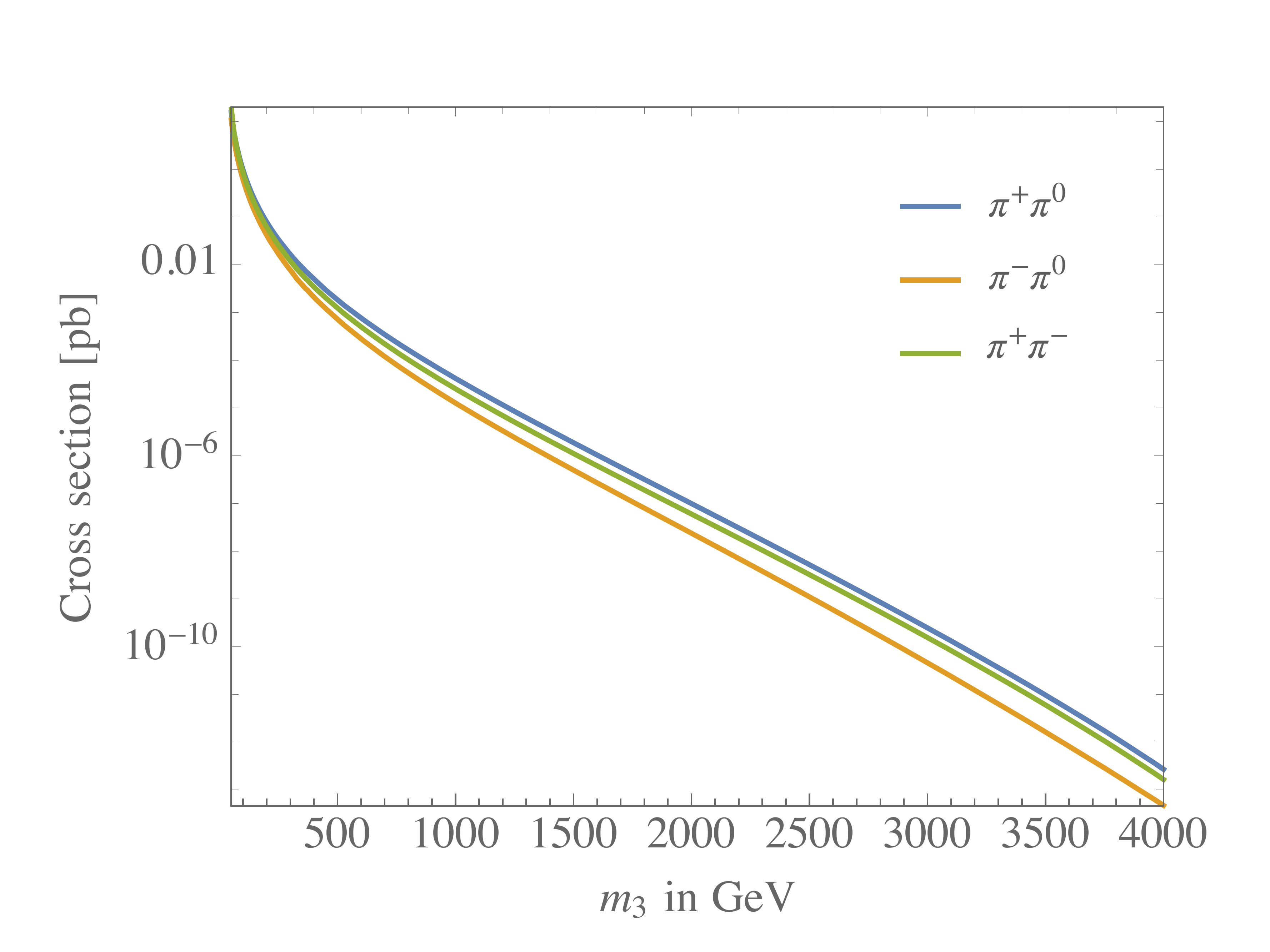}
\caption{Leading-order cross section for $\pi^+ \pi^0$ (blue), $\pi^- \pi^0 $  (orange) and $\pi^+ \pi^- $ (green) pair production at the $13\,$TeV LHC, where $\pi^{\pm,0}$ denotes any of the NGB triplets with electromagnetic charge indicated by the superscript.}
\label{fig:crossprod}
\end{figure}

\begin{figure}[H]
\centering
\includegraphics[width=0.75\textwidth]{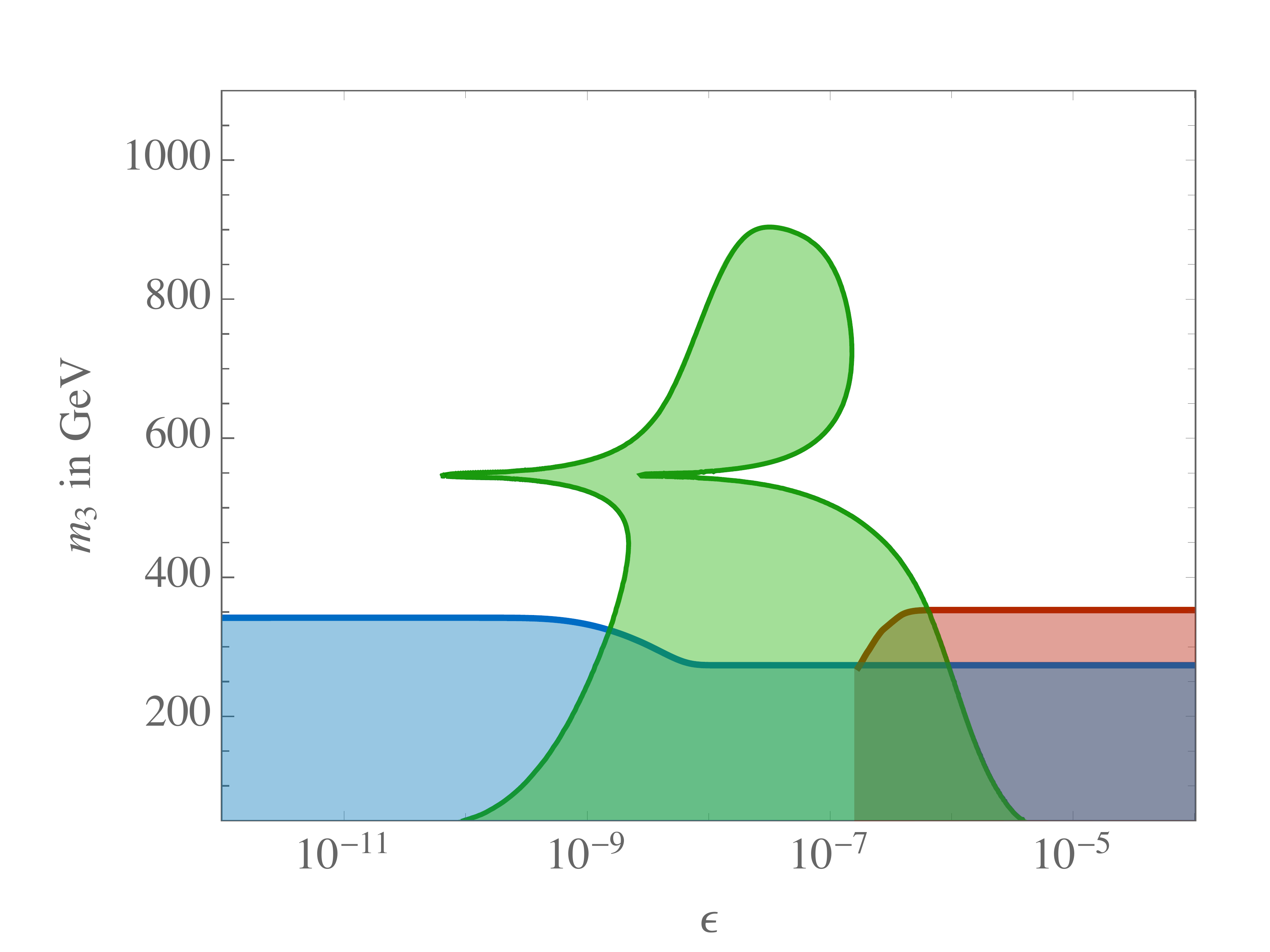}
\includegraphics[width=0.75\textwidth]{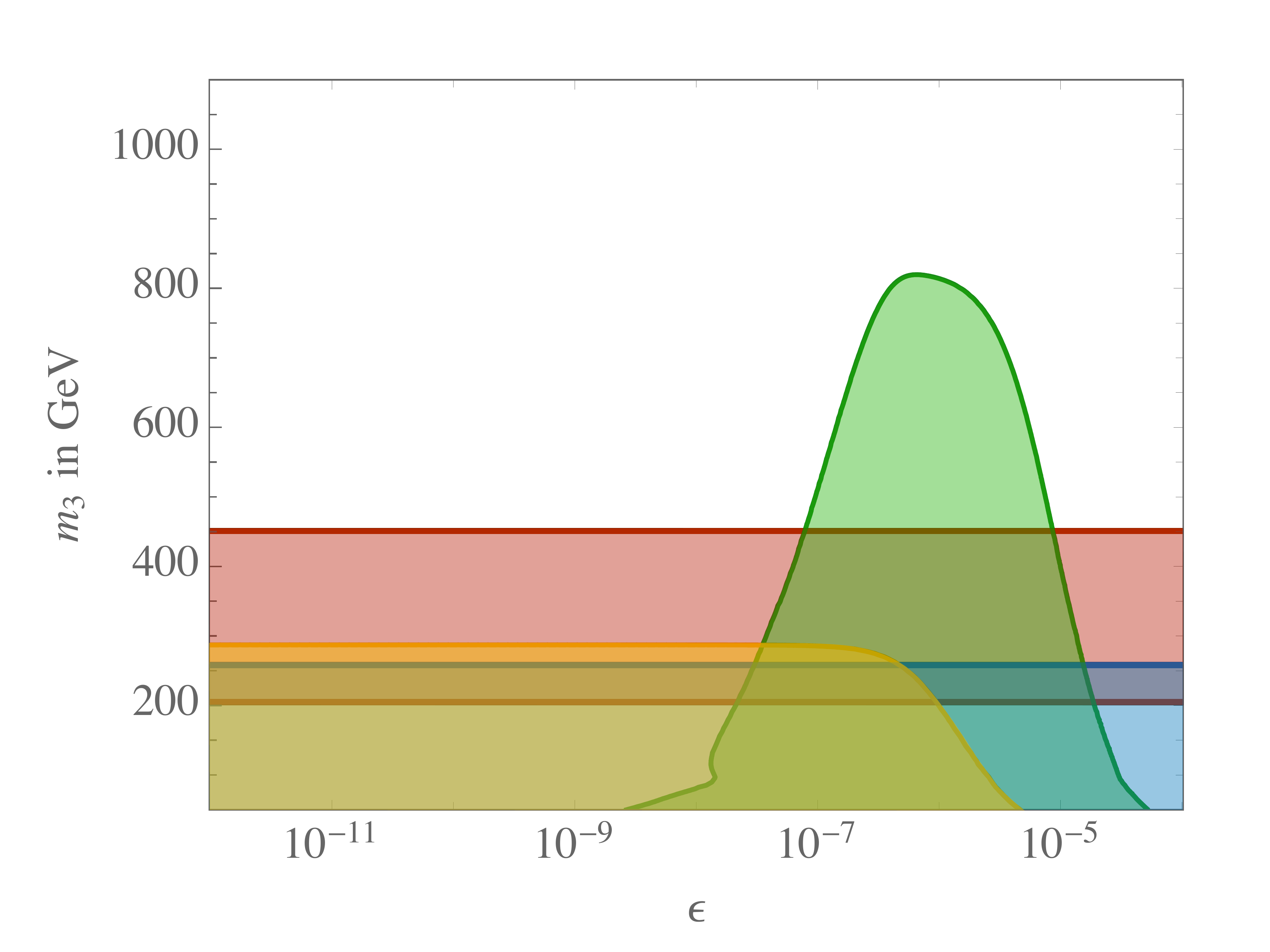}
\caption{Exclusions from four different kinds of collider searches as a function of $\varepsilon$: missing energy events (blue region), displaced decays (green region), resonant decays to jet or lepton pairs (red region), and disappearing tracks (yellow region). The top panel assumes $\mt =3\, \ms$ (normal hierarchy), while in the bottom panel $\mt = \ms/3$ (inverted hierarchy). Both plots are done setting $\ms=2\,m_{\gamma_D}$. The sharp cut-off of the bound from resonant decays at $\varepsilon \sim 2\times 10^{-7}$ in the top panel is due to a lower bound on the dark photon mass imposed in the experimental analysis of Ref.~\cite{CMS-PAS-EXO-19-018}. The two cusps of the DV region in the top panel are a consequence of the resonant growth of the dark photon width close to $m_{\gamma_D}=m_Z$, which for a fixed decay length can be compensated by a decrease in $\varepsilon$. The bound from disappearing tracks has been derived from the analysis of Ref.~\cite{Chiang:2020rcv}.}
\label{fig:comparison}
\end{figure}

In the model with $\SUEW$ doublets, the most promising process to probe the dark sector at colliders seems to be pair production of the NGB triplets; singlets couple either through~$\varepsilon$ or via non-renormalizable operators, and their direct production cross section is correspondingly suppressed. Triplets can be pair produced either via a Drell-Yan process or resonantly through the decay of a spin-1 dark meson. Since $m_{\rho} \sim 7 \,  m_{\pi}$ in our model, the latter contribution is subdominant and will be neglected in the following. For Drell-Yan production, the spin and color-averaged partonic cross sections at leading order are given in Eqs.~(\ref{eq:xsec1})-(\ref{eq:xsec3}) of Appendix~\ref{app:darkpions}. The corresponding hadronic cross sections have been obtained by convoluting those expressions with the 2014 MMHT parton distribution functions (PDFs)~\cite{Harland-Lang:2014zoa}, and the result is shown in Fig.~\ref{fig:crossprod}.
Once produced, the triplets are generally unstable and decay as shown in Tab.~\ref{tab:dec}, where for illustration purposes we set $\ms= 2 m_{\gamma_D}$. 
%%%%%%%%%%%%%%%%%%%%
\begin{table}[t]
\centering
\begin{tabular}{l:l|l:l}
 $m_3 < m_{\gamma_D} = m_1/2$ & $m_3<2 m_{\gamma_D} = m_1$ & $m_3>2 m_{\gamma_D} = m_1$ & $m_3 > \sqrt{3} m_1$\\[0.05cm] \hline %\rule{0pt}{4ex}
&&&\\[-0.4cm]
$3_{\pm}\longrightarrow 1_{\pm}\, V \,\,$ & $3_{\pm}\longrightarrow 1_{\pm}\, V \,\,$ & $3_{\pm}\longrightarrow 1_{\pm}\, V $ & $3_{\pm}\longrightarrow 1_{\pm}\, V\,(\gamma_D \gamma_D)$ \\[0.05cm]
$3_0 \longrightarrow \bar{f} f/VV/Vh$ & $ 3_0 \longrightarrow \bar{f} f/VV/Vh$ & $3_0 \longrightarrow V \,\, \gamma_D \, \gamma_D $ & $3_0 \longrightarrow V \,\, \gamma_D \, \gamma_D $\\[0.05cm]
$3_0' \longrightarrow V \, \bar{f} f$ & $3_0' \longrightarrow V \, \gamma_D$ & $3_0' \longrightarrow V \, \gamma_D $ & $3_0' \longrightarrow V \, \gamma_D $ \\[0.05cm]
$3_0^{\prime\,\pm} \longrightarrow 3_0^{\prime\, 0} \pi^\pm$ &&&
\end{tabular}
\caption{Main triplet decay modes in the various kinematic regimes, assuming $m_1= 2 m_{\gamma_D}$. Here $V$ can be any electroweak gauge boson, depending on the electromagnetic charge of the initial triplet, $h$ is the Higgs boson, while $\bar f f$ denotes a pair of SM fermions. In the case of inverted hierarchy, decays among different components of the $3_0'$ become important for $\varepsilon$ small, when the rate of $3_0' \to V \, \bar{f} f$ is suppressed.}
\label{tab:dec}
\end{table}
%%%%%%%%%%%%%%%%%%%%
Broadly speaking, one can distinguish two main regions of parameter space exhibiting different phenomenologies: 
if $\mt > m_{\gamma_D}$, then (all) the neutral triplets decay by emitting dark photons, otherwise they decay to SM particles. We shall refer to these as the normal and inverted hierarchies respectively. 
While the singlets always escape detection and are recorded as missing energy, the signature of the dark photons produced in the final state depends on their lifetime, which is a function of $\varepsilon$. In the case of an inverted hierarchy, the $3_0'$ is also long lived for small $\varepsilon$. One can thus distinguish four kinds of possible experimental signatures characterizing the final state:
\begin{itemize}
\item {\bf{Missing Energy:}} For very small $\varepsilon$ the dark photons decay outside the detector and, together with the NGB singlets, give rise to missing energy in collider events. The signatures in this case are similar to those of Supersymmetric models (where decaying charginos play the role of the triplets), and SUSY searches can be exploited to derive bounds on our scenario.
\item  {\bf{Displaced vertices:}} In a large portion of parameter space, for small $\varepsilon$, the $\gamma_D$ and $3_0'$ can decay inside the detector far from the interaction point. Due to the extremely low background, events with such displaced vertices lead to the strongest constraints in the region of the parameter space where they apply. 
\item {\bf{Disappearing tracks:}} For $\varepsilon \lesssim 3 \times 10^{-6} (2 \mt/\ms)^{3/2}(1\,\text{TeV}/ \ms)^{1/2}$ and in the case of an inverted hierarchy, the electromagnetically-charged components of $3_0'$ mostly decay into the neutral one by emitting a soft pion. The latter goes undetected at high-energy colliders and the decaying particle manifests itself as a disappearing track. The same signature characterizes minimal DM models.
\item  {\bf{Prompt resonant decays:}} For large $\varepsilon$, dark photons decay promptly in the detector and can be reconstructed as peaks in the invariant mass spectrum of jet or lepton/anti-lepton pairs. A similar resonant signature comes from the decays of $3_0$'s in the case of inverted hierarchy. 
Traditional searches for $Z'$ resonances can be exploited in this case to set constraints on our model.
\end{itemize}
In order to illustrate the relative importance of these signatures in testing our model, we anticipate the results of the analysis performed in the next section and show in Fig.~\ref{fig:comparison} the bounds in the $(\varepsilon,\mt)$ plane for fixed ratios of the masses. We find that displaced decays can give the strongest bounds, followed by prompt resonant decays. The limits set by each of the different experimental signatures are in fact similar in strength, despite the different strategies and backgrounds involved. This can be understood as the result of the strong dependence of the production cross section on the triplet mass, mostly due to the scaling of the proton PDFs. The bounds we obtain are not far from the value of $\mt$ at which the number of signal events becomes of order unity, i.e. from the strongest obtainable bound. For this reason, although our analysis makes use of many simplifying approximations, we believe that its results give a good estimate of the actual constraints.

\subsection{Bounds}

In this section we derive the bounds on the triplet mass from each of the signatures discussed above, using data from ATLAS and CMS. For both dark photon and triplet decays, the final yields of leptons and hadrons are of the same order of magnitude but, except for displaced vertices, hadronic events have always a much higher background.  For this reason, when analyzing missing energy events and searches for promptly-decaying resonances we shall concentrate on final states containing electrons or muons.

%%%%%%%%%%%%%%%%%%%%%%
%%%%%%%%%%%%%%%%%%%%%%

\begin{figure}[tp]
\centering
\includegraphics[width=0.7\textwidth]{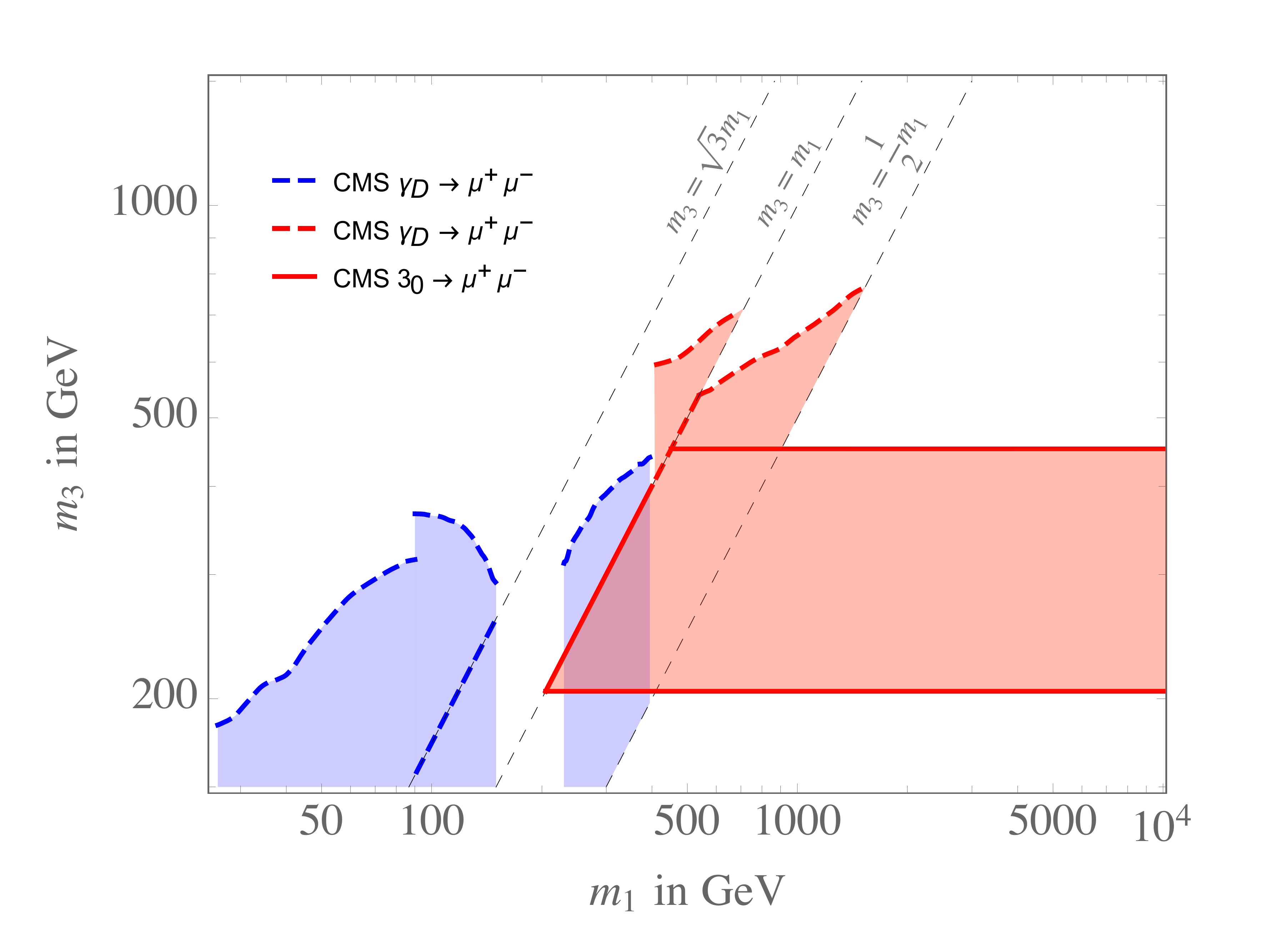}
\caption{Bounds from the $Z'$ searches of Refs.~\cite{CMS-PAS-EXO-19-018,CMS-PAS-EXO-19-019} for $\ms=2 m_{\gamma_D}$ (corresponding to $a\simeq 1/2$ at $\NDC=4$).
The plane is divided into different kinematic regions (see Tab.~\ref{tab:dec}) by the gray thin dashed lines, inside which each constraint is valid. While exclusions denoted by continuous lines correspond to the decays of the $3_0$ and are valid for arbitrary values of~$\varepsilon$, dashed curves refer to $\gamma_{D}$ decays and only apply for $\varepsilon \gtrsim 10^{-7}$, i.e. when dark photons decay promptly.}
\label{fig:prompt}
\end{figure}

The $3_0$ decays promptly into a pair of SM particles if $\mt < 2 m_{\gamma_D}$, while the dark photon will do so for large $\varepsilon$. Thus, both particles may be observed as resonances in the mass spectrum of the final products. 
Traditional $Z'$ searches provide almost model-independent bounds on the production cross section times the branching ratio. It is then simple to recast these bounds into constraints on our model by inverting the triplet production cross section computed in the previous section.~\footnote{Although the limits from $Z'$ searches are strictly valid for spin-1 resonances only, we shall apply them also to the scalar $3_0$.} 
Figure~\ref{fig:prompt} shows the constraints set in the $(\ms,\mt)$ plane by the searches performed by the CMS collaboration into leptonic channels~\cite{CMS-PAS-EXO-19-018,CMS-PAS-EXO-19-019}. 
%%%%%%%%%%%%%%%%%%%%%%%%%%%%%
%%%%%%%%%%%%%%%%%%%%%%%%%%%%%
A similar search performed by ATLAS in Ref.~\cite{Aad:2019fac} sets slightly weaker bounds. Depending on the ratio $\ms/\mt$, the bounds come from the decays of the $\gamma_D$ (if $\varepsilon$ is large), the $3_0$ or both. If the decaying particle is a dark photon, its mass scales proportionally to that of the singlet, and this explains why the bounds grow quickly with $m_1$ (the background decreases with the mass of the decaying particle). The bounds from the decays of $3_0$, instead, are horizontal lines in the $(\ms,\mt)$ plane. Although the same analysis can be carried out with jets as final states, the corresponding bounds are weaker by at least one order of magnitude, because the slightly higher branching ratios are not able to compensate for the larger background (see for example Refs.~\cite{Sirunyan:2018xlo,Sirunyan:2018rlj,Aad:2019hjw}).

Recasting the bounds from missing energy searches at the LHC is not as simple as for resonant decays. This signature is common to many supersymmetric scenarios, and the data are usually interpreted as limits on the masses and couplings of the various SUSY particles, in a way that depends explicitly on the details of the model. To give an estimate of the bounds on our model, it is then simpler to take a step back and compare the total number of observed events with the theoretical prediction.~\footnote{A more refined analysis should take into account the distribution of the number of events with respect to the relevant kinematic variables, such as the transverse missing energy. This requires detailed numerical simulations and,  given the large number of final states, is beyond the scope of this paper. We leave such analysis to a future study.} 
The number of signal events is schematically modelled as
 \begin{equation}
\label{eq:ndm}
 N_{s}(\ms,\mt,\varepsilon) = \mathscr{L} \sum_i \sigma_{pp \rightarrow (\pi \pi)_i}(\ms,\mt) BR((\pi \pi)_i \rightarrow f) P(\ms, \varepsilon)\, \delta_{eff},
 \end{equation}
where $\mathscr{L}$ is the integrated luminosity and the sum runs over $(\pi\pi)_i = 3_0 3_0, 3_0' 3_0', 3_+ 3_-$ with all possible combinations of electromagnetic charges. The  branching ratios $BR((\pi \pi)_i \rightarrow f)$ are given in Tab.~\ref{tab:decays} for the final states $f$ of interest in the case of a normal hierarchy. 
%%%%%%%%%%%%%%%%%%%%%%%%%%%%%%%%%%%
 \begin{table}[tp]
\centering
\begin{tabular}{|c|l|l|l|}
\hline
Production mode & Primary decay products  & Final decay products  & B.R. \\ 
\hline
$3_0^0 \, 3_0^{\pm}$ & $4\gamma_D\,Z\,W^{\pm}\,$ &  $4\gamma_D \,\ell_1^+\,\ell_1^-\,\ell_2^{\pm}\,\overset{\textbf{\fontsize{5pt}{5pt}\selectfont(---)}}{\nu_2}$  & $1.7 \%$ \\
$3_0^0\, 3_0^{\pm}$ & $4\gamma_D\,\gamma\,W^{\pm}$ &  $4\gamma_D\,\ell^{\pm}\,\overset{\textbf{\fontsize{5pt}{5pt}\selectfont(---)}}{\nu}\,\gamma$  & $6.4 \%$\\
$3_0^+ \, 3_0^-$ & $4\gamma_D\,W^-W^+$ &  $4\gamma_D\,\ell_1^-\,\nu_1\,\ell_2^+\,\nu_{2}$  & $3.2\%$ \\
\hline \hline
$3_0^{\prime\, 0} \, 3_0^{\prime\,\pm}$ &  $2\gamma_D\,Z\,W^{\pm}\,$ & $2\gamma_D\,\ell_1^+\,\ell_1^-\,\ell_2^{\pm}\,\overset{\textbf{\fontsize{5pt}{5pt}\selectfont(---)}}{\nu_2}$  & $1.7 \%$ \\
$3_0^{\prime \, 0}\, 3_0^{\prime\,\pm}$ &  $2\gamma_D\,\gamma\,W^{\pm}$ &   $2\gamma_D \,\gamma\,\ell^{\pm}\,\overset{\textbf{\fontsize{5pt}{5pt}\selectfont(---)}}{\nu}$  & $6.4 \%$ \\
$3_0^{\prime\,+}\, 3_0^{\prime\,-}$ &  $2\gamma_D\,W^-W^+$ &  $4\gamma_D\,\ell_1^-\,\bar{\nu}_1\,\ell_2^+\,\nu_{2}$  & $3.2\%$ \\
\hline \hline
$3_{\pm}^0\, 3_{\mp}^{\pm}$ & $1_+1_-\,Z\, W^{\pm}\,(2/4\gamma_D)$ &  $1_+1_-\,\ell_1^+\,\ell_1^-\,\ell_2^{\pm}\,\overset{\textbf{\fontsize{5pt}{5pt}\selectfont(---)}}{\nu_2}\,(2/4\gamma_D)$  & $1.7 \%$ \\
$3_{\pm}^0\, 3_{\mp}^{\pm}$ & $1_+1_-\,\gamma\, W^{\pm}\,(2/4\gamma_D)$ &  $1_+1_-\,\gamma\,\ell^{\pm}\,\overset{\textbf{\fontsize{5pt}{5pt}\selectfont(---)}}{\nu}\,(2/4\gamma_D)$  & $6.4 \%$\\
$3_{\pm}^+\, 3_{\mp}^-$ & $1_+1_-\,W^-\,W^+\,(2/4\gamma_D)$ &  $1_+1_-\,\ell_1^-\,\bar{\nu}_1\,\ell_2^+\,\nu_{2}\,(2/4\gamma_D)$  & $3.2\%$ \\
\hline
\end{tabular}
\caption{Branching ratios into leptonic final states for each pair of triplets produced via Drell-Yan processes, assuming a normal hierarchy. Upper (lower) indices indicate the electromagnetic ($\UoneV$) charge of each particle. Leptons appearing in the third column can belong to any of the three SM families, while the values in the fourth column report the branching ratios into fully leptonic final states and include the branching ratio of taus into lighter leptons.}
\label{tab:decays}
\end{table}
%%%%%%%%%%%%%%%%%%%%%%%%%%%%%%%%%%%
The factor $P(m_{\gamma_D}, \varepsilon)$ corresponds to the probability that the decay of the $\gamma_D$ or $3_0'$ happens outside the detector, while $\delta_{eff}$ is a reconstruction efficiency.
Within a Bayesian framework, we set a 95\% probability limit on the maximum number of signal events compatible with the data, and then translate it into a bound in the $(\ms,\mt)$ plane using Eq.~(\ref{eq:ndm}). We assume a Poissonian distribution for the number of events, and model the background by using a log-normal distribution.
The mass of the triplets must be larger than half the $Z$ mass to pass the constraints on the $Z$ width from LEP. In practice, the particular value of this upper limit is irrelevant because the likelihood is exponentially suppressed for (much) lower values of $N_{s}$. We have thus assumed a flat prior on $N_{s}$ and set $\delta_{eff} = 0.35$ to reproduce the bounds of Ref.~\cite{Aad:2019vnb} on the mass of supersymmetric particles. Using the data from Refs.~\cite{Aad:2019vnb,Aad:2019vvi}, we obtain the $95 \%$ probability bounds on the number of signal events shown in Tab.~\ref{tab:result} in the channels with two or three leptons. 
%%%%%%%%%%%%%%%%%%%%%%%%
%\vspace{0.3cm}
\begin{table}[tp]
\centering
\begin{tabular}{|c|c|c|c|}
\hline
Channel & $N_b$ & $N_{obs}$ &  $N_s^\text{max}$ \\
\hline
3 leptons with OSSF pair ($+j$) & $69 \pm 5.5$ & $81$ & $ 31.5 $ \\
\hline
2 leptons (OSDF) ($+j$) & $172 \pm 17.5 $ & $170$ & $41.0$ \\
\hline
2 leptons (OSSF) ($+j$) & $269 \pm 17 $ & $267$ & $45.5$ \\
\hline
\end{tabular}
\caption{$95 \% $ probability bounds on the number of signal events, $N_{s}^\text{max}$, from the missing energy searches of Refs.~\cite{Aad:2019vnb,Aad:2019vvi} performed with $139\,\rm{fb}^{-1}$ of integrated luminosity collected at the $13\,$TeV LHC. The number of background ($N_b$) and observed ($N_{obs}$) events is reported in the second and third columns respectively. Leptons are either electrons or muons, and the acronyms stand for opposite sign same flavour (OSSF) and opposite sign different flavour (OSDF). The $+j$ in brackets refers to the possible inclusion of one jet, since initial-state radiation might be present.}
\label{tab:result}
\end{table}
%%%%%%%%%%%%%%%%%%%%%%%%
Lepton plus photon searches and searches for hadronic final states have larger backgrounds and give less stringent bounds, see for example Ref.~\cite{Sirunyan:2018psa} and Ref.~\cite{Sirunyan:2019ctn}.

\begin{figure}[tp]
\centering
\includegraphics[width=0.7\textwidth]{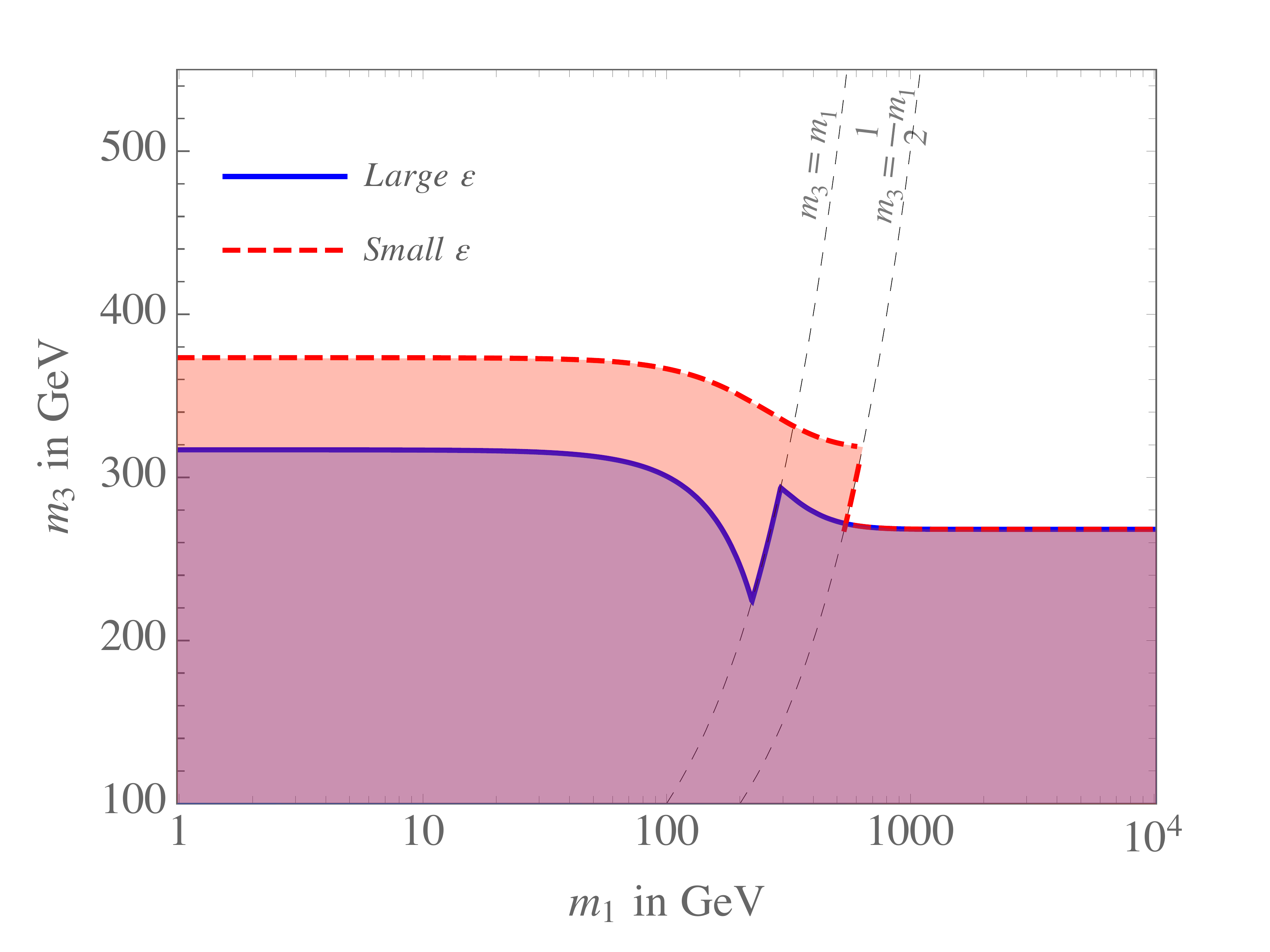}
\caption{$95\%$ probability exclusion regions in the $(\ms,\mt)$ plane from collider searches of events with three leptons and missing energy (first row of Table \ref{tab:result}). The dashed red curve assumes a value of $\varepsilon$ sufficiently small to have dark photon decays outside the detector, while the solid blue one applies for larger values of $\varepsilon$. Both curves assume $\ms=2 m_{\gamma_D}$. The gray thin dashed lines mark the different kinematic regions of Tab.~\ref{tab:dec}.}
\label{fig:met}
\end{figure}

Figure~\ref{fig:met} shows our recast in the plane $(\ms,\mt)$ of the limit from three-lepton events of Tab.~\ref{tab:result}, both for large and small values of $\varepsilon$.
%%%%%%%%%%%%
%%%%%%%%%%%%
Let us first consider the limit obtained assuming that $\varepsilon$ is sufficiently small to let the dark photons decay outside the detector (red dashed line of Fig.~\ref{fig:met}). When $\ms \ll \mt$, all four triplets are produced in the same amount and they all decay to missing energy, contributing to the bound. As $\ms$ increases, the charged triplets become heavier and their production cross section is suppressed, so they gradually become irrelevant. Furthermore, while the decays of $3_0$ always give rise to missing energy (either through dark photons or through neutrinos), when $\mt < m_{\gamma_D} $ the $3'_0$ cannot decay into dark photons and thus also stops contributing.~\footnote{Decays $3_0' \to V \bar f f$ can give missing energy in the form of neutrinos, but in those cases the final state usually contains too many jets or leptons to pass the selection of Refs.~\cite{Aad:2019vnb,Aad:2019vvi}.}
The effective multiplicity of triplets therefore changes from 4 to 1 along the $m_1$ axis, implying the two asymptotes in the plot.
When $\varepsilon$ is large and dark photons decay inside the detector, the only missing energy can come from singlets or neutrinos. Thus only charged triplets contribute for low $\ms$, until the $3_0$ switches from dark-photon final states to leptons and neutrinos (see Tab.~\ref{tab:dec}). This corresponds to an effective change of multiplicity from 2 to 1, as shown by the blue curve in the plot.

\begin{figure}[tp]
\centering
\includegraphics[width=0.7\textwidth]{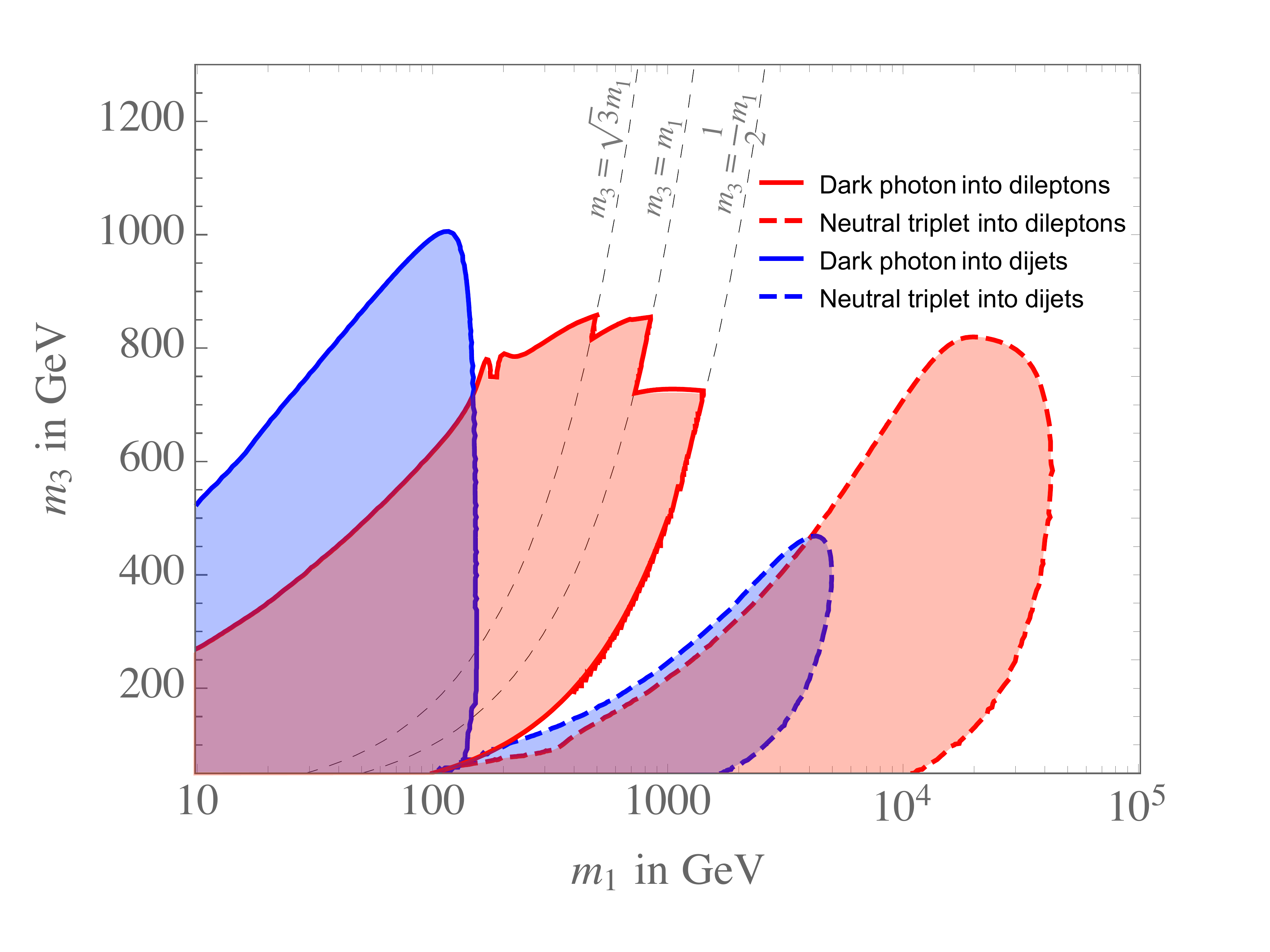}
\caption{Bounds in the triplet-singlet mass plane for $\varepsilon = 10^{-8}$ and $\ms = 2m_{\gamma_D}$ from searches performed by ATLAS that look for displaced dileptons in the inner tracker~\cite{Aad:2019tcc} and dijets in the muon spectrometer~\cite{Aaboud:2018aqj}. Regions with solid contours are excluded by the decay of  the dark photons, while those with thick dashed contours are excluded by the decays of the~$3_0'$. Since the latter has a smaller decay rate compared to the dark photon, its exclusion region extends to larger values of $\ms$. The thin dashed lines identify the different kinematic regions of Table \ref{tab:dec}, and can coincide with discontinuities in the value of the bounds.}
\label{fig:disp}
\end{figure}

Finally, it is possible to use the same approach to investigate the case where the dark photons or the $3_0'$ decay inside the detector but far from the interaction point. Their signature in this case is that of two visible tracks originating from a displaced vertex.  Such displaced decays occur in a range of values of $\varepsilon$ that depends on $\ms$ and $\mt$ (and only for an inverted hierarchy in the case of the $3_0'$). Since a single displaced vertex from any of the $2-4$ dark photons produced in each collision is enough to yield a measurable signal, the reach in $\varepsilon$ is slightly larger than the naive expectation, but still confined to a relatively narrow window of values.
We made use of the searches for displaced dileptons and jets performed by the ATLAS collaboration in Refs.~\cite{Aad:2019tcc,Aaboud:2018jbr,Aaboud:2018aqj,Aaboud:2019opc}. Searches performed by CMS lead to similar or slightly stronger results but they rely on tighter event selections or trigger requirements that are not necessarily satisfied by our signal events.~\footnote{For example, the CMS search of Ref.~\cite{Sirunyan:2019gut} makes use of the timing information from the ECAL to identify long-lived particles and sets strong bounds. Deriving the constraints on our model from this and other searches would require a dedicated Montecarlo simulation and is beyond the scope of this work.}
Each search looks for decays occurring in different parts of the detector (inner tracker, calorimeter and muon spectrometer), at distances ranging from millimeters to several meters away from the primary vertex, and is thus sensitive to different lifetimes of the long-lived particle. 
Table~\ref{tab:resultV} shows the number of observed and background events -- or the number of displaced vertices, in the case of Ref.~\cite{Aaboud:2018jbr} -- together with the upper bound on the signal derived within our Bayesian approach.
%%%%%%%%%%%%%%%%%%%%%%%%%%%%%%%%%%%%%%%%%%%%
\begin{table}[tp]
\centering
\begin{tabular}{|c|c|c|c|c|}
\hline
Search & Integrated Luminosity & $N_b$ & $N_{obs}$ &  $N_s^\text{max}$ \\
\hline
ATLAS Dilepton  \cite{Aad:2019tcc} & $32.8\,\rm{fb}^{-1}$  $@\,13\, \rm{TeV}$ & $0.27 \pm 0.17$ & $0$ & $3.0$ \\
\hline
ATLAS Dimuon \cite{Aaboud:2018jbr} & $32.9\,\rm{fb}^{-1}$ $@\,13\, \rm{TeV}$ & $0.5 ^{+1.4}_{-0.0} $ & $2$ & $6.0$  \\
\hline
ATLAS Dijet  \cite{Aaboud:2018aqj} & $36.1\,\rm{fb}^{-1}$ $@\,13\, \rm{TeV}$ & $ 0.027\pm 0.011 $ & $0$ & $3.0$ \\
\hline
ATLAS Dijet \cite{Aaboud:2019opc} & $35.9\,\rm{fb}^{-1}$ $@\,13\, \rm{TeV}$ & $8.5 \pm 2.2 $ & $10$ & $10.0$ \\
\hline
\end{tabular}
\caption{$95 \% $ probability bounds on the number of signal events, $N_s^\text{max}$, obtained from searches for displaced dilepton or dijet vertices performed by ATLAS. The number of expected background ($N_b$) and observed ($N_{obs}$) events is reported in the third and fourth columns respectively.}
\label{tab:resultV}
\end{table}
%%%%%%%%%%%%%%%%%%%%%%%%%%%%%%%%%%%%%%%%%%%%
Figure~\ref{fig:disp} shows the corresponding constraints in the $(\ms,\mt)$ plane obtained for $\varepsilon = 10^{-8}$ by computing the number of signal events through a formula analogous to Eq.~(\ref{eq:ndm}), where $P$ now corresponds to the probability for the decay(s) to occur in the relevant part of the detector.
%%%%%%%%%%%%%%%%%%%%%%%%%%%%%%%%%%%%%%%%%%%%
%%%%%%%%%%%%%%%%%%%%%%%%%%%%%%%%%%%%%%%%%%%%%
We display only the exclusions set by the dilepton and dijet searches of Refs.~\cite{Aad:2019tcc, Aaboud:2018aqj}, since they give the strongest bounds; they were obtained by assuming an efficiency equal to $\delta_{eff} = 0.4$ and $\delta_{eff} = 0.03$ respectively. Notice that, differently from missing-energy searches, in this case final states with jets are competitive with leptons since they have equally suppressed backgrounds. 
The plot suggests that values of $\mt$ as large as the TeV can be excluded with displaced decays. The range of excluded $\ms$ depends on the value of $\varepsilon$, as a combination of these two parameters controls the lifetime of the decaying particle, while the yield of signal events largely depends on $\mt$. Increasing (decreasing) $\varepsilon$, in particular, would deform the excluded regions and shift them towards smaller (larger) values of $\ms$.

\section{Discussion and Outlook}
\label{sec:summary}

We have analyzed a class of dark sector theories characterized by a chiral $\GDC\times \UoneD\times \GSM$ gauge group where the vector-like factor $\GDC$ confines at energies higher than the EW scale, while $\UoneD$ remains weak and is spontaneously broken. We assumed that the SM fermions are neutral under $\GDC\times \UoneD$ and that the dark fermions transform as non-trivial vector-like representations of the SM gauge group $\GSM$. The minimal models of this kind are listed in Tab.~\ref{tab:minimal_theories}. They have four dark fermion multiplets and contain an accidental $\UoneV\times \UoneB$ invariance that makes some of the NGBs and the lightest dark baryon cosmologically stable. Among minimal theories with SM irreducible representations, we found that only those with $\SUEW$ doublets or $\SUC$ triplets are realistic. In these theories the DM abundance is reproduced by a pair of NGBs, the $\pi^\pm$, that are charged under $\UoneV$ but neutral under the SM, with a subdominant component in the form of dark baryons. We focused on the model with EW doublets, which predicts four additional EW triplets in the NGB spectrum besides the $\pi^\pm$. We analyzed its cosmological history and the constraints set by direct and indirect DM searches, as well as by collider data. 

\begin{figure}[tp]
\centering
\includegraphics[width=0.7\textwidth]{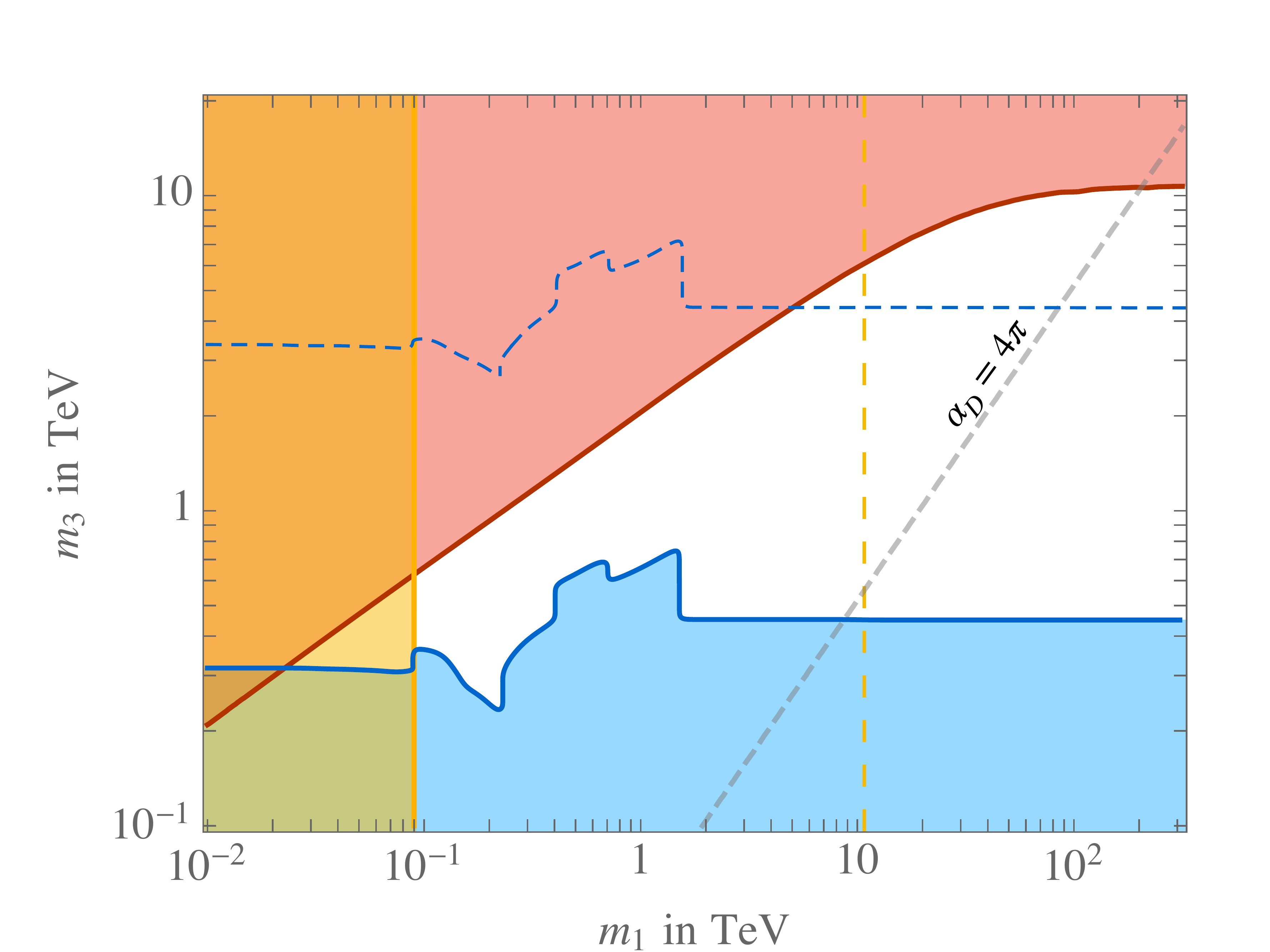}
\caption{Isocurve of thermal relic density corresponding to the observed DM abundance (red curve), in the theory with EW doublets with $a=1/2$ and $\NDC =4$. Colored regions are excluded by current data from LHC searches (blue area), observations of gamma rays from dwarf spheroidal galaxies (yellow area) and by requiring that dark particles do not overclose the Universe (red area). The dashed blue curve shows the expected exclusion reach of a $100\,$TeV FCC with $20\,\text{ab}^{-1}$, while the dashed yellow curve shows the sensitivity of the CTA observatory from observations of the Galactic Center, assuming an Einasto DM profile. Our assumption of a weakly-coupled $\UoneD$ gauge factor breaks down naively for points on the right of the dashed gray line. The plot assumes $\varepsilon = 10^{-6}$, which corresponds to prompt dark photon decays at high-energy colliders.}
\label{fig:summaryplot-largeeps}
\end{figure}

\begin{figure}[tp]
\centering
\includegraphics[width=0.7\textwidth]{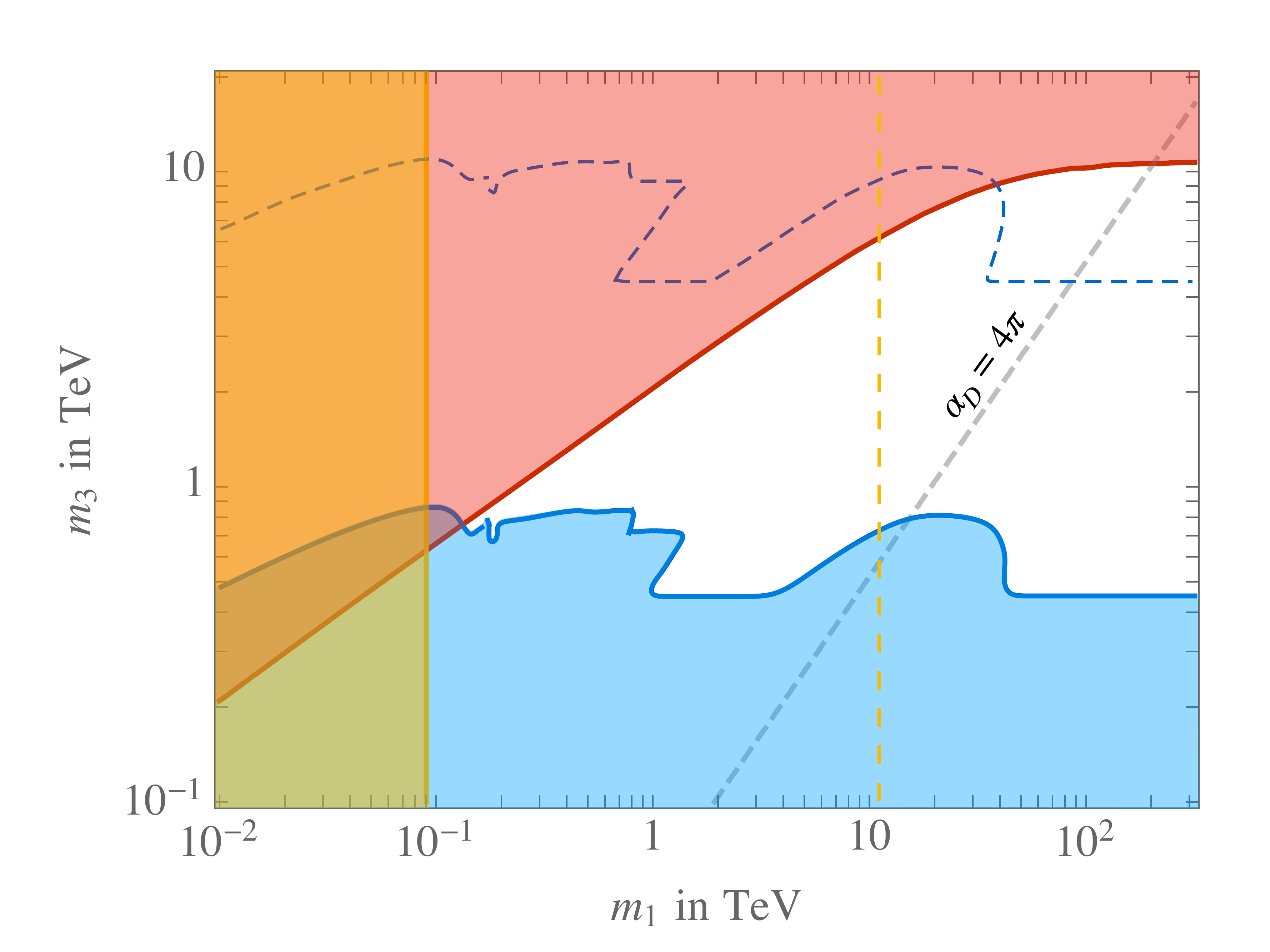}
\caption{Same as Fig.~\ref{fig:summaryplot-largeeps} for $\varepsilon = 10^{-8}$, which corresponds to displaced dark photon decays at high-energy colliders.}
\label{fig:summaryplot-mediumeps}
\end{figure}

\begin{figure}[tp]
\centering
\includegraphics[width=0.7\textwidth]{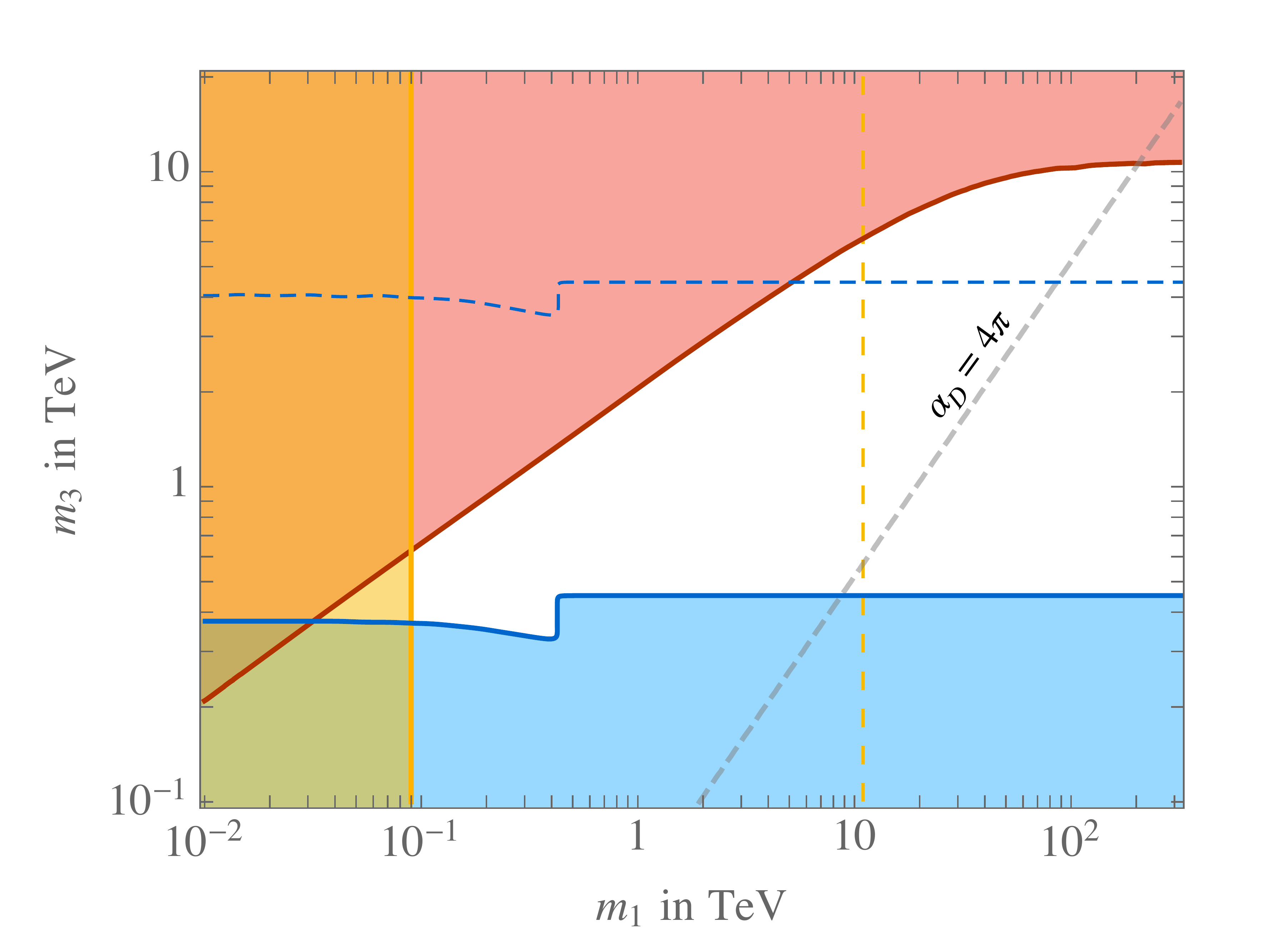}
\caption{Same as Fig.~\ref{fig:summaryplot-largeeps} for $\varepsilon = 10^{-10}$, which corresponds to dark photons decaying outside the detector at high-energy colliders.}
\label{fig:summaryplot-smalleps}
\end{figure}

Our results are summarized by the plots of Figs.~\ref{fig:summaryplot-largeeps},~\ref{fig:summaryplot-mediumeps} and~\ref{fig:summaryplot-smalleps}.
%%%%%%%%%%%%%%%%%%%%%%%%%%%%%%%%%%%%%%%%%
%%%%%%%%%%%%%%%%%%%%%%%%%%%%%%%%%%%%%%%%%
These show the allowed parameter space in the singlet-triplet mass plane for values of the kinetic mixing parameter equal to $10^{-6}$ (Fig.~\ref{fig:summaryplot-largeeps}), $10^{-8}$ (Fig.~\ref{fig:summaryplot-mediumeps}) and $10^{-10}$ (Fig.~\ref{fig:summaryplot-smalleps}). These benchmark values are chosen as representative of the three possible scenarios that characterize the production of the lightest dark states at high-energy colliders, i.e. those where dark photons decay promptly, with displaced vertices or outside the detector respectively. The white region is allowed by current data, while the remaining parameter space is excluded at 95\% probability by collider searches (blue region), observations of gamma rays from dwarf spheroidal galaxies (yellow region), and by requiring that the thermal abundance of dark pions and baryons does not exceed the observed DM density (red region). One can further restrict the parameter space to the region where $\UoneD$ is weakly coupled at the dark confinement scale. Naively, these points lie on the left of the dashed gray line shown in the plots. The resulting parameter space has a finite extension and has been already probed significantly by the current experimental data. Notice, in particular, that the mass of the triplets cannot be much larger than $\sim 10\,$TeV, otherwise dark sector particles would overclose the Universe. 
It is thus relevant to ask whether all of the physically sensible parameter space can be tested at future high-energy colliders or by future astrophysical observations. The dashed blue curves in the plots show the expected reach of a proton-proton FCC operating at a $100\,$TeV center-of-mass energy with $20\,\text{ab}^{-1}$ of integrated luminosity, and have been obtained by performing a naive rescaling of the expected LHC exclusions with the Collider Reach$^\beta$ tool~\cite{Salam:colliderreach}. Similarly, the yellow dashed lines show the projected sensitivity of the CTA observatory from the observation of the Galactic Center~\cite{Acharyya:2020sbj}. Although points with the highest values of $\mt$ and an inverted hierarchy will remain inaccessible, most of the parameter space can be probed both by a $100\,$TeV collider and by future observations of the Galactic Center by CTA. The joint observation of a signal by the two classes of experiments will allow a detailed test of the model thanks to the correlated prediction of the particles' masses, dark matter annihilation cross section and collider production cross section. We expect that a similar or stronger conclusion can be drawn for the model with $\SUC$ triplets.

The existence of SM-charged partners of the DM is the key prediction that distinguishes our theories from the chiral model of Refs.~\cite{Harigaya:2016rwr,Co:2016akw}, where dark fermions are SM singlets. Both kinds of theories, on the other hand, lead to a similar DM phenomenology in terms of one SM-singlet scalar field (the DM candidate $\pi^\pm$), plus a massive dark photon. These are the lowest lying states in the spectrum of new particles in a large portion of our parameter space, and in fact characterize the low-energy limit of a larger class of DM theories studied in the literature. It is interesting to analyze how much of the results we obtained on the phenomenology of the DM relies on the properties of these infrared degrees of freedom, and which are instead the aspects distinctive of our UV completion. 

The most general effective lagrangian which describes one scalar field $\pi^\pm$ (the DM candidate) plus a massive spin-1 field (the dark photon), and which is invariant under a global $\UoneV$, has the following form:
\begin{equation}
\label{eq:darkem}
\begin{split}
\mathcal{L}_{eff} = 
&- \dfrac{1}{4}F^{2}_{\mu\nu,D} + \dfrac{1}{2} m_{\gamma_{D}}^{2} A_{\mu,D}^{2}+ \varepsilon \, F_{\mu\nu,D} B^{\mu\nu}+ 
\partial_{\mu} \pi^{+} \partial^{\mu} \pi^{-} - \ms^{2} \pi^{+}\pi^{-} \\[0.2cm]
&+i a_3 e_D A^{D}_{\mu}(\pi^- \partial^{\mu} \pi^+ -\pi^+ \partial^{\mu} \pi^- ) +a_4 e_D^2 A^{2}_{\mu,D} \pi^+ \pi^- + \dots \, ,
\end{split}
\end{equation}
where $a_3$, $a_4$  are arbitrary dimensionless coefficients and the dots stand for higher-dimensional operators. For generic values of $a_3$ and $a_4$, this theory becomes strongly coupled at the scale 
\begin{equation}
\Lambda_S \sim \frac{4\pi}{e_D} \frac{m_{\gamma_D}}{\sqrt{|a_4 - a_3^2|}}\, .
\end{equation}
The easiest way to see this is by introducing the Stueckelberg field corresponding to the longitudinal polarization of the dark photon by means of the field redefinition $A_\mu^D \to A_\mu^D - \partial_\mu \pi^0/(e_D f)$. The new basis makes the $\U(1)$ gauge invariance of the theory manifest, and at the same time uncovers the terms responsible for the strong coupling scale, namely the derivative interactions between $\pi^0$ and $\pi^\pm$. For the special choice $a_4 = a_3^2$ these terms can be redefined away by a local phase shift of $\pi^\pm$, and the theory becomes UV complete. This is very much analogous to the case of the electroweak chiral lagrangian plus a Higgs boson, which becomes UV complete for values of the Higgs couplings $c_V = c_{2V} =1$ (see for example Ref.~\cite{Contino:2010mh}). A relative strength $a_4/a_3^2 =1$ between quartic and cubic couplings is in fact what scalar QED would predict. Indeed, the limit of scalar QED can be recovered by fixing $a_4 = a_3^2$ and letting $m_{\gamma_D}\to 0$. 

The theory described by Eq.~(\ref{eq:darkem}) with $a_4 = a_3^2$ and vanishing higher-dimensional operators has been analyzed in the literature as an example of dynamics with a scalar DM candidate plus a dark photon, see for example Ref.~\cite{Berlin:2016gtr}. In that case the mass ratio $m_{\gamma_D}/\ms$ is an arbitrary parameter. For $m_{\gamma_D} < 2\ms$ and sufficiently small $\varepsilon$, the theory goes through an early phase of matter domination, which in turn leads to a dilution of the DM relic density.
The models studied in this work and in Refs.~\cite{Harigaya:2016rwr,Co:2016akw} lead to different values for the coefficients of Eq.~(\ref{eq:darkem}). They predict
\begin{equation}
a_3 = 1+a\, , \quad a_4 = 4a\, , \quad m_{\gamma_D} = \sqrt{2} (1-a) e_D f\, .
\end{equation}
In this family of points of the parameter space the effective theory has an approximate enhanced global $\SU(2)_L\times \SU(2)_R$ symmetry, broken down spontaneously to $\SU(2)_V$ at the scale~$f$. This symmetry is broken explicitly by the $\pi^\pm$ mass term and by the gauging. The $\pi^\pm$ can be thus thought of as two additional (pseudo) NGBs, together with $\pi^0$.  The value of $\ms$ is still arbitrary from the low-energy viewpoint, i.e. it depends on the UV completion of Eq.~(\ref{eq:darkem}). For example, it remains a free parameter if the UV theory is a linear sigma model with fundamental scalar fields. Hence, while the relation among $a_3$, $a_4$ and $m_{\gamma_D}$ is a consequence of the global symmetries of the low-energy theory and of their weak chiral gauging, the prediction $\ms \sim e_D f$ is specific to UV theories -- such as those analyzed in this work and in Refs.~\cite{Harigaya:2016rwr,Co:2016akw} -- where $\pi^{\pm,0}$ are bound states of new strongly-coupled dynamics at the scale $f$, and where the only spurion is the gauge coupling $e_D$. The fact that the DM and the dark photon have comparable masses (unless $a\to 1$) has specific implications in the cosmological evolution and in the DM phenomenology. For example, it implies that effects due to Sommerfeld enhancement and bound-state formations are negligible in determining the DM relic density. 
Another prediction specific to our UV completion is the role of higher-dimensional operators, such as $A_{\mu,D}^{2} W_{\rho\sigma}^{2}$ and $\pi^{2} W_{\mu\nu}^{2}$, generated by loops of triplets. They are the main interactions between the dark and SM sectors for small $\varepsilon$, and in this case play a key role in the cosmological evolution. For example, as discussed in Sec.~\ref{sec:kineq}, they keep the dark sector in thermal equilibrium with the SM bath until low temperatures. If, on the other hand, the UV dark dynamics interacts with the SM only through the kinetic mixing, like in the model of Refs.~\cite{Harigaya:2016rwr,Co:2016akw}, then the dark and SM sectors might be thermally decoupled throughout their entire cosmological histories. Predicting the DM thermal abundance in this case requires knowing the ratio of entropies set during the reheating epoch.  The dim-6 operator $A_{\mu,D}^{2} W_{\rho\sigma}^{2}$ is also crucial to ensure efficient annihilations of the dark photons into SM particles. As discussed in Secs.~\ref{sec:chemeq} and~\ref{sec:dilution}, this process plays an important role during the freeze out and controls the amount of dilution induced by long-lived dark photons (see Fig.~\ref{fig:dilution_BBN}).

These considerations make clear the importance of having a UV complete description of the dynamics rather than just an effective theory. Crucial aspects of the cosmological evolution might not be captured by the lowest-lying degrees of freedom. On one hand, this is true because the cosmological relevance of a given process depends on its rate, which can be large despite a small cross section if the number density is large enough. This is the case of dark photon annihilation into SM vector bosons in our theory. On the other hand, heavy degrees of freedom can contribute to the DM density if they are stable and have a thermal abundance. Dark baryons are an example in our case. A top-down approach in the search for a DM theory is thus justified and has its own advantages over strategies based on effective models. Obviously, the landscape of UV theories is larger and much more difficult to explore than that of low-energy ones. We tried such path by taking as a guidance the following two criteria: the theory should generate dynamically all the new scales, including the mass of the DM candidate, and the stability of the latter should be explained by an accidental symmetry. This led us to consider strongly-coupled chiral gauge theories, which represent an incredibly rich playground but, at the same time, are subject to highly non-trivial theoretical constraints. 
The models analyzed in this work are interesting and particularly attractive since they lead to correlated predictions for experiments at high-energy colliders and for astrophysical observations.
It is not clear, however, if a model of this kind can be compatible with the unification of SM forces at high energies. Grand Unification could in fact be taken as the third criterion to guide our search for the theory of DM. It would be very interesting if a chiral gauge theory with all these properties could be constructed.

\newpage
\acknowledgments

We would like to acknowledge helpful discussions with Dario Buttazzo, Fabrizio Caola, Graham Kribs, Jesse Liu, Rashmish Mishra, Paolo Panci, Michele Papucci, Riccardo Rattazzi, Josh Ruderman, Alessandro Strumia and Luca Vecchi. We especially thank Andrea Mitridate, Michele Redi and Filippo Sala for important comments and suggestions.
We thank the participants of the workshop \textit{New Directions in Heavy Dark Matter} at DESY for useful comments and questions.
A.P. also would like to thank the audience of a seminar at the University of Maryland for comments and questions and the Columbia University Physics Department for its kind hospitality during part of this work. 
The work of R.C. and A.P. has been partly supported by the PRIN grants 2015P5SBHT\_007 and 2017FMJFMW of the Italian MIUR.
F.R. acknowledges support from the Dalitz Graduate Scholarship, jointly established by the Oxford University Department of Physics and Wadham College.

\appendix

\section{Useful formulas for dark pions}
\label{app:darkpions}

In this Appendix we collect two useful results on dark pions.

The first concerns the vertices with three NGBs and one dark photon that appear in the chiral Lagrangian.
They have the following expression:
\begin{equation}
\begin{split}
\mathcal{L}_{0} \supset \frac{e_D  (a-1)}{\fD} \bigg[  
&
gA^D_{\mu} \varepsilon_{abc} \big( s_+ \pi_-^a  + s_- \pi_+^a  \big)\pi'^b W^{\mu}_c 
-\frac{ig}{\sqrt{2}}  A^D_{\mu} W^{\mu}_a \big( \pi_+^a \pi_-^b -\pi_-^a \pi_+^b  \big) \pi^b \\
&
+\frac{1}{3} A^D_{\mu}  \partial_{\mu}\Big( \pi'^a(\pi^a_+ s_- +\pi^a_- s_+)\Big)
+ \frac{i}{3\sqrt{2}} A^D_{\mu} \varepsilon_{abc} \partial_{\mu} (\pi_+^a \pi_-^b \pi^c  )  \\
& 
-A^D_{\mu} \partial^{\mu} \pi'_a (\pi_+ s_-^a+  \pi_-s_+^a)
+ \frac{i}{3\sqrt{2}} A^D_{\mu} \varepsilon_{abc}  \pi_+^a \pi_-^b \partial_{\mu} \pi^c \bigg]\, ,
\end{split}
\end{equation}
where $\SUEW$ indices are denoted by lower case Latin letters ($a,b,c$) and, only for this formula, $s_{\pm}$,  $\pi^a_{\pm}$, $\pi^a$ and $\pi'^a$ denote respectively the $1_{\pm}$, $3_{\pm}$, $3_{0}$ and $3_{0}'$.

The second result are the (tree-level) spin- and color-averaged partonic cross sections for Drell-Yan production of pairs of dark pions.
We find (here superscripts indicate the electromagnetic charges and $\pi$ denotes any of the NGBs)
\begin{align}
\label{eq:xsec1}
\hat{\sigma}_{u_i \bar{d}_j \rightarrow \pi^{+}\, \pi^0}   & =  \hat{\sigma}_{\bar u_i d_j \rightarrow \pi^{-}\, \pi^0}=
\frac{g^4}{576 \pi}  \frac{|V_{ij}|^2}{\sqrt{\hat{s}}(\hat{s}-m_W^2)^2}(\hat{s}-4M^2)^{\frac{3}{2}}, \\[0.25cm]
\label{eq:xsec2}
\hat{\sigma}_{{u_i \bar{u}_j \rightarrow \pi^{+}\, \pi^-}} & = 
\begin{aligned}[t]
\frac{g^4}{144 \pi}  \frac{(\hat{s}-4M^2)^{\frac{3}{2}}}{\sqrt{\hat{s}}}
  \Bigg[ &\frac{1}{8} \frac{1}{(\hat{s}-m_Z^2)^2} +\frac{4}{9} \sin^4\!\theta_W \left(\frac{1}{\hat{s}-m_Z^2}-\frac{1}{\hat{s}} \right)^2 \\ 
&+\frac{1}{12} \sin^2\!\theta_W \frac{1}{\hat{s}-m_Z^2} \left(\frac{1}{\hat{s}}-\frac{1}{\hat{s}-m_Z^2} \right) \Bigg] ,
\end{aligned}\\[0.25cm]
\label{eq:xsec3}
\hat{\sigma}_{{d_i \bar{d}_j \rightarrow \pi^{+}\, \pi^-}} & =
\begin{aligned}[t]
\frac{g^4 }{144 \pi } \frac{(\hat{s}-4M^2)^{\frac{3}{2}}}{\sqrt{\hat{s}}}
 \Bigg[ &\frac{1}{8} \frac{1}{(\hat{s}-m_Z^2)^2} +\frac{1}{9} \sin^4\!\theta_W \left(\frac{1}{\hat{s}-m_Z^2}-\frac{1}{\hat{s}} \right)^2
 \\ 
&+\frac{1}{24} \sin^2\!\theta_W \frac{1}{\hat{s}-m_Z^2} \left(\frac{1}{\hat{s}}-\frac{1}{\hat{s}-m_Z^2}\right) \Bigg],
\end{aligned}
\end{align}
in agreement with the results of Ref.~\cite{Cirelli:2005uq}. Here $M$ denotes the mass of the particle that is being produced, which can be either a $\UoneV$-charged or neutral triplet. For $\theta_W = 0$ the first cross section reduces to the sum of the other two, as expected by $\SUEW$ invariance.

\section{Model with $\SUC$ triplets}
\label{app:modelwithtriplets}

The other viable minimal model identified by the analysis of section~\ref{sec:minimalmodels} is a Type I theory with SM color triplets (hence $r$ in Tab.~\ref{tab:minimal_theories} is a fundamental of $\SUC$). 
From a qualitative viewpoint, the analysis of this model closely parallels the one carried out in the main text for the model with electroweak doublets, since the SM strong interactions are perturbative at energies of the order of the dark confinement scales we are interested in. The approximate global symmetry breaking pattern gets enlarged to 
\begin{equation}
 \SU(6)_L \otimes \SU(6)_R \otimes \UoneB \rightarrow \SU(6)_V \otimes \UoneB \,  , 
\end{equation}
so there are $35$ pseudo-NGBs, one of which is eaten by the dark photon. They can be classified as:
\begin{itemize}
 \item Two color singlets, charged under $\UoneD$, the $\mathbf{1^{\pm}}$. These have exactly the same properties as the corresponding states in the model with electroweak doublets, and thus constitute stable dark matter candidates.
 \item Two neutral octets charged under $\SUC$ only, the $\mathbf{8_0,8'_0}$. These are the analog of the $3_0,3_0'$.
 \item Two charged octets, the $\mathbf{8^{\pm}}$, charged under $\SUC$ and $\UoneD$. These are the analog of the $3^{\pm}$.
\end{itemize}
The phenomenology is very similar to the electroweak case. Dark matter is still made of SM singlets and the calculation of the relic abundance is unmodified. Heavier pseudo-NGBs now transform as SM color octects (i.e. still in the adjoint of the SM group), and the constraints from colliders are expected to be more severe. We leave the analysis of this model to a future work.

\section{Boltzmann equations for long-lived dark photons}
\label{app:boltzmann}

A complete description of the evolution of the dark sector during the freeze-out epoch would include, in principle, all number-changing processes involving dark photons and SM-singlet dark pions. Since  the underlying theory is strongly coupled, every effective interaction allowed by the symmetries is expected to be generated at some level. This leads to a complicated system of coupled Boltzmann equations with many processes. It is possible to obtain a simplified yet accurate description by identifying the leading effects and focus our attention on them. We can distinguish the following classes of leading processes:~\footnote{Processes with dark photon absorption and conversion, such as $\gamma_{D} \psi_{\rm SM} \longrightarrow V \psi'_{\rm SM}$, where $V$ is a standard model gauge boson, can be safely neglected. Their rate is suppressed by $\varepsilon^{2}$ due to $\mathcal{C}_{D}$ and turns out to be negligible when $\varepsilon$ is small enough to have long-lived dark photons.}
\begin{itemize}
\item Pion annihilations into two dark photons, with cross section given by Eq.~(\ref{eq:pion_ann});
\item Dark photon decays, suppressed by an $\varepsilon$ insertion due to $\CD$ protection;
\item Dark photon annihilations into EW gauge bosons, mediated by a loop of triplet pions. This process is allowed by $\CD$ and has a cross section given by Eq.~(\ref{eq:gaDgaDtoWW});
\item Pion scatterings with initial or final state radiation: $\pi^{\pm} \pi^{\pm} \to \pi^{\pm} \pi^{\pm} \gamma_{D}$ or $\pi^{\pm} \gamma_{D} \to \pi^{\pm} \gamma_{D} \gamma_{D}$, with thermally-averaged cross sections $\langle \sigma_{\pi\pi\to\pi\pi\gamma_D} v \rangle \sim \langle \sigma_{\pi\gamma_D\to\pi\gamma_D\gamma_D} v \rangle \sim \alpha_{D}^{3}T^{4} e^{-m_{\gamma_{D}}/T}/ \ms^{6}$. This estimate relies on the assumption that $m_{\gamma_{D}} \sim m_{1}$.~\footnote{For $m_{\gamma_{D}} \ll m_{1}$, instead, $2\rightarrow 3$ processes are no longer Boltzmann suppressed and can be efficient, keeping dark photons in chemical equilibrium until much lower temperatures -- similarly to what happens to CMB photons after recombination.}
\end{itemize}
Assuming that the kinetic decoupling of the SM and dark sectors occurs at sufficiently small temperatures and that entropy is conserved,~\footnote{This is a good approximation as long as the entropy from the dark photon decay products is negligible, that is, as long as the dark photons' energy density is subleading in the energy budget of the Universe. If dark photons  dominate the energy density, they can give rise to a dilution of cold relics due to entropy injection, as described in Sec.~\ref{sec:dilution}.} the set of Boltzmann equations is thus given by ($x = \ms/T$):
\begin{equation}
\label{eq:boltzmann-full}
\begin{split}
\dfrac{\mathrm{d}Y_{\pi}}{\mathrm{d}x} & =-\dfrac{1}{x^{2}} \dfrac{s(m)}{H(m)} \dfrac{1}{2} \langle \sigma_{\pi\pi\to\gamma_D \gamma_D} v \rangle \left(Y_{\pi}^{2}-\frac{Y_{\pi , eq}^2}{Y_{\gamma_{D} , eq}^2}\, Y_{\gamma_{D}}^2\right)\,, 
\\[0.4cm]
\dfrac{\mathrm{d}Y_{\gamma_{D}}}{\mathrm{d}x} & = \dfrac{1}{x^{2}} \dfrac{s(m)}{H(m)} \Bigg[ 
\dfrac{1}{2} \langle \sigma_{\pi\pi\to\gamma_D \gamma_D} v \rangle \left(Y_{\pi}^{2}-\frac{Y_{\pi , eq}^2}{Y_{\gamma_{D} , eq}^2}\, Y_{\gamma_{D}}^2\right) 
\\ & \hspace{65pt} 
 + \dfrac{1}{4}\langle \sigma_{\pi\pi\to\pi\pi\gamma_D} v \rangle Y_{\pi}^{2} \left(1-\frac{Y_{\gamma_{D}}}{Y_{\gamma_{D} , eq}}\right) 
 \\& \hspace{65pt} 
 + \dfrac{1}{2}\langle \sigma_{\pi\gamma_D\to\pi\gamma_D\gamma_D} v \rangle Y_{\pi} Y_{\gamma_{D}} \left(1-\frac{Y_{\gamma_{D}}}{Y_{\gamma_{D} , eq}}\right)
 \\& \hspace{65pt} 
  -2 \langle \sigma_{\gamma_{D}\gamma_{D}\rightarrow SM} v \rangle \left( Y_{\gamma_{D}}^{2}- Y_{\gamma_{D} , eq}^{2}\right)
 \Bigg] 
\\ & \hspace{15pt} 
- x \dfrac{1}{H(m)} \Gamma_{\gamma_{D}\rightarrow SM} \Bigg(Y_{\gamma_{D}} - Y_{\gamma_{D} , eq} \Bigg)\,.
\end{split} 
\end{equation}

\begin{figure}[tp]
\centering
\includegraphics[width=0.65\textwidth]{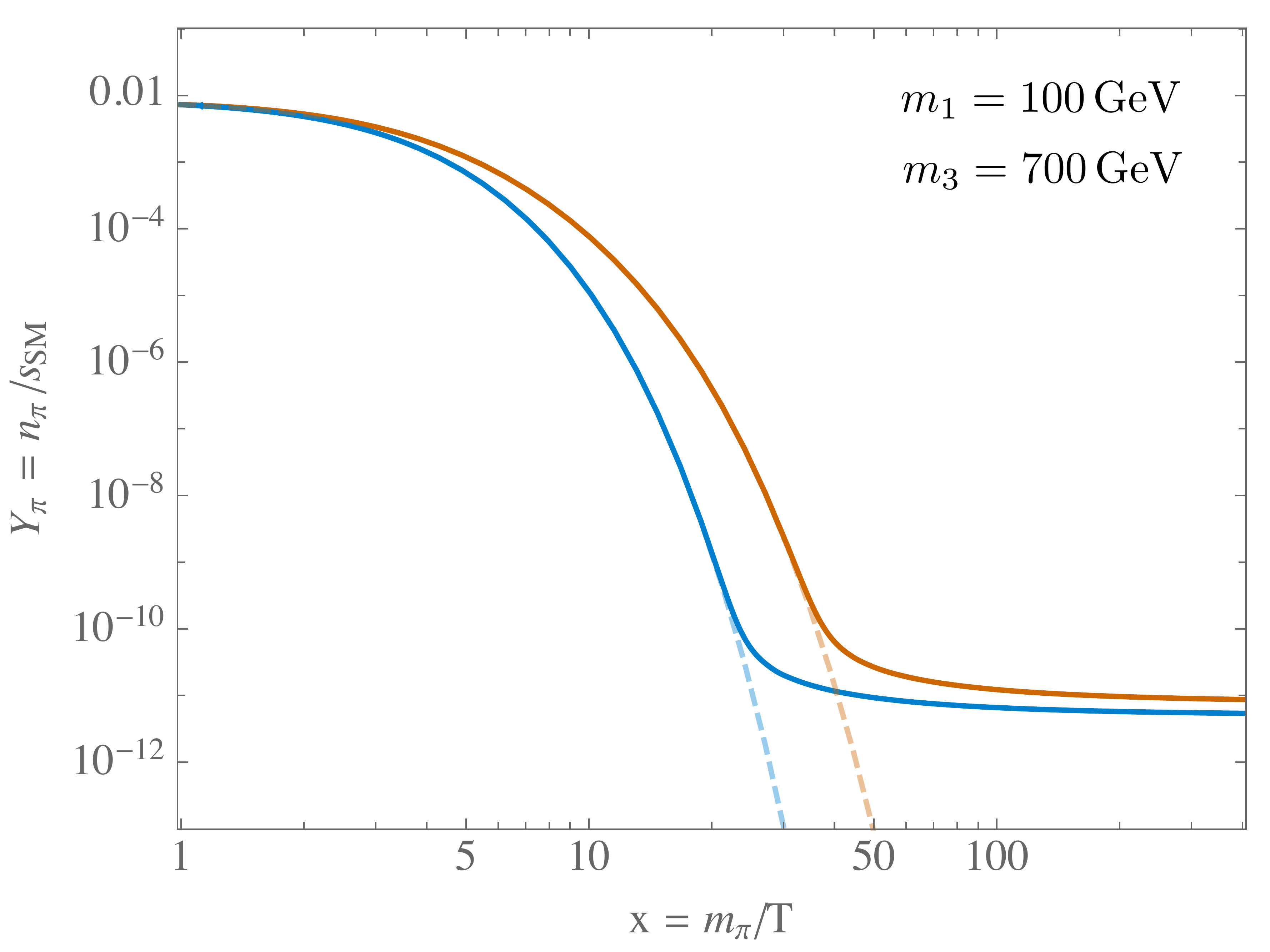}
\caption{Thermal evolution of the dark pion abundance obtained by solving numerically the system~\eqref{eq:boltzmann-full} with $a=1/2$ and $\NDC=4$. The numerical solution for $\varepsilon \lesssim 10^{-8}$ (solid orange line) traces the combination $(Y_{\pi , eq}^2 / Y_{\gamma_{D} , eq}^2 )\, Y_{\gamma_{D}}^2$ (dashed orange line) until the freeze out.
For comparison, we also show the thermal abundance of dark pions for $\varepsilon > 10^{-8}$ (solid blue line), corresponding to the ordinary freeze-out evolution of Eq.~(\ref{eq:boltzmann}), and its thermal equilibrium density $Y_{\pi , eq}$ (dashed blue line).}
\label{fig:Ypi}
\end{figure}

\begin{figure}[pt]
\centering
\includegraphics[width=0.65\textwidth]{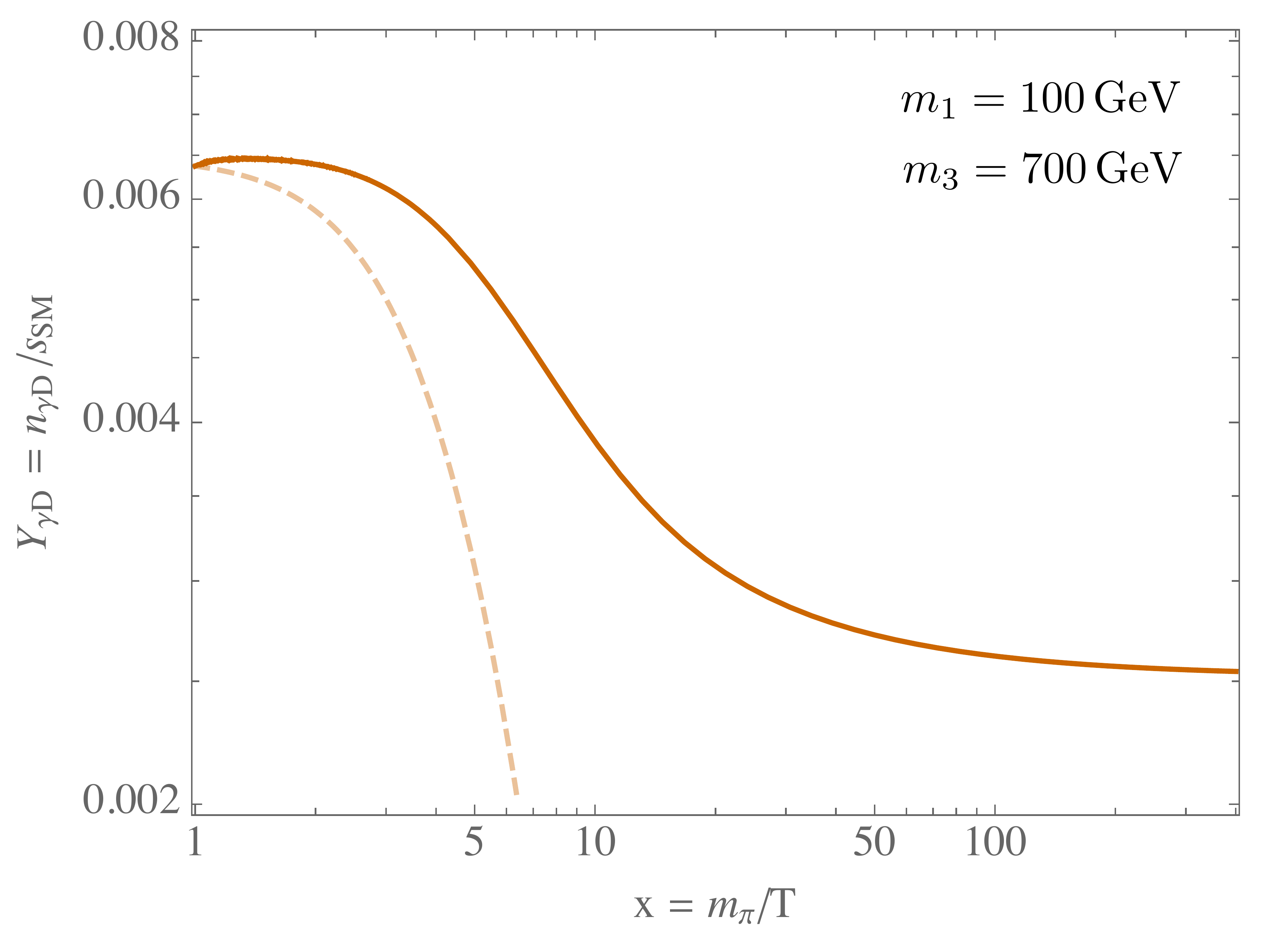}
\caption{Thermal evolution of the dark photon abundance obtained by solving numerically the system~\eqref{eq:boltzmann-full} with $a=1/2$ and $N_{\rm DC}=4$. The solid orange line is the numerical solution for $\varepsilon \lesssim 10^{-8}$, while the dashed orange line corresponds to the thermal equilibrium density $Y_{\gamma_{D} , eq}$.}
\label{fig:YgD}
\end{figure}

We have checked numerically, by performing the corresponding thermal average and including the kinematic cut to emit a massive particle, that $2\rightarrow 3$ processes are subleading and give a relative corrections of order $10^{-3}$ or smaller.
The last term in Eq.~(\ref{eq:boltzmann-full}), corresponding to the decay of dark photons, is the only one that grows with $x$. It can be neglected as long as the dark photon energy density is subleading in the energy budget of the Universe and $x \,\Gamma_{\gamma_{D}\rightarrow SM} \ll H(m)$.  We discuss the effect of the dark photon decays in section~\ref{sec:dilution}.
Neglecting these terms, we are left with the simplified system of Eq.~\eqref{eq:boltzmann_simplified}, which describes the out-of-equilibrium evolution of two species. Its numerical solution is shown in Figs.~\ref{fig:Ypi} and~\ref{fig:YgD} for a benchmark point with $\ms=100\,$GeV and $\mt=700\,$GeV. 
%%%%%%%%%%%%%%%%%%%%%%%%
%%%%%%%%%%%%%%%%%%%%%%%%
For this choice, $\sigma_{\gamma_{D}\gamma_{D}\rightarrow SM} \ll \sigma_{\pi\pi \rightarrow \gamma_{D} \gamma_{D}}$ and dark photons are out of equilibrium during freeze out, see Fig.~\ref{fig:YgD}. The dark pion abundance tracks $(Y_{\pi , eq}/Y_{\gamma_{D} , eq})\, Y_{\gamma_{D}}$ and is much larger than the value predicted by a standard evolution until freeze out occurs, see Fig.~\ref{fig:Ypi}. Despite the different thermal evolutions and freeze-out temperatures, the asymptotic abundances differ by less than $30\%$ compared to the standard case in all the relevant parameter space.

\section{The Dark Photon}
\label{app:gammad}

In this Appendix we summarise some useful properties of the dark photon, providing formulas valid to all orders in $\varepsilon$ and also at resonance $m_{\gamma_{D}}=m_{Z}$. These equations confirm and extend the computations of Refs.~\cite{Cline:2014dwa, Curtin:2014cca, Evans:2017kti}, where similar results were provided to first order in $\varepsilon$ and away from resonance. 

We start from a lagrangian written in terms of the ordinary SM gauge bosons $\hat{W}_{\mu}^{3}, \hat{B}_{\mu}$ and a massive $\hat{Z}'$ with a kinetic mixing term with the hypercharge gauge boson:
\begin{equation}
\label{eq:interactingL}
\begin{split}
\mathcal{L}= &-\dfrac{1}{4} \hat{W}_{\mu\nu}^{3}\hat{W}^{\mu\nu,3}-\dfrac{1}{4} \hat{B}_{\mu\nu}\hat{B}^{\mu\nu}-\dfrac{1}{4} \hat{A}_{\mu\nu,D}\hat{A}_{D}^{\mu\nu}-\dfrac{\varepsilon}{2} \hat{A}_{\mu\nu,D}\hat{B}^{\mu\nu}\\
&+\dfrac{1}{2} c_{W}^{2} \hat{m}_{Z}^2\, \hat{W}_{\mu}^{3}\hat{W}^{\mu,3} +\dfrac{1}{2} s_{W}^{2} \hat{m}_{Z}^2\, \hat{B}_{\mu}\hat{B}^{\mu}- c_{W} s_{W} \hat{m}_{Z}^2\, \hat{W}_{\mu}^{3}\hat{B}^{\mu} +\dfrac{1}{2} \hat{m}_{\gamma_{D}}^2\, \hat{A}_{\mu,D}\hat{A}_{D}^{\mu} \, .
\end{split}
\end{equation}
All the symbols with a hat (such as $\hat{B}_{\mu}$) refer to fields and parameters in the interacting basis of Eq.~(\ref{eq:interactingL}).
Symbols without a hat will refer to the same quantities in the mass eigenbasis.

The mass and kinetic mixing can be diagonalized to all orders in $\varepsilon$ through the field redefinition
\begin{equation}
\begin{pmatrix}
\hat{W}^{\mu,3}\\
\hat{B}^{\mu}\\
\hat{A}_{D}^{\mu}
\end{pmatrix}
=
\begin{pmatrix}
s_{W} &\quad c_{W} c_{\xi} &\quad -c_{W} s_{\xi} \\
c_{W} &\quad -s_{W} c_{\xi}- \dfrac{\varepsilon}{\sqrt{1-\varepsilon^{2}}} s_{\xi} & \quad s_{W} s_{\xi}- \dfrac{\varepsilon}{\sqrt{1-\varepsilon^{2}}} c_{\xi} \\
0 &\quad \dfrac{1}{\sqrt{1-\varepsilon^{2}}} s_{\xi} & \quad \dfrac{1}{\sqrt{1-\varepsilon^{2}}} c_{\xi} \\
\end{pmatrix}
\begin{pmatrix}
A^{\mu}\\
Z^{\mu}\\
A_{D}^{\mu}
\end{pmatrix},
\end{equation}
where $c,s$ are shorthands for cosines and sines, the subscript $W$ identifies the weak mixing angle $\theta_{W}$, and $\xi $ refers to a new mixing angle defined by the relation:
\begin{equation}
\tan(2\xi)=\dfrac{2 \hat{m}_{Z}^2 \, \varepsilon  \sqrt{1-\varepsilon ^2} s_{W}}{\hat{m}_{Z}^2-\hat{m}_{\gamma_D}^2-\varepsilon ^2 \hat{m}_{Z}^2 (1+ s_{W}^{2})}\, .
\end{equation}
We are left with a diagonal mass matrix with:
\begin{equation}
\begin{split}
m_{\gamma}& = 0\\[0.1cm]
\dfrac{m_{Z}^{2}}{2} & =\dfrac{\hat{m}_{Z}^2}{2} c_{\xi}^{2}+ \dfrac{\varepsilon c_{\xi} s_{\xi} s_{W}\hat{m}_{Z}^{2}}{\sqrt{1-\varepsilon^{2}}} +\dfrac{1}{2} \dfrac{\hat{m}_{\gamma_D}^{2} s_{\xi}^{2}+\hat{m}_{Z}^{2}\varepsilon^{2} s_{W}^{2} s_{\xi}^{2}}{1-\varepsilon^{2}}  \\[0.1cm]
\dfrac{m_{\gamma_{D}}^{2}}{2} & = \dfrac{\hat{m}_{Z}^2}{2} s_{\xi}^{2} - \dfrac{\varepsilon c_{\xi} s_{\xi} s_{W}\hat{m}_{Z}^{2}}{\sqrt{1-\varepsilon^{2}}} +\dfrac{1}{2} \dfrac{\hat{m}_{\gamma_D}^{2} c_{\xi}^{2}+\hat{m}_{Z}^{2}\varepsilon^{2} s_{W}^{2} c_{\xi}^{2}}{1-\varepsilon^{2}} \, .
\end{split}
\end{equation}
It is useful to express the currents to which the physical vector bosons are coupled in terms of the currents for the original photon, $Z$ and dark photon fields:
\begin{equation}\label{eq:curr}
\begin{pmatrix}
J_{A}^{\mu}\\
J_{Z}^{\mu}\\
J_{A_{D}}^{\mu}
\end{pmatrix}
=
\begin{pmatrix}
1 & 0 & 0 \\
\quad - \dfrac{\varepsilon}{\sqrt{1-\varepsilon^{2}}} c_{W} s_{\xi} & 
\quad c _{\xi} + \dfrac{\varepsilon}{\sqrt{1-\varepsilon^{2}}} s_{W} s_{\xi} & 
\quad \dfrac{1}{\sqrt{1-\varepsilon^{2}}} s_{\xi} 
\\
\quad - \dfrac{\varepsilon}{\sqrt{1-\varepsilon^{2}}} c_{W} c_{\xi}&
\quad - s_{\xi} + \dfrac{\varepsilon}{\sqrt{1-\varepsilon^{2}}} s_{W} c_{\xi}&
 \quad \dfrac{1}{\sqrt{1-\varepsilon^{2}}} c_{\xi}\\
\end{pmatrix}
\begin{pmatrix}
J_{\hat{A}}^{\mu}\\
J_{\hat{Z}}^{\mu}\\
J_{\hat{A}_{D}}^{\mu}
\end{pmatrix}.
\end{equation}

\subsection*{Decays}

By virtue of its coupling to the SM sector, the dark photon is unstable and decays to SM particles. Restricting to two-particle final states, that are more important than other ones due to phase space suppression, the possible channels are
\begin{equation}
\label{eq:channels}
\gamma_D \rightarrow f \,\bar{f},   \quad \quad \gamma_D \rightarrow Z h, \quad \quad \gamma_D \rightarrow W^+\, W^-,
\end{equation}
where $f$ can be any SM fermion.  The decays to fermions dominate the total width and are mediated by the following interactions
\begin{equation}
\mathcal{L} \supset A_{\mu,D} \sum g_{L,i} \bar{\psi}_{L,i} \partial^{\mu} \psi_{L,i} +
A_{\mu,D} \sum g_{R,i} \bar{\psi}_{R,i} \partial^{\mu} \psi_{R,i} \, ,
\end{equation}
where $g_{L/R}$ can be extracted to all orders in $\varepsilon$ from Eq.~(\ref{eq:curr}):
\begin{equation}
g_i= (T_{3L}+Y) \frac{\varepsilon c_{W} c_{\xi} }{\sqrt{1-\varepsilon^2}}- \frac{T_{3L}c_{W}-Ys_{W}}{c_{W} s_{W}} \Big( \frac{\varepsilon s_{W} c_{\xi} }{\sqrt{1-\varepsilon^2}} -s_{\xi} \Big).
\end{equation}
At leading order in $\varepsilon$, this formula reduces to the expressions reported in Ref.~\cite{Curtin:2014cca}, up to a factor of $c_{W}$ that is reabsorbed in the definition of $\varepsilon$. The tree-level decay width is given by~\cite{Curtin:2014cca}:
\begin{equation}
\Gamma_{\gamma_{D} \rightarrow f\bar{f}  }=\frac{ \alpha_{em}  N_{c}}{6 \pi m_{\gamma_{D}}} \sqrt{1-\frac{4 m_{f}^{2}}{m_{\gamma_{D}}^{2}}}\left(m_{\gamma_{D}}^{2}\left(g_{L}^{2}+g_{R}^{2}\right)-m_{f}^{2}\left(-6 g_{L} g_{R}+g_{L}^{2}+g_{R}^{2}\right)\right).
\end{equation}
This expression is a good approximation above the $b \bar{b}$ threshold, below which QCD corrections must be taken into account. As regards the second channel of Eq.~(\ref{eq:channels}), the kinetic mixing induces a coupling of the form
\begin{equation}
\mathcal{L} \supset c_{h Z \gamma_{D}} h Z^D_{\mu} Z^{\mu},
\end{equation}
where 
\begin{equation}
 c_{h Z \gamma_{D}} = \frac{2 m_Z^2}{v} \Big( c_{\xi}+\frac{\varepsilon s_{W} s_{\xi} }{\sqrt{1-\varepsilon^2}}\Big)  \Big(-s_{\xi}+\frac{\varepsilon s_{W} c_{\xi} }{\sqrt{1-\varepsilon^2}} \Big)\, .
\end{equation}
The corresponding decay rate is 
\begin{equation}
\begin{split}
\Gamma_{\gamma_{D} \rightarrow Z h  } = & \, c_{h Z \gamma_{D}}^2 \frac{\big(m_{\gamma_D}^2 - m_h^2 - m_z^2\big)^2 + 12 m_{\gamma_D}m_Z^2 - 4 m_Z^2 m_h^2}{192 \pi m_{\gamma_D}^5 m_Z^2}  \\[0.1cm]
& \times \sqrt{\big(m_{\gamma_D}^2 - m_h^2 - m_z^2\big)^2 - 4 m_Z^2 m_h^2}\, .
\end{split}
\end{equation}
The couplings mediating the third channel in Eq.~(\ref{eq:channels}) are of the form 
\begin{equation}
\begin{split}
\mathcal{L} \supset \,
&i \frac{c_{W} s_{\xi}}{s_{W}}\Big[ \partial_{\mu} Z^D_{\nu}\left(W_{\mu}^{+} W_{\nu}^{-}-W_{\nu}^{+} W_{\mu}^{-}\right) \\
&+Z^D_{\nu}\left(-W_{\mu}^{+} \partial_{\nu} W_{\mu}^{-}+W_{\mu}^{-} \partial_{\nu} W_{\mu}^{+}+W_{\mu}^{+} \partial_{\mu} W_{\nu}^{-}-W_{\mu}^{-} \partial_{\mu} W_{\nu}^{+} \right) \Big],
\end{split}
\end{equation}
and the rate is~\cite{Bhattacharya:1988tw}
\begin{equation}
\Gamma_{\gamma_{D} \rightarrow W^+ W^-  } = \frac{\alpha s_{\xi}^2 c_{W}^2 \, m_{\gamma_D}}{48 s_{W}^2} \, \frac{m_{\gamma_{D}}^4+20m_{\gamma_{D}}^2 m_W^2+12m_W^4}{m_W^4} \Bigg(1-4 \frac{m_W^2}{m_{\gamma_{D}}^2} \Bigg)^{\frac{3}{2}}.
\end{equation}

\end{document}